%% file: main.tex
\documentclass[
11pt, 
oneside, 
english, 
singlespacing, 
headsepline, 
]{MastersDoctoralThesis} 

\usepackage[utf8]{inputenc} 
\usepackage[T1]{fontenc} 

\usepackage{mathpazo} 
\usepackage{amsmath} 
\usepackage{graphicx} 
\usepackage{xcolor}
\usepackage{url}
\usepackage{hyperref}

\usepackage{listings} 
\usepackage[autostyle=true]{csquotes} 

\newcommand{\harpoon}{\overset{\rightharpoonup}} 

\definecolor{x11gray}{rgb}{0.9, 0.9, 0.9}

\thesistitle{Real-Time Global Illumination Using OpenGL And Voxel Cone Tracing} 
\supervisor{Prof. Wolfgang \textsc{Mulzer}} 
\degree{Bachelor of Science} 
\author{Benjamin \textsc{Kahl}} 

\university{\href{https://www.fu-berlin.de/}{Freie Universität Berlin}} 
\department{\href{https://www.mi.fu-berlin.de/en/index.html}{Department of Mathematics and Computer Science}} 
\group{\href{http://researchgroup.university.com}{Research Group Name}} 
\faculty{\href{https://www.mi.fu-berlin.de/en/inf/index.html}{Institute of Computer Science}} 

\AtBeginDocument{
\hypersetup{pdftitle=\ttitle} 
\hypersetup{pdfauthor=\authorname} 
}

\begin{document}

\frontmatter 

\pagestyle{plain} 


\begin{titlepage}
\begin{center}

\vspace*{.06\textheight}
{\scshape\LARGE \univname\par}\vspace{1.5cm} 
\textsc{\Large Bachelor's Thesis}\\[0.5cm] 

\HRule \\[0.4cm] 
{\huge \bfseries \ttitle\par}\vspace{0.4cm} 
\HRule \\[1.5cm] 
 
\begin{minipage}[t]{0.4\textwidth}
\begin{flushleft} \large
\emph{Author:}\\
\href{https://github.com/Helliaca}{\authorname} 
\end{flushleft}
\end{minipage}
\begin{minipage}[t]{0.4\textwidth}
\begin{flushright} \large
\emph{Supervisor:} \\
\href{http://page.mi.fu-berlin.de/mulzer/}{\supname} 
\end{flushright}
\end{minipage}\\[3cm]
 
\vfill

\large \textit{A \textbf{revised and corrected} version of a thesis submitted in fulfillment of the requirements\\ for the degree of \degreename}\\[0.3cm] 
\textit{in the}\\[0.4cm]
\deptname\\[0.4cm] 
\textit{on} {\large February 19, 2019}
 
\vfill

{\large \today}\\[0cm] 
 
\vfill
\end{center}
\end{titlepage}

\begin{abstract}
\addchaptertocentry{\abstractname} 
Building systems capable of replicating global illumination models with interactive frame-rates has long been one of the toughest conundrums facing computer graphics researchers.

Voxel Cone Tracing, as proposed by Cyril Crassin et al. in 2011, makes use of mipmapped 3D textures containing a voxelized representation of an environments direct light component to trace diffuse, specular and occlusion cones in linear time to extrapolate a surface fragments indirect light emitted towards a given photo-receptor.

Seemingly providing a well-disposed balance between performance and physical fidelity, this thesis examines the algorithms theoretical side on the basis of the rendering equation as well as its practical side in the context of a self-implemented, OpenGL-based variant.

Whether if it can compete with long standing alternatives such as radiosity and raytracing will be determined in the subsequent evaluation.
\end{abstract}


\tableofcontents 


\mainmatter 

\pagestyle{thesis} 


\include{Chapters/Chapter1}
\include{Chapters/Chapter2} 
\include{Chapters/Chapter3}
\include{Chapters/Chapter4} 
\include{Chapters/Chapter5} 
\include{Chapters/Chapter6}






\nocite{c1}
\nocite{c2}
\nocite{c3}
\nocite{c4}
\nocite{c5}
\nocite{c6}
\nocite{c7}
\bibliographystyle{siam}
\bibliography{biblio}


\end{document}

%% file: Chapters/Chapter1.tex

\chapter{Preface} 

\label{Chapter1} 


\newcommand{\keyword}[1]{\textbf{#1}}
\newcommand{\tabhead}[1]{\textbf{#1}}
\newcommand{\code}[1]{\texttt{#1}}
\newcommand{\file}[1]{\texttt{\bfseries#1}}
\newcommand{\option}[1]{\texttt{\itshape#1}}


\section{Introduction}
The last decades have brought forth an ever-increasing need of photorealistic image synthesis within various fields such as virtual reality (VR), visual effects (VFX) and video games.
The Rendering Equation, formulated by James Kajiya in 1989, describes a generalized mathematical model that has served as an underlying basis for a wide range of shader-based rendering algorithms such as Ray Tracing and Photon Mapping.

The challenge in solving the rendering equation through approximation originates from the plethora of light-based phenomena that can occur in the real world, such as caustics, light-scattering or refraction. The particular phenomenon grappled in this thesis is {\it{indirect light}}.

Going mostly unnoticed in people's everyday life, indirect light surrounds every aspect of environment illumination. Without it, all the parts of a room not facing a window would be in absolute darkness. Mirrors would simply display a blank, silver color with no reflection image and a simple umbrella would plunge the area underneath into a total twilight.


\section{Problem and Objective}
As outlined above, any photorealistic rendering algorithms must incorporate the vital component of indirect light.

Unfortunately, simulating light with such physical verity is computationally very expensive and cannot be done on present consumer-grade computers while maintaining interactive frame-rates.

Thus, the approach to any real-time global illumination algorithm is to employ a fair amount of reductionism in order to shed performance-weight wherever possible while maintaining a reasonable degree of real world fidelity.
\\

In 2011 {\it{voxel cone tracing}} was introduced by Cyril Crassin et al.\cite{Crassin} as a novel global illumination algorithm that avoids expensive precomputation and enables interactive frame-rates.

Generally heralded as the next big breakthrough in real-time rendering approaches, the technology made some impressive initial headway with {\it{Nvidia}} releasing the {\it{VXGI}} framework and {\it{Unreal Engine}} picking up on its trail. However, some eight years later, the industry still lacks major software applications making use of this approach, defaulting instead to far older methods such as {\it{radiosity}}. 

Having seemingly failed to meet its expectations this far, this thesis sets out to investigate these claims and find out whether if voxel cone tracing can indeed compete with its peers.

\begin{figure}[th]
\centering
\includegraphics[scale=0.2]{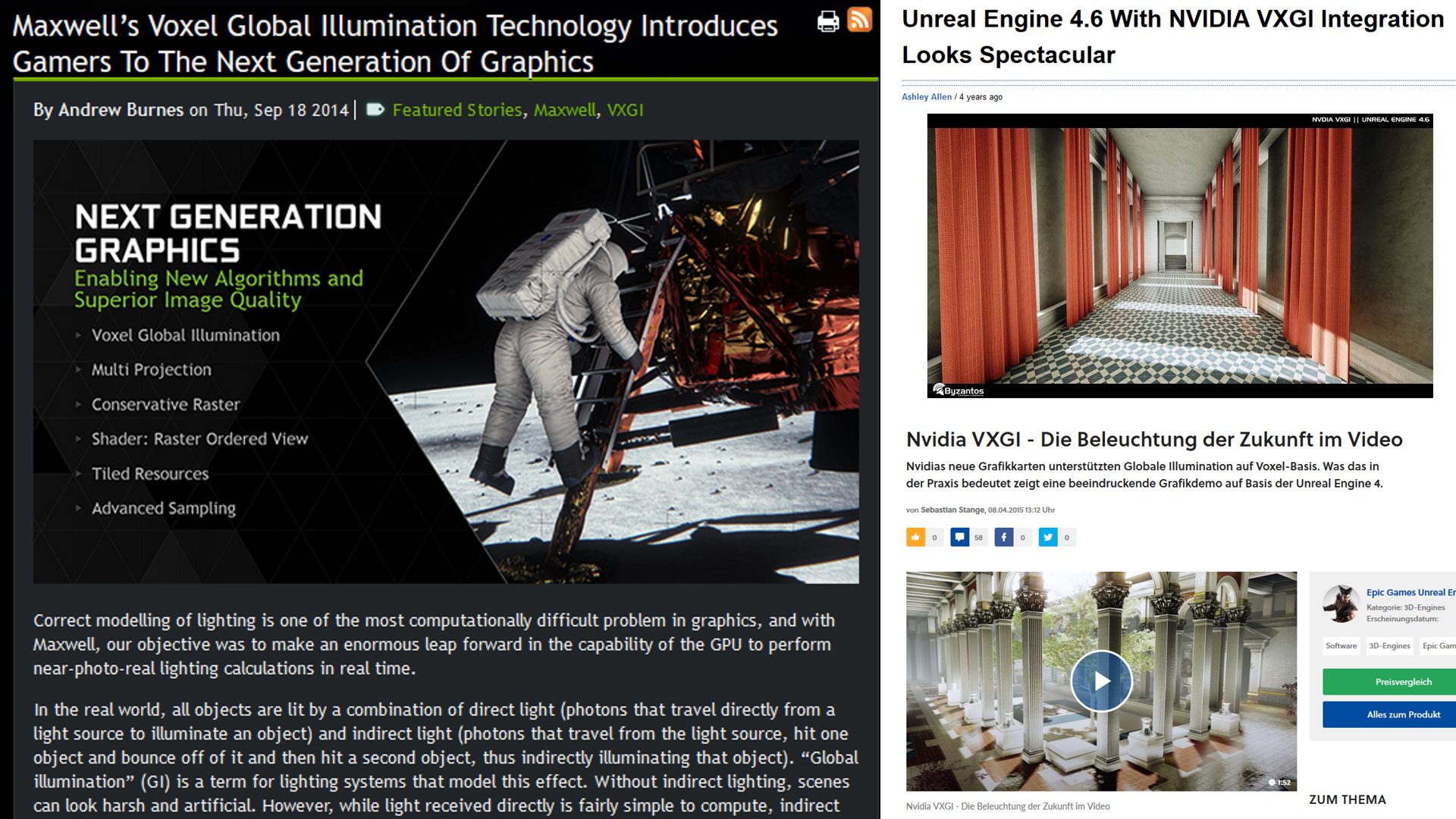}
\decoRule
\caption[Solid Angle]{Example articles and demos putting the advantages of voxel cone tracing on display}
\end{figure}

\section{Thesis Structure}

This chapter has provided a rough outline of the problems surrounding global illumination models. In turn, chapter \ref{Chapter2} will purvey a brief introduction into the field of computer graphics as well as establish a well-defined, mathematical basis for the objective in question. As will be further explained, the goal in mind will be to approximate the rendering equation.

Subsequently, chapter \ref{Chapter3} will describe the necessary context of the proposed solution. The working patterns of OpenGL will, in essence, be the tools available to achieve the objective defined in chapter \ref{Chapter2}.

With the available tools and environment as well as the posed problem defined, chapter \ref{Chapter4} will introduce the solution proposed by Crassin et al.\cite{Crassin} on a theoretical basis.

Afterwards, chapter \ref{Chapter5} will investigate how the proposed theory translates into practice by presenting a written implementation of the algorithm.

%% file: Chapters/Chapter2.tex

\chapter{Introduction} 

\label{Chapter2} 


Given a specification of a scene, a classical image-synthetization process computes how light scattered into this environment translates into pixel colors on a given retina.

Otherwise functioning similarly to the bitmap rasterization of scalable vector graphics (SVG), this process is subject to a vast range of additional intrinsics such as the physical properties of the materials encountered as well as the geometric arrangements of the objects in question.

In reality, the color an item adopts on a camera's retina can be traced back to the spectral absorption of certain wavelengths by the surface's material in relation to the wavelength distribution emitted by any present {\it{light sources}} in the first place. However, the largest complexity hurdles that have long confronted computer graphics researches specifically stem from the broadened definition of what makes a light source.

Different materials can reflect light in a variety of manners and, due to interreflection, every instance of reflected light can qualify as a separate light-source itself, further impacting any surfaces visible to it.

The immense complexity of the task implies that the rasterization of a retina's captured radiation cannot be computed linearly and instead need to be approximated through the estimation of a scene's light distribution.

Thus, the problems of realistic image rendering are inextricably linked to those posed by the simulation of light propagation in an environment. The physical accuracy of the simulation techniques employed in turn determine the realism of the resulting raster, but also heavily affect the required computation time.

The contents of this chapter aim to provide a deeper understanding of the physical phenomena involved in light-surface interactions and how these are mathematically modeled in the field of computer graphics.


\section{Basic Optics}

Visible light is a sub-spectrum of electromagnetic radiation that is visible to the human eye and is thus the only spectrum relevant to the domain of realistic rendering.

In the field of quantum optics, light is modelled as a series of discrete bundles, called {\it{photons}}, which carry electromagnetic energy proportional to the radiations wavelength by

\begin{equation}
E = hf = \frac{hc}{\lambda}
\end{equation}

where {\it{h}} is Planck's constant, $\lambda$ is the wavelength and {\it{c}} is the speed of light.

Most light sources encountered in the natural world produce a wide range of different wavelengths that propagate through the environment incoherently. \cite{radiosity}

Within the field of computer graphics, the limitless possible combinations of wavelengths need to be mapped onto a finite amount of numerical color values. The most frequently used approach is the 24-bit {\it{True Color}} format, which provides eight bits of color depth for each component of an RGB-triplet. In many cases an additional alpha-value component is included to represent the opacity of the pixel color. The same format will be employed  throughout this thesis.

\subsection{Fundamental Radiometric Quantities}

The physics domain most closely related to computer graphics is the field of {\it{geometrical optics}} as it heavily focuses on predicting the macroscopic propagation behaviour of electromagnetic radiation. It is commonly subdivided into the sciences of {\it{radiometry}} and {\it{photometry}}. Radiometry concerns itself with the raw, physical measurement of electromagnetic energy, while photometry focuses more on the  human based perception of the visible spectrum.

Unlike in fields where recorded images are directly handled by algorithms, such as the field of {\it{robotics}}, the here computationally synthesized images are conceivably meant to undergo human inspection. For this reason, differing wavelength sensitivities of human eye photoreceptors do not need to be individually accounted for, making the field of radiometry the ideal building brick for physically faithful rendering algorithms.

\subsubsection{Radiant Flux or Radiant Power}
The total amount of radiant energy emitted per unit time is described in terms of {\it{radiant flux}}, or power, $\Phi$ and is usually quantified in units of Watts or Joules per second:

\begin{equation}
\Phi = \frac{\partial Q}{\partial t} [W]
\end{equation}

where $Q$ is the radiant energy emitted, transmitted or reflected.

\subsubsection{Irradiance or Illuminance}

The total flux incident on a surface per unit surface area is defined as {\it{irradiance}} $E$. The respective photometric quantity is called {\it{illuminance}}.

\begin{equation}
E = \frac{d\Phi_i}{dA} [W/m^2]
\label{eqn:Irradiance}
\end{equation}

\subsubsection{Radiosity or Radiant Exitance}

Contrary to irradiance, the {\it{radiant exitance}} $B$ is defined as the total flux per surface area {\it{leaving}} or being emanated from a surface. 

\begin{equation}
B = \frac{d\Phi_e}{dA} [W/m^2]
\label{eqn:Radiosity}
\end{equation}

\subsection{Solid Angle}\label{SolidAngle}

When handling differential incident flux values on surface points, it is useful to define a measurement for the amount of {\it{field of view}} from that particular point in a given direction.

Similarly to how a two-dimensional angle is proportional to the length of the arc it covers on a unit circle, a {\it{solid angle}}  (measured in  {\it{steradians}}) is equal to a corresponding surface area intersected on a unit sphere centered around the point of origin. As a result, a solid angle value can be a simple way of quantifying how large, or bright, an object looks to a given observer.

The direction of the solid angles center is indicated by a three-dimensional vector. This representation can be simplified by utilizing only unit vectors and thus regarding them as points on a unit sphere. Any location on a unit sphere can, in turn, be represented by a pair angles:

$\theta$ corresponds to the angle between said vector and the coordinate systems upwards axis (or north pole) and $\phi$ corresponds to the vector's angle of rotation around the upwards axis itself.

\begin{figure}[th]
\centering
\includegraphics[scale=0.5]{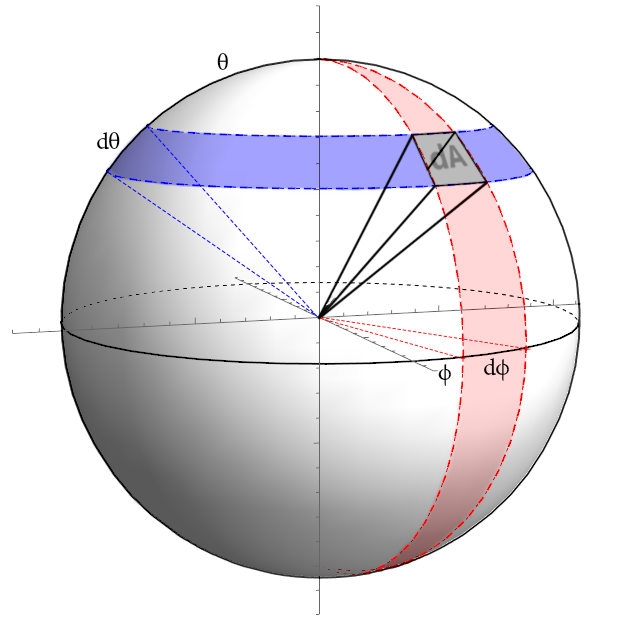}
\decoRule
\caption[Solid Angle]{Solid angle on a unit sphere}
\label{fig:SolidAngle}
\end{figure}

By defining $d\theta$ and $d\phi$ as the differential latitudinal and longitudinal angles, as seen in fig. \ref{fig:SolidAngle}, we can calculate the differential surface element $dA$ on a sphere with a radius $r$ intersected by a solid angle $(\phi, \theta, d\theta, d\phi)$ as follows:

\begin{equation}
dA = r^2\sin (\theta) * d\theta * d\phi
\label{eqn:SolidAngleSurface}
\end{equation}

Analogously to a circle having a circumference of $2\pi r$ and subtending an angle of $2\pi$ radians, a sphere has a surface area of $4\pi r^2$ and thus subtends a total solid angle of $4\pi$ steradians. 

As a result, a differential solid angle $d\omega$ corresponds to

\begin{equation}
d\omega = \frac{dA}{r^2} = \sin (\theta) * d\theta * d\phi
\label{eqn:SolidAngle}
\end{equation}

The field of view that an arbitrarily rotated surface $dA$ at point $x'$ subtends to an observer at point $x$ corresponds to the area that surface projects onto a unit sphere around $x$.

Let $\harpoon n$ be the normal vector of the surface in question and $\gamma$ be the angle between $\harpoon n$ and the vector $\harpoon r$ pointing from $x$ to $x'$. 

\begin{figure}[th]
\centering
\includegraphics[scale=0.35]{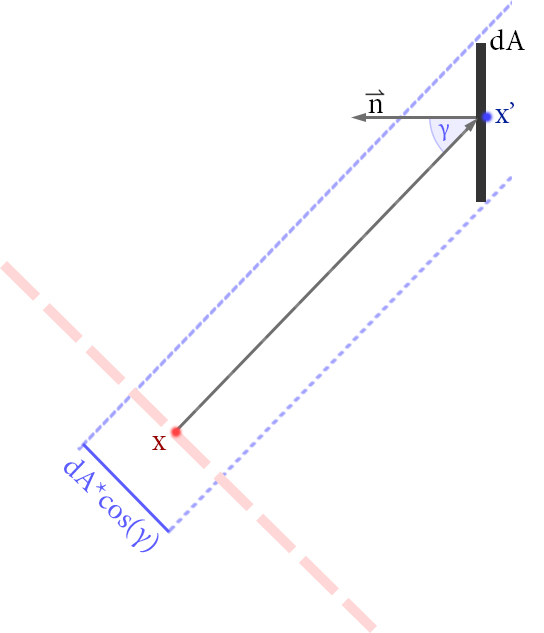}
\decoRule
\caption[Solid Angle]{Projected surface area $dA$ at point $x'$ onto a plane perpendicular to $x'-x$}
\label{sa_proj_area}
\end{figure}

As can be seen in fig. \ref{sa_proj_area}, projecting the surface onto a plane that is perpendicular to $\harpoon r$ results in a total projected surface area equal to $\cos(\gamma)*dA$.

\begin{figure}[th]
\centering
\includegraphics[scale=0.4]{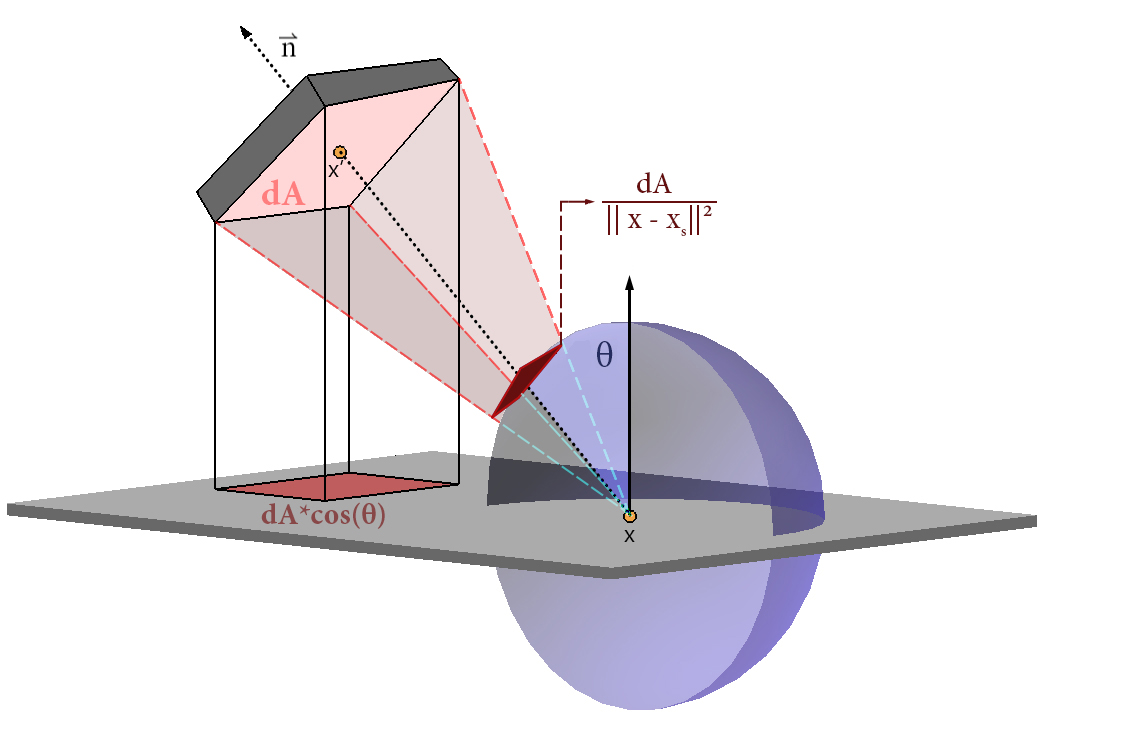}
\decoRule
\caption[Solid Angle]{Projecting a surface onto a plane results in a projected surface area of $dA\cos \theta$. Projecting a perpendicular surface towards the center of a sphere results in a projected surface area on the sphere of $dA / \left||x-x'\right||^2$.}
\label{solid_angle_adv}
\end{figure}

If the upwards axis of a solid angle is equated with$\harpoon r$, then $\theta$ is becomes equal to $\gamma$, as can be observed in the downward projection portrayed in fig. \ref{solid_angle_adv}.

The second projection illustrated in fig. \ref{solid_angle_adv} (from $x'$ towards $x$) demonstrates the change in area when a perpendicular surface is projected along a field of view frustum.

In fact, the surface $dA'$ a surface $dA$ perpendicular to $\harpoon r$ projects onto a unit sphere surrounding $x$, corresponds to the original surface area divided by the distance squared:

\begin{equation}\label{solid_angle_perp_proj}
dA' = \frac{dA}{\left |  x'-x \right |^2}
\end{equation}

The distance factor scales by an exponent of 2, due to the {\it{inverse-square law}}, which will be discussed in \ref{inverse_square}.

Combining the results from fig. \ref{sa_proj_area} with the ones obtained in equation (\ref{solid_angle_perp_proj}) allows for the calculation of the solid angle that encompasses an arbitrarily rotated surface at $x'$ from point $x$:

\begin{equation}\label{SolidAngleProjectionRelation}
\omega = \frac{\cos(\theta) dA}{\left |  x'-x \right |^2}
\end{equation}

\subsection{Radiance}

Radiance describes the volumetric radiant energy and is a fundamental quantity to describe light flow in an environment.
Given the volume density of photons $p(x, \harpoon\omega, \lambda)$ with a wavelength $\lambda$ at position $x$ that are travelling in direction $\harpoon\omega$, the radiance at point $x$ propagating in direction $\harpoon \omega$ equates to the product of said photon density and the energy of a single photon $\frac{hc}{\lambda}$.

\begin{equation}
L(x, \harpoon \omega) = \int_{\lambda } p(x, \harpoon \omega, \lambda) \frac{hc}{\lambda}
\label{eqn:RadiancePhotons}
\end{equation}

When dealing with problems linked to computer graphics, the quantum nature of light tends to be discarded, as most rendering methods utilize quantities relating to radiant power rather than number of photons.

Given its definition, radiance can thus also be written as the radiant power entering or exiting a surface per unit solid angle per unit projected area.

\begin{equation}
L = \frac{d\Phi}{d\omega dA_{\perp}} = \frac{d\Phi}{d\omega dA\cos \theta}
\label{eqn:Radiance}
\end{equation}

The quantity of radiance is of particular importance within this field as it can indicate how much of the radiant power a surface emits is received by a camera pointing towards it and, by extension, how bright the surface appears.

In this case, $\omega$ is the solid angle correspondent with the camera's field of view, $A_{\perp}$ is the surface area perpendicular to the camera's forward vector and $\Phi$ is the total flux emitted by said surface.

Furthermore, the in (\ref{eqn:Irradiance}) given definition for irradiance can be expanded by integrating the solid angle over the hemisphere of directions $\Omega$ above the given surface:

\begin{equation}
E = \frac{d\Phi}{dA} = \cos \theta * d\omega * L_i = \int_{\Omega} L(\omega_i) * \cos \theta * d\omega
\label{eqn:IrradianceExpanded}
\end{equation}

\subsection{Types of Lightsources}\label{lightsource_types}

The emanation of light can occur in different manners depending on the light sources' geometrical properties.

\begin{figure}[th]
\centering
\includegraphics[scale=0.185]{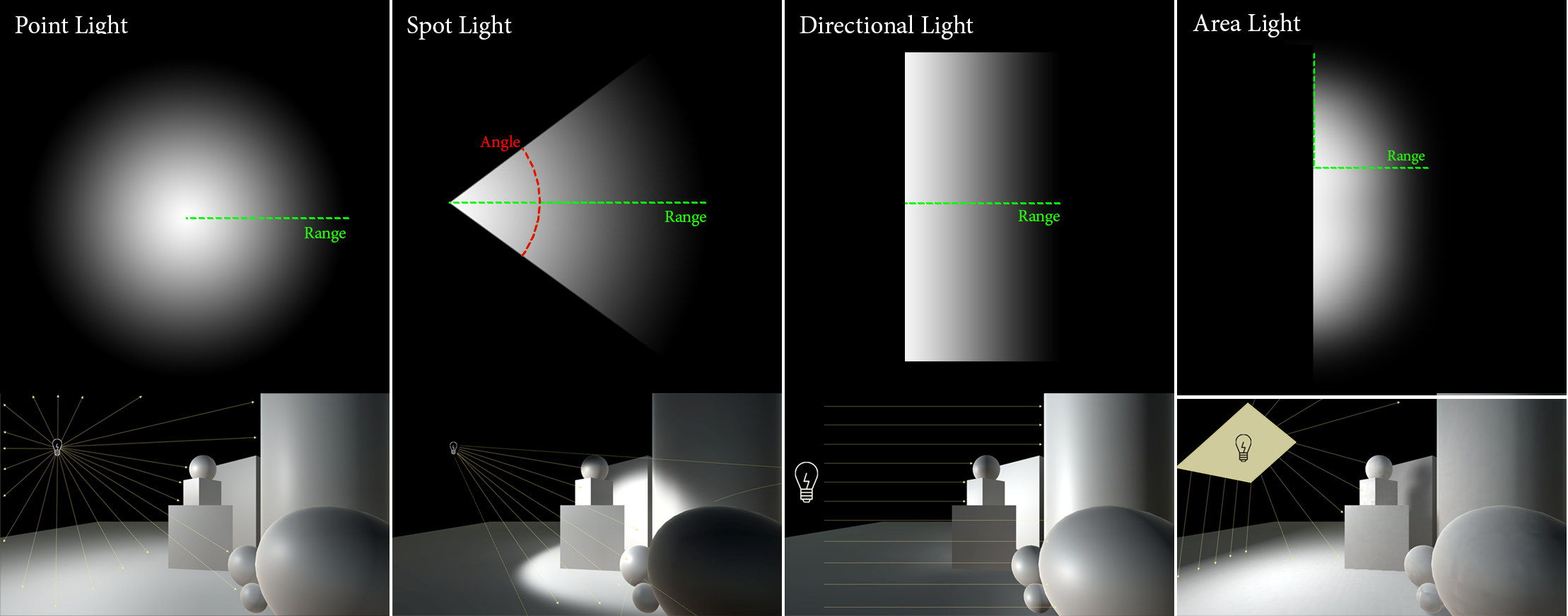}
\decoRule
\caption[LightSourceTypes]{Types of lightsources, as seen in the Unity 5.6.1 Engine}
\label{lightsources_pic}
\end{figure}

The field of computer graphics typically differentiates between four categories of light sources which are described below as well as depicted in fig. \ref{lightsources_pic}:

\begin{itemize}
    \item \textbf{Point Light}\\
    A point light occupies an infinitesimal volume in space from which it scatters light outwards isotropically.
    
    In most 3D simulations, point lights are used as light sources that only illuminate a limited volume surrounding the source. For this purpose, the intensity of its radiance is modelled to diminish with distance. The exact nature of this decay in intensity will be defined in \ref{Attenuation}.
    
    The effects of localized light emitters such as incandescent light bulbs or candles are easily simulated using point lights.
    \item \textbf{Spot Light}\\
    A spot light behaves similarly to a point light, in that it has a specified location-vector and fall-off distance. The difference lies within the directional and angular constraints imposed upon a spot light.
    
    These result in a cone-shaped volume of illumination that can be used to simulate the behaviour of conic light sources such as flashlights or headlights.
    \item \textbf{Directional Light}\\
    Directional light sources cast parallel light-rays uniformly along a specified direction vector.
    Implementations of this model commonly relinquish the positional attribute, meaning that the light source does not have an identifiable source position. In this view, the distance between the light source and target becomes undefined, meaning that the radiance does not diminish over distance.
    
    The above given properties make directional lights a simple, yet convenient method of accurately modelling distant light sources such as the sun, given how the light rays originating from these are very close to parallel to each other when hitting objects on planet Earth's surface.
    \item \textbf{Area Light}\\
    Area lights are one-sided, polygon-shaped areas that consist of an emissive material that sends out light in all directions uniformly across the surface. The resulting effect resembles the one produced by a multitude of spotlights scattered around the surface with an aperture angle of 180 degrees and a directional vector corresponding to the surface normal of the area light.
    
    Examples of rectangular area lights are LCD-screens or LED-billboards.
\end{itemize}

\subsection{Attenuation}\label{Attenuation}

The reduction of light-intensity over distance is generally termed {\it{light attenuation}} or {\it{fall-off}}.
In physics, the gradual loss of flux intensity due to geometric dilution is given by the {\it{inverse-square law}}.

\subsubsection{Inverse-square law}\label{inverse_square}

In the real world, a light's brightness appears to diminish rapidly for close ranges and far slower for further distances.
This phenomenon is an observable effect of the {\it{inverse-square law}} which, when applied to light sources, dictates that the light intensity at any given location produced by a point light is inversely proportional to square of the distance between the two.

The justification for the given inverse square proportionality results from the equivalently proportional surface area of a sphere in relation to its radius ($4\pi r^2$). Flux emanating from a point light spreads out across a spherical surface, hence why the light intensity at any given point is derivable through the inverse-square law.

\begin{figure}[th]
\centering
\includegraphics[scale=0.25]{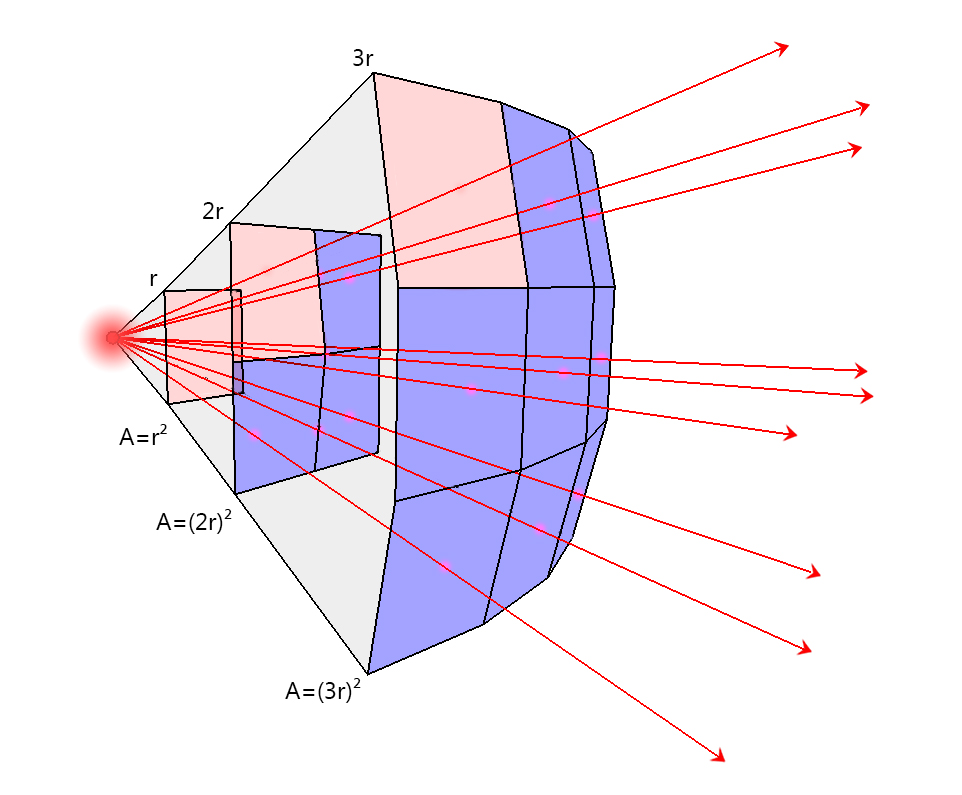}
\decoRule
\caption[]{Inverse-square law}
\end{figure}

Applying this law to the definition of irradiance given in (\ref{eqn:Irradiance}), allow the computation of the irradiance of a differential surface at $x$ caused by a single point light at $x_s$. 

To accomplish this, the total flux $\Phi$ is multiplied with the solid angle fraction of flux heading towards said differential surface. This fraction equates to the solid angle $d\omega$ it occupies from the point of view of the lightsource divided by the total amount of solid angle the flux is emanated into ($4\pi$).

In accordance with the definition of irradiance, the resulting value is divided by the differential surface area:

\begin{equation}
E =\Phi * \frac{d\omega}{4\pi} * \frac{1}{dA}
\end{equation}

With use of the inverse square law, $d\omega$ can now be substituted by term from (\ref{SolidAngleProjectionRelation}), resulting in the following equation:

\begin{equation}
E = \frac{\Phi}{4\pi} * \frac{\cos \theta}{\left | x-x_s \right |^2}
\label{eqn:IrradianceFalloff}
\end{equation}

\subsubsection{Constant-Linear-Quadratic Falloff}
While the inverse-square law applies for real-world geometric dilution, the fast rate of intensity-diminishment can often lead to unnatural glares within virtual environments.

To provide an improved degree of adjustability to the desired effects of a light source, it is common practice to define three configurable constants $K_c$, $K_l$ and $K_q$, which set the proportions of constant, linear and quadratic fall-off respectively.

Given a distance value $d = \left | x-x_s \right |$, the final attenuation factor would result from a combination of all three constants:

\begin{equation}
Att = \frac{1}{K_c+K_l*d+K_q*d^2}
\end{equation}

In a blend of quadratic and linear attenuation, the light intensity will decay in a mostly linear manner until the distance becomes large enough for the quadratic effect to supersede, leading to a more rapid decrease in intensity.

\begin{figure}[!htb]
\minipage{0.32\textwidth}
  \includegraphics[width=\linewidth]{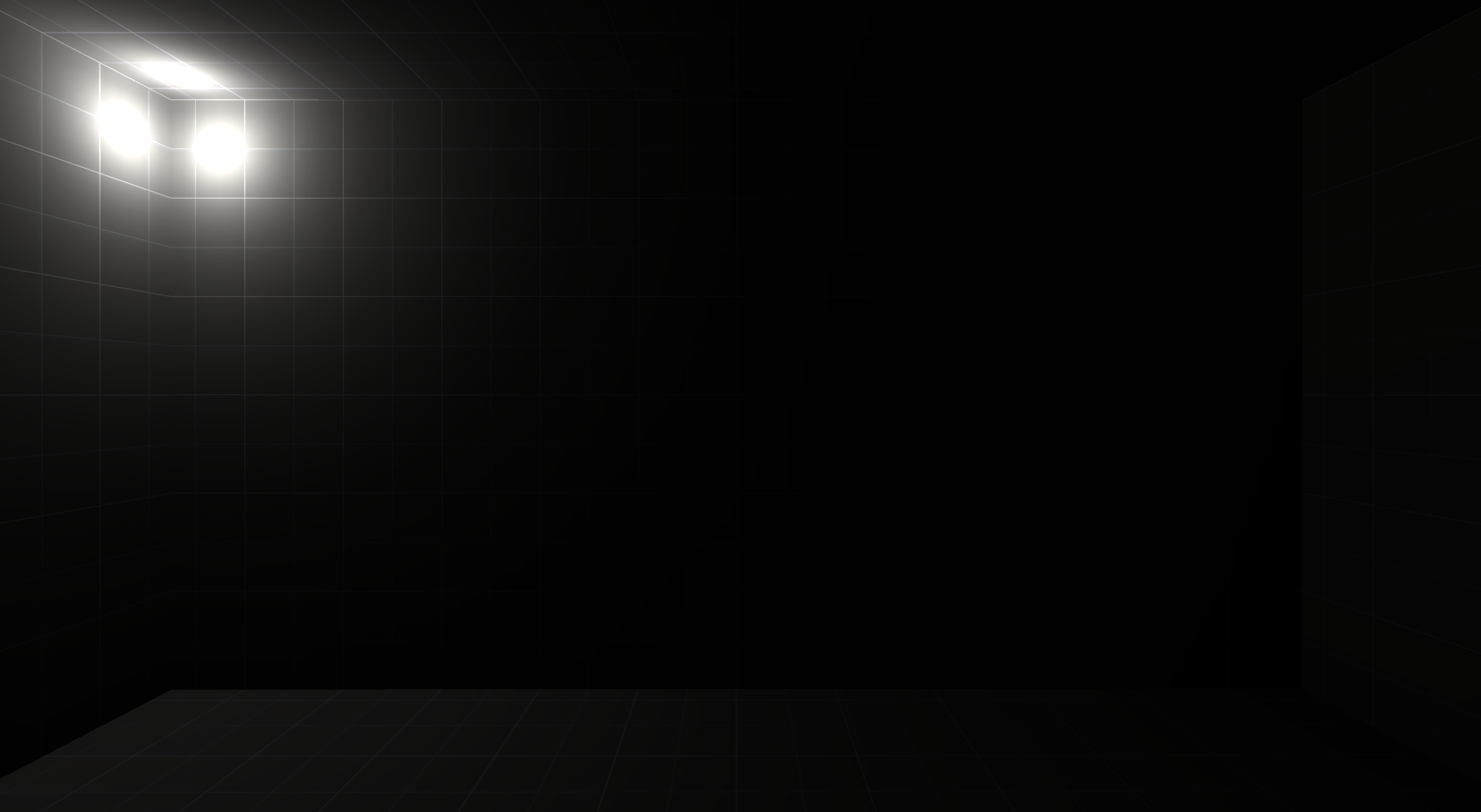}
\endminipage\hfill
\minipage{0.32\textwidth}
  \includegraphics[width=\linewidth]{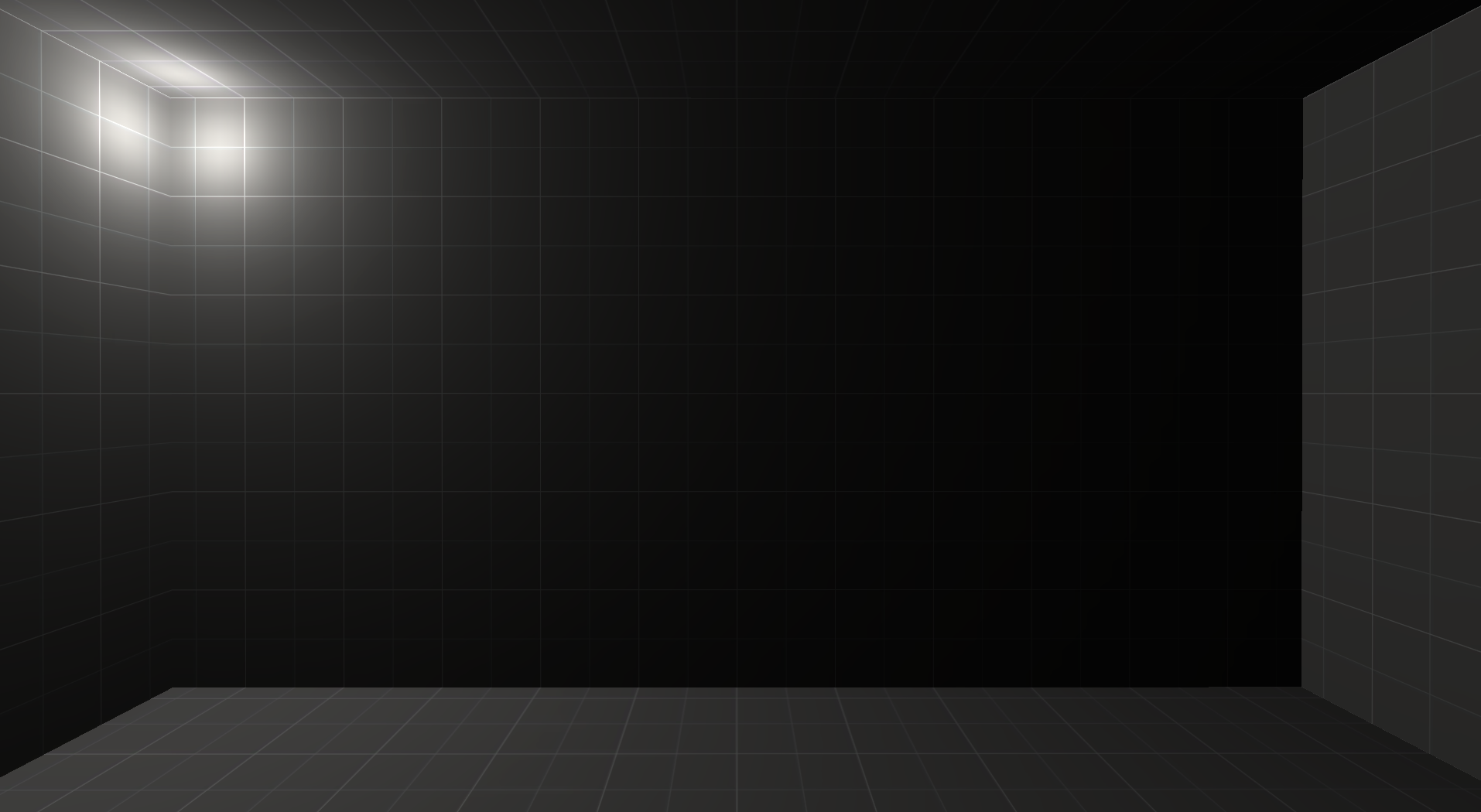}
\endminipage\hfill
\minipage{0.32\textwidth}%
  \includegraphics[width=\linewidth]{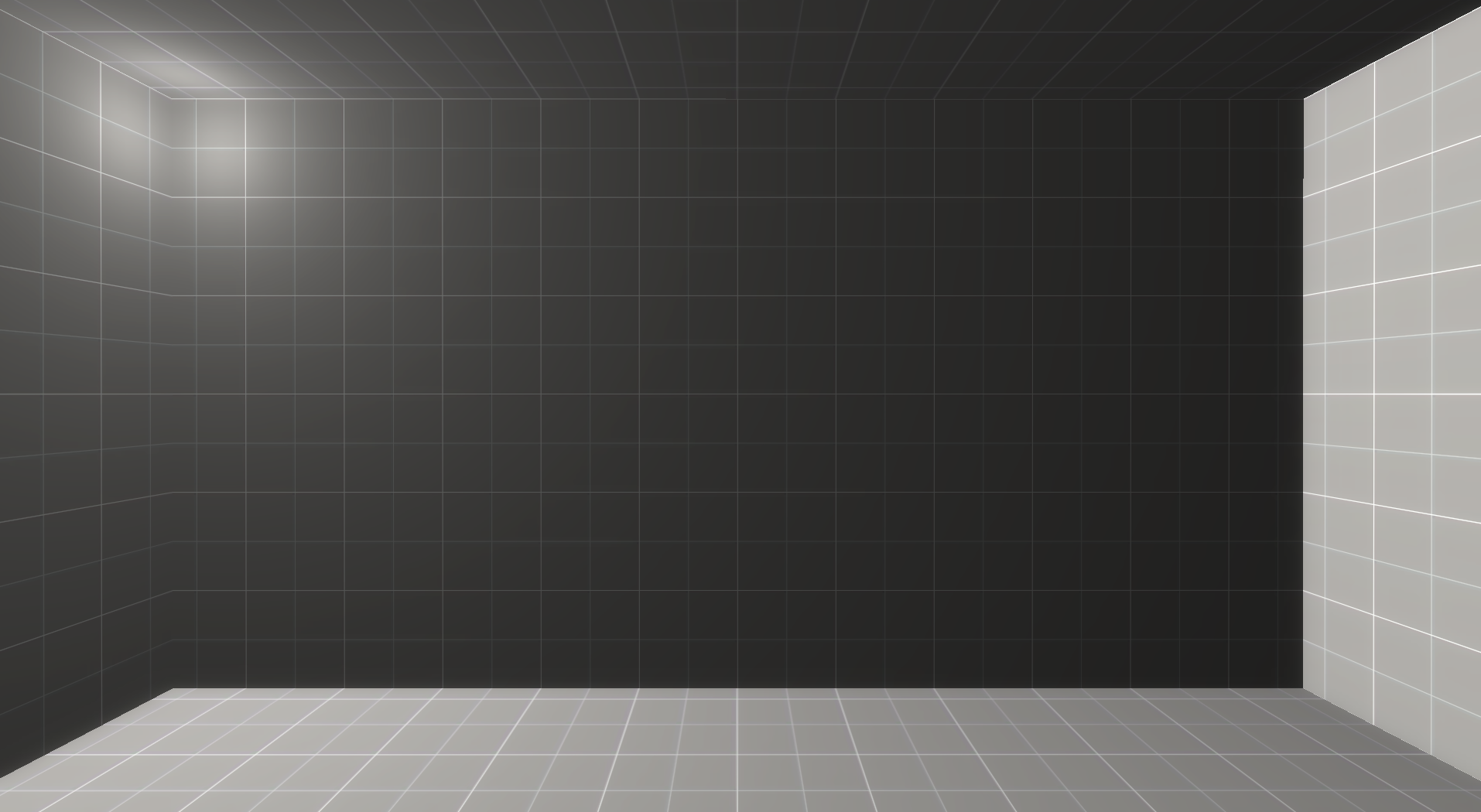}
\endminipage
\caption{From left to right: Quadratic, linear and constant attenuation as seen in the 2009 Source Engine}
\end{figure}
\pagebreak

\section{Reflection Modelling}

\subsection{Reflectance Components}\label{reflectanceComponents}

A reflection occurs when light changes its direction after hitting a surface.

Depending on the interface-materials' physical properties, reflections can happen either specularly or diffusely.
In specular reflections, the incident light's trajectory vector is mirrored along the  normal of the surface hit.
On the contrary, diffuse reflections will relinquish the original image, meaning that the direction of diffusely reflected light is independent of the light's origin.

However, real materials rarely possess a surface finish smooth enough to reflect light in a perfectly specular manner. The reflections produced by non-planar materials with rough surfaces do not translate directly into a single reflection ray, but instead a complex distribution of outgoing rays. 
This intermediate variant between specular and diffuse is often called a {\it{glossy}} reflection.

To portray the aggregate reflected light of a given surface element, the tendency in computer graphics is to sum up the three individually computed components (diffuse, specular and glossy) in various proportions to one another which depend on the material's physical properties.

\begin{figure}[th]
\centering
\includegraphics[scale=0.5]{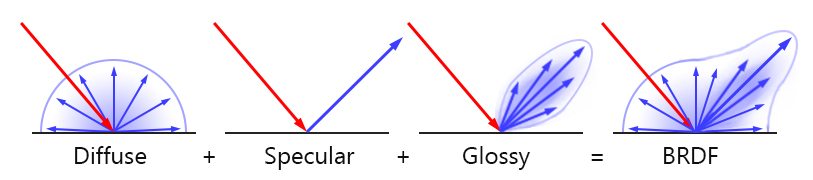}
\decoRule
\caption[]{Reflectance components as portrayed by Cohen et al.\cite{radiosity}}
\end{figure}

The characteristics of specific reflection types are modelled by a so-called bidirectional reflectance distribution function.

\subsection{Bidirectional Reflectance Distribution Function}\label{brdf}

For any given pair of small, differential solid angles $\harpoon \omega_i$ and $\harpoon \omega_r$, the radiance reflected into $\harpoon \omega_r$ is proportional to the irradiance incident from $\harpoon \omega_i$.

\begin{equation}
dL_r(\harpoon \omega_r) \propto dE(\harpoon \omega_i)
\end{equation}

The coefficient of proportionality $f$ in dependence of $\harpoon \omega_i$ and $\harpoon \omega_r$ is called a {\it{bidirectional reflectance distribution function}} (BRDF).

\begin{equation}
f(\harpoon\omega_i \rightarrow \harpoon\omega_r) = \frac{L_r(\harpoon\omega_r)}{E(\harpoon\omega_i)} = \frac{L_r(\harpoon\omega_r)}{L_i(\harpoon\omega_i) \cos \theta d\omega_i}
\end{equation}

It provides the ratio of flux concentration per steradian reflected into a given direction in consequence of the differential irradiance incident from another given direction.

Solving this equation for $L_r$ and performing the same hemispherical integral over the set of all incident directions $\Omega$ as in (\ref{eqn:IrradianceExpanded}) yields the total amount of light reflected by a surface in a specified direction. This formula is also known as the {\it{reflectance equation}}:

\begin{equation}
\begin{aligned}
& f(\harpoon \omega_i \rightarrow \harpoon \omega_r) = \frac{L_r(\harpoon\omega_r)}{E(\harpoon\omega_i)} \\
\Leftrightarrow   & L_r(\harpoon\omega_r) = \int_{\Omega_i} f(\harpoon\omega_i \rightarrow \harpoon\omega_r) L_i(\harpoon \omega_i) \cos \theta_i d\omega_i
\end{aligned}
\end{equation}

In summary, the radiance $L_r$ reflected in a particular direction $\harpoon\omega_r$ equates to the BRDF-weighted radiance incident from all directions above the surface.

\subsubsection{Specular Reflections}\label{SpecularReflection}

In the case of a perfectly specular reflection, the outgoing radiance is not scattered, but merely mirrors its trajectory across the normal vector of the surface hit.

In terms of solid angles as defined in \ref{SolidAngle}, the angle between the surface normal vector $\theta$ remains identical after a reflection has occurred, while the angle of rotation around the surface normal $\phi$ is flipped by 180 degrees.

\begin{equation}
\begin{aligned}
& \theta_r = \theta_i \\
& \phi_r = \phi_i \pm \pi \\
& \rightarrow L_r(\theta_r, \phi_r) = L_i(\theta_r, \phi_r \pm \pi)
\end{aligned}
\end{equation}

This characteristic can be modelled with a dirac delta function:

\begin{equation}
f_{r,m} = \frac{\delta(\cos \theta_i - \cos \theta_r)}{\cos \theta_i} \delta(\phi_i - (\phi_r \pm \pi))
\end{equation}

Since the delta function yields zero for any non-zero parameter, the given BRDF yields zero (no light reflected) for all cases in which the conditions listed above do not apply.

\subsubsection{Diffuse Reflections}\label{DiffuseReflection}

A Lambertian diffuse reflection is defined as a reflection in which the apparent brightness of a surface is independent from the observer's and the light source's position (disregarding attenuation). The outgoing radiance follows an isotropic distribution leading to an equal effluxing radiance in all directions, regardless of the incident angle.

\begin{equation}
\forall \alpha, \beta \quad L(x \rightarrow \alpha) = L(x \rightarrow \beta)
\end{equation}

This implies that the BRDF of a diffuse reflection yields the same value for all possible parameters. As a result, $f(\harpoon \omega_i \rightarrow \harpoon \omega_r)$ can be regarded as a constant and consequently separated from the integrand:

\begin{equation}
\begin{aligned}
L_r(\harpoon\omega_r) & = \int_{\Omega_i} f_r L_i(\harpoon \omega_i) \cos \theta_i d\omega_i \\
& = f_r \int_{\Omega_i} L_i(\harpoon \omega_i) \cos \theta_i d\omega_i \\
& = f_r E
\end{aligned}
\end{equation}

The BRDF constant $f_r$ can be intuitively parameterized by defining a material-dependent value $\rho$, which matches the fraction of irradiance reflected into radiosity by said material.

\begin{equation}
\rho = \frac{B}{E} = \frac{\int_{\Omega_r} L_r(\harpoon\omega_r)\cos \theta_r d\omega_r}{E}
\end{equation}

And since the reflected radiance $L_r$ is equal in all directions, the term can be simplified to the following:

\begin{equation}
\begin{aligned}
\rho & = \frac{\int_{\Omega_r} L_r(\harpoon\omega_r)\cos \theta_r d\omega_r}{E} \\
& = \frac{L_r \int_{\Omega_r} \cos \theta_r d\omega_r}{E} \\
& = \frac{L_r \pi }{E} \\
& = \pi f_r
\end{aligned}
\end{equation}

Which results in a reflectance equation of the form

\begin{equation}
L_r(\harpoon\omega_r) = \frac{\rho}{\pi} \int_{\Omega_i} L_i(\harpoon \omega_i) \cos \theta_i d\omega_i \\
\end{equation}

where $\frac{\rho}{\pi}$ represents the BRDF function of a lambertian diffuse reflection.

\subsubsection{Glossy Reflections}\label{GlossyReflection}

As laid out in \ref{reflectanceComponents}, a common approach to model specular reflections on rough surfaces, is to produce a wider distribution of rays around the direction of a perfect  mirror reflection.

For this purpose, a broader version of the delta function used in \ref{SpecularReflection} can be expressed by taking the dot product of two normalized vectors and raising it to some high exponent. In this case, we are regarding the outgoing vector of a perfect mirror-reflection $\harpoon R$ and the vector pointing from the surface to the retina ${\harpoon\omega_r}$. 

If the dot product is negative, the reflection angle is too large to consider for glossy reflections, and a value of 0 will be adopted instead. The result is then raised to the power of $\alpha$, which determines the narrowness of the reflection and is termed the {\it{shininess}} of the material.

\begin{equation}
L_r(\harpoon\omega_r) = L_{i}d\omega_{i}(\max(\harpoon R\cdot\harpoon\omega_r), 0)^{\alpha}
\end{equation}

Note that in this instance, ${\harpoon\omega_r}$ is taken as a regular directional vector, rather than as a solid angle.

The result of $\max(\harpoon R\cdot\harpoon\omega_r), 0)$ will increase if $\harpoon R$ and ${\harpoon\omega_r}$ are of similar angle to the surface normal, with $\alpha$ providing a pseudo-delta function of adjustable narrowness.

\begin{figure}[th]
\centering
\includegraphics[scale=0.65]{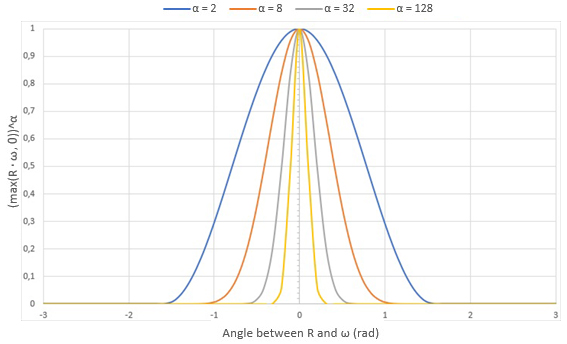}
\decoRule
\caption[]{Glossy BRDF for different shininess ($\alpha$) values}
\label{alpha_graph}
\end{figure}

The effects produced by different $\alpha$ values can be seen in fig. \ref{alpha_graph}.

\section{Rendering Equation}

The reflectance equation provides the reflected light distribution as a contingency of a specified incident light distribution.
By computing said incident light and thereafter applying it to the reflectance equation, a general description of radiance distribution in a scene can be defined.

Introduced by Kajiya\cite{Kajiya} in 1986, the {\it{rendering equation}} generalizes the rendering problem in an integral equation well-suited for computer graphics applications. 

With it, the outgoing radiance at point $x$ in direction $\harpoon\omega$ can be expressed as the following recursive term:

\begin{equation}\label{RenderingEquation_general}
\begin{aligned}
L_o(x, \harpoon\omega) & = L_e(x, \harpoon\omega) + L_r(x, \harpoon\omega) \\
& = L_e(x, \harpoon\omega) + \int_{\Omega} f_r(\harpoon\omega_i, \harpoon\omega, x) L_i(\harpoon \omega_i, x) (\harpoon \omega_i \cdot \harpoon {n_x}) d\omega_i
\end{aligned}
\end{equation}

with $L_e$ being the self-emitted spectral radiance from point $x$ itself, $f_r$ being the BRDF at point $x$ and $\harpoon{n_x}$ being surface normal at $x$.

The rendering equation can, however, be subject to further restrictions and parameters depending on the underlying {\it{illumination model}}. The generalized equation above does not specify the inner workings of $L_i(\harpoon \omega_i, x)$ and thus might not factor in phenomena such as occlusion or interreflection.

\begin{figure}[th]
\centering
\includegraphics[scale=0.22]{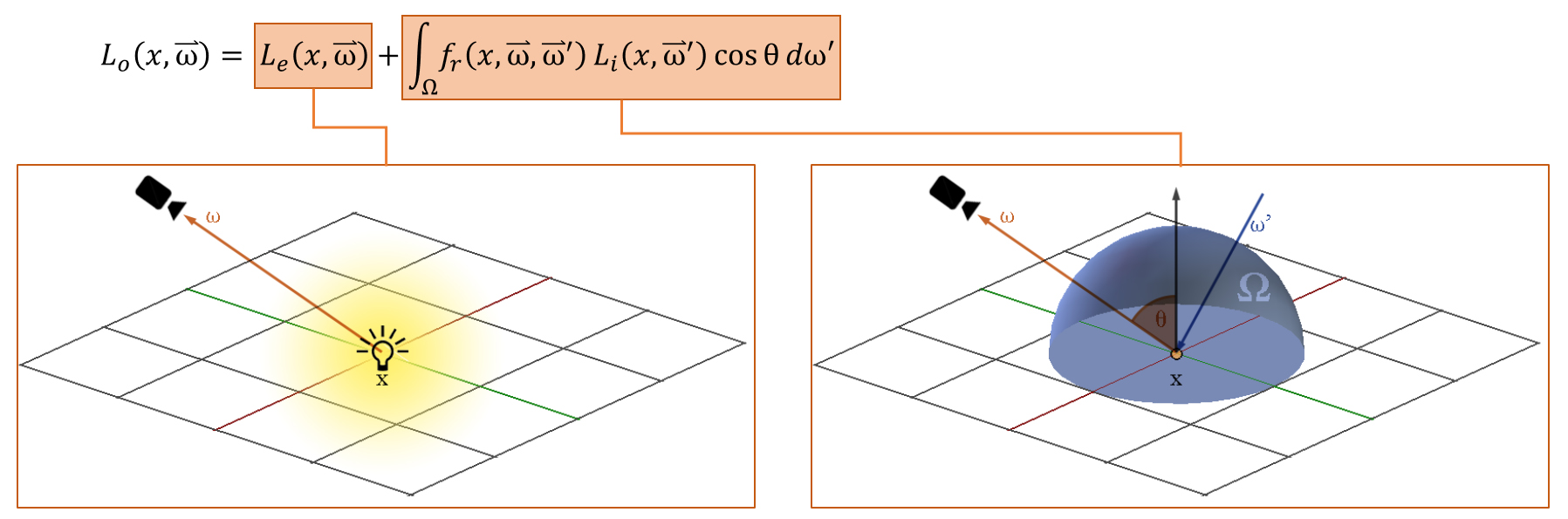}
\decoRule
\caption[]{Components of the rendering equation illustrated.}
\end{figure}

\subsection{Local Illumination Model}

The simplest illumination model to consider is one that only evaluates direct illumination from simple point lights and discards shadow casting in its entirety.

The direct irradiance caused by a single point light, as derived in (\ref{eqn:IrradianceFalloff}), can now be applied to a delta function in order to compute the radiance from said point light. Let $\harpoon\omega_s$ be a small differential solid angle pointing from $x$ to $x_s$:

\begin{equation}
\begin{aligned}
L_i(\harpoon \omega_i) & = \frac{\Phi}{4 \pi \left|x-x_s\right|^2}\delta(\harpoon\omega_s \cdot \harpoon\omega_i) \\
& = \frac{\Phi}{4 \pi \left|x-x_s\right|^2}\delta(\cos \theta_i - \cos \theta_s)\delta(\cos \phi_i - \cos \phi_s)
\end{aligned}
\end{equation}

Applying this radiance to the BRDF of the given material and the perpendicular surface area designated by $\cos \theta_i$ results in a reflectance equation for a single point light:

\begin{equation}
L_r(\omega_r) = \frac{\Phi}{4\pi \left | x - x_s \right |^2} f_r(\omega_r, \omega_s)*\cos \theta_s
\end{equation}

For a model of this type, the hemispherical integral in (\ref{RenderingEquation_general}) collapses into a sum over the amount of point lights $n$ present in the scene, resulting in a complete rendering equation for indirect illumination models:

\begin{equation}\label{RenderingEquation_direct}
L_o(x, \harpoon\omega)  = L_e(x, \harpoon\omega) + \sum_{s=0}^{n} \frac{\Phi}{4\pi \left | x - x_s \right |^2} f_r(\harpoon\omega, \harpoon\omega_s, x)*\cos \theta_s
\end{equation}

\begin{figure}[th]
\centering
\includegraphics[scale=0.3]{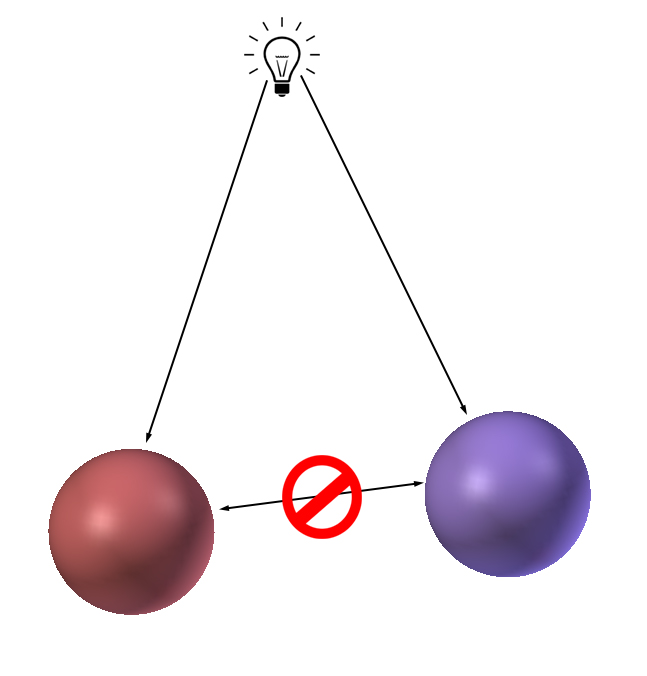}
\decoRule
\caption[]{Local illumination model restrictions. All light captured by the retina either comes straight from the light source or is directly reflected by a surface.}
\end{figure}

\subsection{Global Illumination}

A {\it{global illumination}} model incorporates interreflection and occlusion into our lighting representation.

The hurdles this type of model presents over a simple, direct illumination model are manifold. Many of which, however, can be overcome by introducing a visibility function $V(x, x')$ that yields 1 if $x$ and $x'$ are mutually visible and 0 if they are not.

Assuming total invariance of radiance along a ray, the implication is such, that the incident radiance at $x'$ resulting from the radiance emanating from $x$ equates to the following:

\begin{equation}\label{VisibilityFunction}
L_i(x', \harpoon\omega'_r) = L_o(x, \harpoon\omega_o)V(x, x')
\end{equation}

To complete the global illumination rendering equation the hemispherical integral over all incident directions needs to be replaced by an area integral over all other surfaces in the environment. Let $S=\bigcup S_i$ be the union of all surfaces in the scene and recall the relation of a solid angle and its projected surface area laid out in (\ref{SolidAngleProjectionRelation}).

Substituting (\ref{VisibilityFunction}) into (\ref{RenderingEquation_general}) and permuting the solid-angle integral into a surface area integral results in the global illumination rendering equation originally proposed by Kajiya\cite{Kajiya}:

\begin{equation}\label{RenderingEquation_global}
L_o(x, \harpoon\omega) = L_e(x, \harpoon\omega) + \int_{S} f_r(x, \harpoon\omega, \harpoon\omega') L_o(x', \harpoon \omega') V(x, x') \frac{\cos \theta \cos \theta'}{\left | x-x' \right |^2}dA
\end{equation}

\begin{figure}[th]
\centering
\includegraphics[scale=0.42]{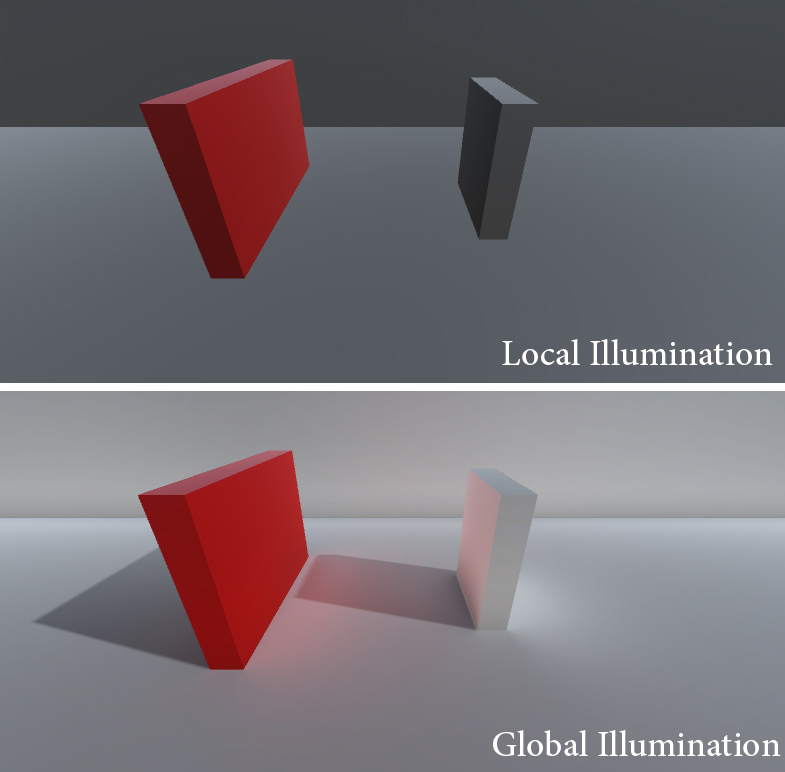}
\decoRule
\caption[]{Local and global illumination models as seen in the Unity 5.6.1 Engine. Note in particular the red light inter-reflected between the two surfaces.}
\end{figure}

%% file: Chapters/Chapter3.tex

\chapter{OpenGL and Illumination Models} 

\label{Chapter3} 

The most commonly used approach to storing and rendering 3D objects heavily relies on an algorithmic strategy that bears great similarities to the display-process of 2D scalable vector graphics. A 3D model contained within a graphical data file contains the object's blueprint in the form of a set of polygons, usually triangles. This model is merely a set of 3D coordinates and does not become a viewable graphic until displayed as a 2D image through a process called 3D rendering.

A large part of the rendering procedure consists primarily of said transformation process from Cartesian coordinates into an array of pixels. When rendered with the graphics library OpenGL, this transformation process is managed by a so-called graphics pipeline consisting of several stages, each with their respective purposes.


\section{OpenGL}\label{opengl}

Regularly released by the Khronos Group\cite{openglspec}, the Open Graphics Library (OpenGL) is an API specification that spans across multiple programming languages and platforms. It describes an abstract programming layer that can used to interact with a graphics processing unit (GPU) in order to achieve hardware-accelerated image synthesis.

An OpenGL context can be created by a variety of programming libraries, which then allows the developer to perform operations on the GPU, using the universally defined OpenGL functions, without having to adjust to different a API for each different GPU model.

OpenGL operates by enlarge like a state machine. A large set of changeable variables determines the modus operandi of subsequently used functions which produce the coveted images on the screen.

\section{OpenGL Graphics Pipeline}\label{pipeline}

The above mentioned fixed-function OpenGL graphics pipeline is comprised of a series of highly specialized steps which allow the system to efficiently draw 3D primitives in a given perspective. Each of these steps executes a concrete task with the output of the previous step as its input parameters. Given the consecutive nature of the pipeline, the individual steps can easily be executed in parallel for successive frames to be rendered.

The small programs run on the GPU that delineate each of these steps are commonly termed {\it{shaders}}, some of which are easily configurable by a developer. In the context of OpenGL, shaders are written in the OpenGL Shading Language (GLSL), which highly resembles the syntax used in C based languages.

\subsection{Overview}

The individual steps of the OpenGL graphics pipeline are manifold, but can usually be associated with one of three subsequent parts: Application, Geometry and Rasterization.

\begin{figure}[th]
\centering
\includegraphics[scale=0.55]{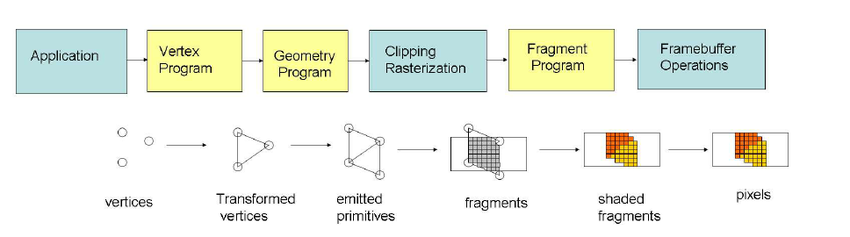}
\decoRule
\caption[]{OpenGL graphics pipeline steps, as depicted by Vettner et al.\cite{vettner}}
\end{figure}

\subsubsection{Application}

The application part is commonly classified as the operations performed on the CPU prior to rendering. Possible changes to the scene due to time-related physics-simulations or user input are computed here.

This phase culminates in the specification of an ordered list of vertices defining primitive shapes that are sent to the GPU to be drawn in the subsequent parts. On GPU memory, the vertex data is sorted into a vertex array object (VAO), a vertex buffer object (VBO) and an element buffer object (EBO).

Summed up, the VBO contains the actual vertex data copied over, while the VAO points to the given vertex attributes such as position and normal vector with respect to the specified strides. The EBO contains the index data, where three consecutive indices correspond to a triangle.

The process of {\it{vertex specification}} may also set the values of specific variables, called {\it{uniforms}}, contained in the forthcoming shaders. These may represent anything from the material of the object to the position of the camera and will affect the calculations performed by the shaders.

\subsubsection{Vertex Shader}

The initial step of the geometry part is the vital {\it{vertex shader}} which is used to process each vertex individually.
The most common operations performed in the vertex shader are the applications of the model-, view- and projection- matrices.

The model matrix is a common transformation matrix which converts the coordinates of the object from local space into world space by applying scale, rotation and translation to each vertex. 

The view matrix will thereafter transform the coordinates into view space, which represents how each vertex is seen from the camera's point of view. 

Lastly, the projection matrix projects the coordinates from view space into a $[-1, 1]$-ranged clip space coordinate-system. This process determines which vertices will end up on the screen as well as applying the perspective-distortion of non-orthographic cameras.

Other uses of the vertex shader include vertex-based animations such as morphing or waves on an ocean.

\begin{figure}[th]
\centering
\includegraphics[scale=0.4]{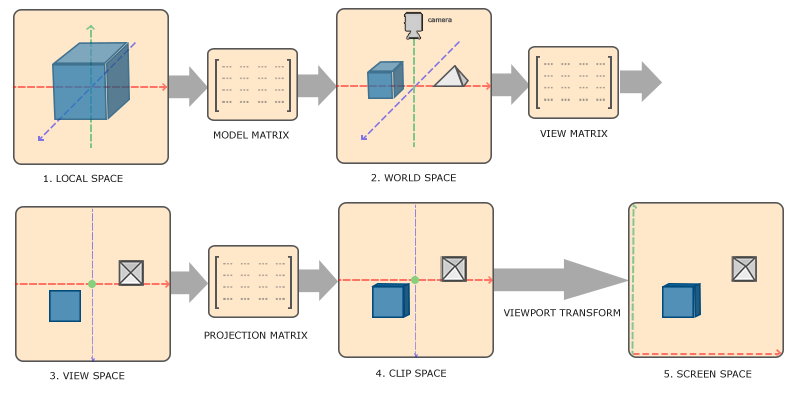}
\decoRule
\caption[]{Coordinate system transformations in the vertex shader as pictured by Joey de Vries\cite{learnopengl}}
\end{figure}

\subsubsection{Geometry Shader}

A further, optional step within the geometry stage is an extension of the {\it{primitive assembly}} process, which divides the polygons relayed from the vertex shader into a sequence of individual primitives, commonly triangles. This developer-defined process, called the {\it{geometry shader}}, can remove, subdivide or otherwise transform primitives in any way desired.

\begin{figure}[th]
\centering
\includegraphics[scale=0.4]{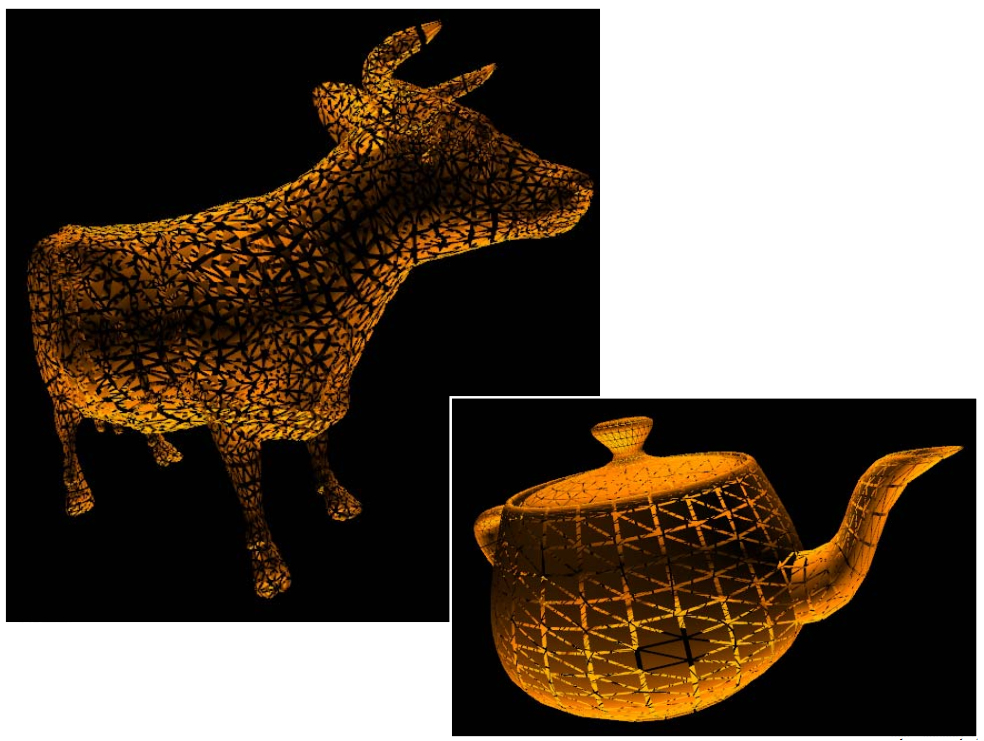}
\decoRule
\caption[]{Example: Shrinking triangles in the geometry shader. (Bailey \cite{geom})}
\label{geom_example}
\end{figure}

It is often used for tessellation, smoothing or vertex based effects such as enlarging or shrinking (see fig. \ref{geom_example}). Further into this thesis, the geometry shader will play a vital part in the process of scene voxelization.

\subsubsection{Vertex Post-processing}

The geometry phase concludes with a number of fixed-function post-processing steps, namely {\it{clipping}} and {\it{face culling}}.

Clipping discards any vertices that lie outside of the viewing volume boundaries. The geometric shape commonly used to perform these calculations is the so-called {\it{view frustum}} that originates from the camera position and expands from a near plane to a far plane. The removal of vertices outside this shape can greatly increase performance for the rendering of scenes with high polygon counts.

Face culling removes any triangles that are facing away from the viewer. Which sides of a triangle, if any, this feature should cull, can be specified with the OpenGL \verb|glCullFace(...)| function.

\subsubsection{Rasterization}

The rasterization process converts the primitives obtained from the geometry stage into a series of {\it{fragments}}. Each fragment represents a sampled point of the given primitive and relates to the pixel that covers said sampled point.

To accomplish effects such as anti-aliasing, multisampling parameters can allow OpenGL to interpolate multiple fragments to form a single pixel.

\subsubsection{Fragment Shader}\label{frag_shader_general}

The fragments computed in the rasterization step are then passed on to the {\it{fragment shader}}, which calculates the final color of the pixel.

Most lighting calculations tend to be done in this stage, with the pixels color value representing the light emitted and reflected by the corresponding fragment towards the location of the camera. 
The mathematical models outlined in chapter \ref{Chapter2} can be used to compute the pixel brightness and color composition equating to the radiance and wavelengths emanated from the fragment position towards the retina.

In order for these calculations to take place, information about the 3D scene must be present within the shader, including the location of the fragment within the scene.

Following the fragment shader, the graphics pipeline performs a depth test and blending stage.

The depth test simply verifies whether if the fragment is hidden behind another object and discards it if it is not visible. However, if the given object is translucent, the corresponding, overlaying fragments are blended accordingly instead.

\section{Phong Model for Local Illumination}\label{PhongModel}

One of the primary techniques used to achieve real-time, direct illumination in the realm of shader-based rendering is the Phong model. 

Developed by Bui Tuong Phong\cite{phong} in 1975, this method has become the de facto baseline approach for many rendering techniques. Similarly to the presented components in \ref{reflectanceComponents}, the Phong model combines various types of lighting by adding their respective fragment values together.

\subsection{Phong Reflection Model}

Given that the Phong model only calculates direct illumination, occlusion and interreflection are not regarded.

Recall the rendering equation for direct lighting from (\ref{RenderingEquation_direct}):

\begin{equation}
L_o(x, \harpoon\omega)  = L_e(x, \harpoon\omega) + \sum^{n}_{s=0} \frac{\Phi}{4\pi \left | x - x_s \right |^2} f_r(\harpoon\omega, \harpoon\omega_s, x)*\cos \theta_s
\end{equation}

The Phong model treats light sources and scene objects as different entities, thus excluding the emission-component $L_e$. Additionally, the present term for point-light irradiance is simply replaced by an adjustable intensity variable $i_s$:

\begin{equation}
L_{phong}(x, \harpoon\omega)  = \sum^{n}_{s=0} i_s* f_r(\harpoon\omega, \harpoon\omega_s, x)*\cos \theta_s
\end{equation}

Note that due to the apparent simplifications, the rendering equation is no longer physically sound, thus $L_{phong}$ no longer reflects the genuine radiance flowing from $x$ in direction $\harpoon \omega$, but rather the approximated color and brightness of $x$ observed by camera located in direction $\harpoon \omega$ from $x$.

The diffuse and glossy BRDF types, as derived in \ref{brdf}, are considered on an individual basis with their respective intensity variables $i_{s,d}$ (diffuse) and $i_{s,g}$ (glossy):

\begin{equation}
L_{phong}(x, \harpoon\omega) = \sum^{n}_{s=0} i_{s,d} f_d(\harpoon\omega, \harpoon\omega_s, x)*\cos \theta_s + i_{s,g} f_s(\harpoon\omega, \harpoon\omega_s, x)*\cos \theta_s
\label{phong_eq_brdfs}
\end{equation}

As established in \ref{DiffuseReflection}, the BRDF of a diffuse reflection is constant. This implies that $f_d$ can be incorporated into $i_{s,d}$ or simply viewed as having a value of 1. The diffuse term  thus only depends on $i_{s,d}$ and the angular predicament $\theta_s$ of the surface in respect to the light source.

$\cos \theta_s$ represents the factor of perpendicularity of the surface which equals the dot product of the point's surface normal $\harpoon N$ and the direction vector $\harpoon L_s$ pointing from $x$ towards the corresponding light source.

The glossy BRDF $f_s$ follows essentially the same pattern as was put forth in \ref{GlossyReflection}. This component is meant to simulate specular reflections on rough surfaces, and since these are unaffected by projected surface area, the $\cos \theta_s$ factor is simply discarded for this term.

Let $\harpoon R_s$ be the directional vector of a perfectly, mirror-reflected ray, $\alpha$ be the shininess of the material and assuming all vectors to be normalized, then the exiting illumination of point $x$ in direction $\harpoon \omega$ is the following:

\begin{equation}\label{phong_equation_base}
\begin{aligned}
L_{phong}(x, \harpoon\omega) & = \sum^{n}_{s=0} i_{s,d} *\cos \theta_s + i_{s,g} f_s(\harpoon\omega, \harpoon\omega_s, x) \\
& = \sum^{n}_{s=0} i_{s,d} (\harpoon L_s \cdot \harpoon N) + i_{s,g} (\harpoon R_s \cdot \harpoon \omega )^{\alpha}
\end{aligned}
\end{equation}

Due to interreflection of light, all parts of an object tend to appear at least somewhat lit as long as a light source is present.

Based on this empirical observation, the Phong model attempts to increase the realism of the resulting images by adding a constant {\it{ambient}} value $i_a$ to the reflected light. This is meant to approximate the effects of indirect light in the most inexpensive way possible.

Furthermore, the Phong model defines a material of an object as being a set of the following parameters:

\begin{equation}\label{material_components_phong}
material = \{c_s, c_d, c_a, \alpha\}
\end{equation}

where:
\begin{description}
\item[$c_s$] is the ratio of specularly reflected light,
\item[$c_d$] is the ratio of diffusely reflected light,
\item[$c_a$] is the ratio of reflected ambient light,
\item[$\alpha$] is the shininess of the material.
\end{description}

Incorporating the ambient light and material parameters into (\ref{phong_equation_base}), yields the equation originally proposed by Phong\cite{phong}:

\begin{equation}\label{RenderingEquation_phong_final}
L_{phong}(x, \harpoon\omega) = c_a i_a + \sum^{n}_{s=0} c_d i_{s,d} (\harpoon L_s \cdot \harpoon N) + c_s i_{s,g} (\harpoon R_s \cdot \harpoon \omega )^{\alpha}
\end{equation}

\begin{figure}[th]
\centering
\includegraphics[scale=0.18]{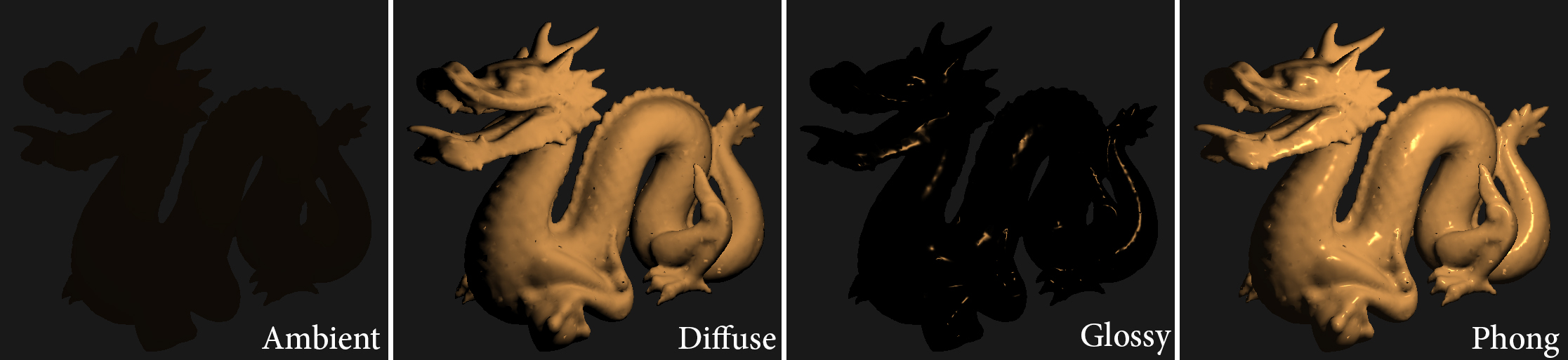}
\decoRule
\caption[]{Phong shading applied to the Stanford 3D Scanning Repository Dragon\cite{stanford}}
\end{figure}

\subsection{Phong Reflection Model with GLSL}\label{phong_pseudo}

An incorporation this model into a OpenGL graphics pipeline mostly takes place in the fragment shader.

It is common practice to draw no distinction between $i_a$, $i_{s,g}$ and $i_{s,d}$ and utilize the RGB color value emitted by the light-sources instead.

Additionally, while not present in the Phong model, incorporating the light attenuation model as presented in \ref{Attenuation} is a common way of including distance based light diminishment.

With \verb|pos_fs| being the fragment position in and \verb|nrm| being the surface normal vector of length 1, a Phong fragment shader for a singular point light implemented in GLSL may look as simple as following:

\begin{lstlisting}[backgroundcolor = \color{x11gray}, language = C, xleftmargin = 0.0cm, framexleftmargin = 0.5em]
// Ambient component
vec3 ambient = material.ambient_str * light.color;
  	
// Diffuse component
vec3 L_s = normalize(light.position - pos_fs);
float brdf_diffuse = max(dot(nrm, L_s), 0.0);
vec3 diff = material.diffuse_str*brdf_diffuse * light.color;
    
// Glossy component
vec3 omega = normalize(viewPos - pos_fs);
vec3 R_s = reflect(-L_s, nrm);  
float brdf_spec = pow(
    max(dot(omega, R_s), 0.0),
    material.shininess );
vec3 spec = material.specular_str * brdf_spec * light.color;  

// Attenuation
float distance = length(light.position - pos_fs);
float attenuation = 1.0f / (light.att_constant
    + light.att_linear * distance 
    + light.att_quadratic * (distance * distance));

vec3 result = (ambient + diff + spec) * material.color;
FragColor = vec4(result * attenuation, 1.0);
\end{lstlisting}

\section{Models for Global Illumination}

No rendering algorithm has established itself as the cardinal approach to global illumination in the same way as Phong has done for local illumination models.

This section will briefly present some of the primary methods used to solve the global illumination rendering equation as well as their apparent advantages and shortcomings.

In contrast, chapter \ref{Chapter4} will introduce a technique that attempts to replicate the fidelity of the methods presented here while dramatically reducing the impact of their drawbacks.

\subsection{Raytracing}\label{raytracing}

The well-trodden ground that constitutes raytracing is a long-standing method of numerically solving the rendering equation (\ref{RenderingEquation_global}).

The core concept consists of tracing rays from an imaginary retina position through each pixel of a virtual screen.
Upon intersection with a surface, further rays can be traced recursively to achieve an approximation for the global illumination model. 

These rays typically pertain to one of three categories: specular, diffuse and occlusion.

The depth of recursion is stopped either at an explicit cap or once the ray encounters a light source.

This technique can generate highly realistic images but comes with an equally high computational cost. The tracing process of each ray carries out intersection tests with each surface of the scenes' objects. Additionally, the algorithm needs to estimate the incoming radiance $L_i$ at the intersection point $x$, which requires further rays to be traced in compliance with the respective BRDFs.

The tremendous amount of intersection calculations performed, make this approach unsuitable for real-time rendering applications on consumer-grade hardware.

\begin{figure}[th]
\centering
\includegraphics[scale=0.2]{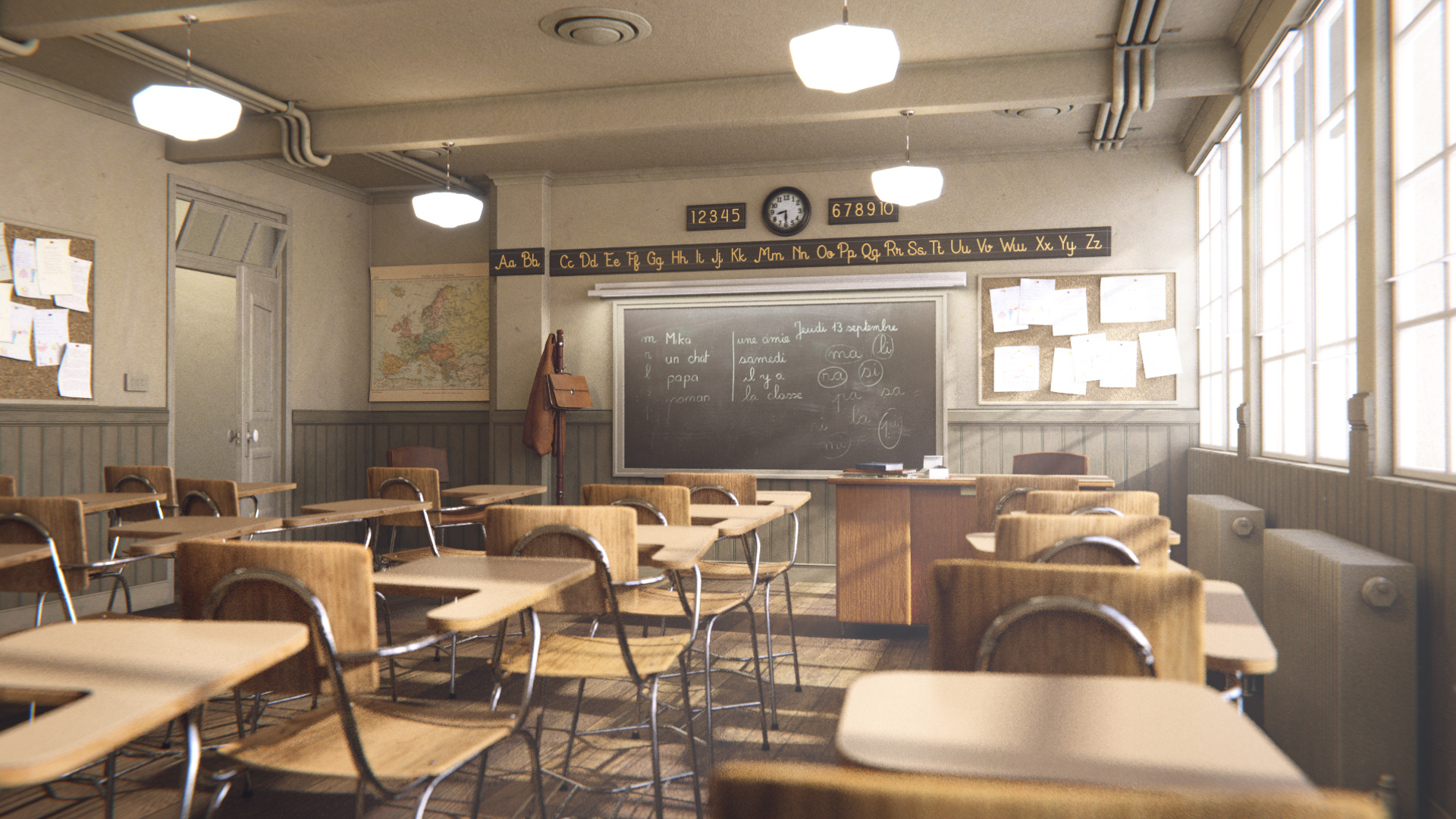}
\decoRule
\caption[]{Photorealism achievable with raytracing: "Classroom" by Christophe Seux rendered with Blender 2.79 Cycles}
\end{figure}

\subsection{Radiosity}

The radiosity approach makes use of the {\it{finite element method}} by subdividing the environment's geometry into a series of smaller, subdivided surfaces. A coefficient, the {\it{view factor}}, is calculated and assigned to each pair of surfaces. It describes the degree of their mutual visibility.

Recall the rendering equation for a global illumination model from (\ref{RenderingEquation_global}):

\begin{equation}
L_o(x, \harpoon\omega) = L_e(x, \harpoon\omega) + \int_{S} f_r(x, \harpoon\omega, \harpoon\omega') L(\harpoon \omega', x) V(x, x') \frac{\cos \theta \cos \theta'}{\left | x-x' \right |^2}dA
\end{equation}

The radiosity method only solves this equation for surfaces that reflect light exclusively in a diffuse manner. This means that the BRDF can be replaced by the constant defined in \ref{DiffuseReflection}:

\begin{equation}
L_o(x, \harpoon\omega) = L_e(x, \harpoon\omega) + \frac{\rho(x)}{\pi} \int_{S} L(\harpoon \omega', x) V(x, x') \frac{\cos \theta \cos \theta'}{\left | x-x' \right |^2}dA
\end{equation}

As the scene is subdivided into a finite set of small surfaces, the integral can be transformed into a sum over these patches, yielding the discrete radiosity equation:

\begin{equation}
L_o(i) = L_e(i) + \frac{\rho(i)}{\pi} \sum_{j=1}^m L_o(j) F(i, j)
\end{equation}

where $i$ and $j$ index their corresponding patch and $F(i,j)$ is the {\it{view factor}} mentioned above, incorporating both the visibility and projected area.

Given the recursive nature of the equation, the calculations for a single bounce of light need to be applied repeatedly in a number of passes. The amount of passes will determine the brightness and fidelity of the scene but also impact the required computation time.

The great advantage of the radiosity approach, is that pre-computed radiance values for each patch can be embedded into lightmaps and subsequently rendered cheaply into a scene. The most adamant limitations of this method arise from the simple fact, that only static scenery is accounted for. Should any object within the scene excluding the camera be moved, all lighting values would have to be calculated anew. 

Additionally, as noted above, the radiosity equation does not encompass specular and glossy reflectivity. Some approaches attempt to integrate the remainder of the rendering equation by affixing specular raytraces atop the view-independent radiosity lightmaps \cite{wallace}.

\begin{figure}[th]
\centering
\includegraphics[scale=0.35]{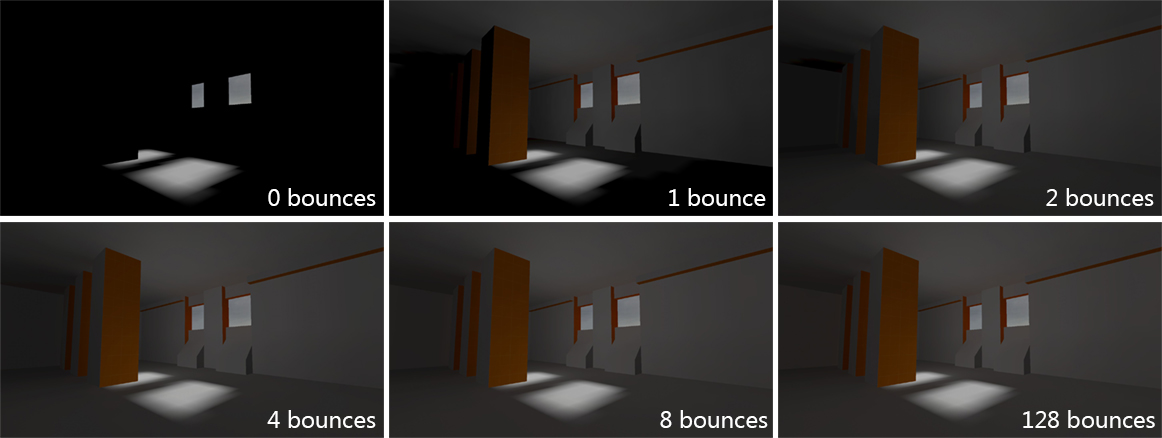}
\decoRule
\caption[]{Radiosity after a specified amount of additional light bounces, as done by the 2014 VRAD tool and rendered by the 2009 Source Engine}
\end{figure}

\subsection{Beam Tracing}\label{BeamTracing}

Proposed by Heckbert and Hanrahan\cite{Heckbert} in 1984, beam tracing ameliorates the amount of samples necessary for raytracing by bundling volumes of rays together into convex, polyhedral beams. 

The algorithm commences by casting a vast beam into the scene that encompasses the whole of the viewing volume and is thus equivalent with the camera's view frustum. From closest to furthest, every visible polygon intersected is removed from the shape of the beam and further beams are cast for reflection, refraction and occlusion measurements.

Through this process a so-called {\it{beam tree}} is built, the nodes of which represent the reflection paths traversed by the light that reaches the camera.

This method solves some aliasing and sampling issues typically encountered in traditional raytracing approaches. However, "the beams might become rather complex and the implementation of a robust and fast beam casting algorithm is difficult."\cite{Bittner}

\begin{figure}[th]
\centering
\includegraphics[scale=0.44]{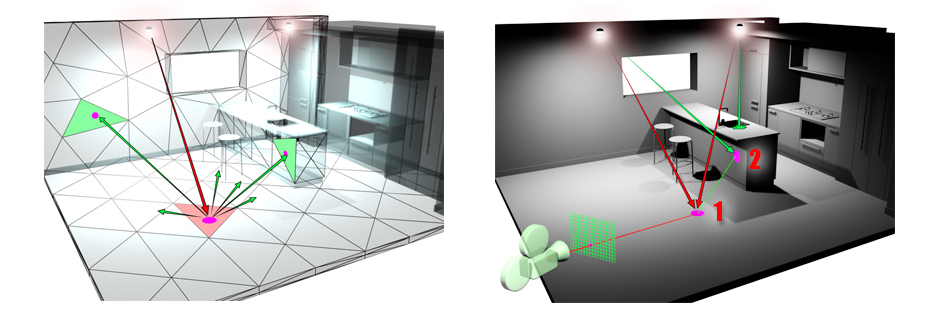}
\decoRule
\caption[]{Radiosity (left) and Raytracing (right) (3DS MAX 2016 Manual)\cite{maya}}
\end{figure}

\begin{figure}[th]
\centering
\includegraphics[scale=0.45]{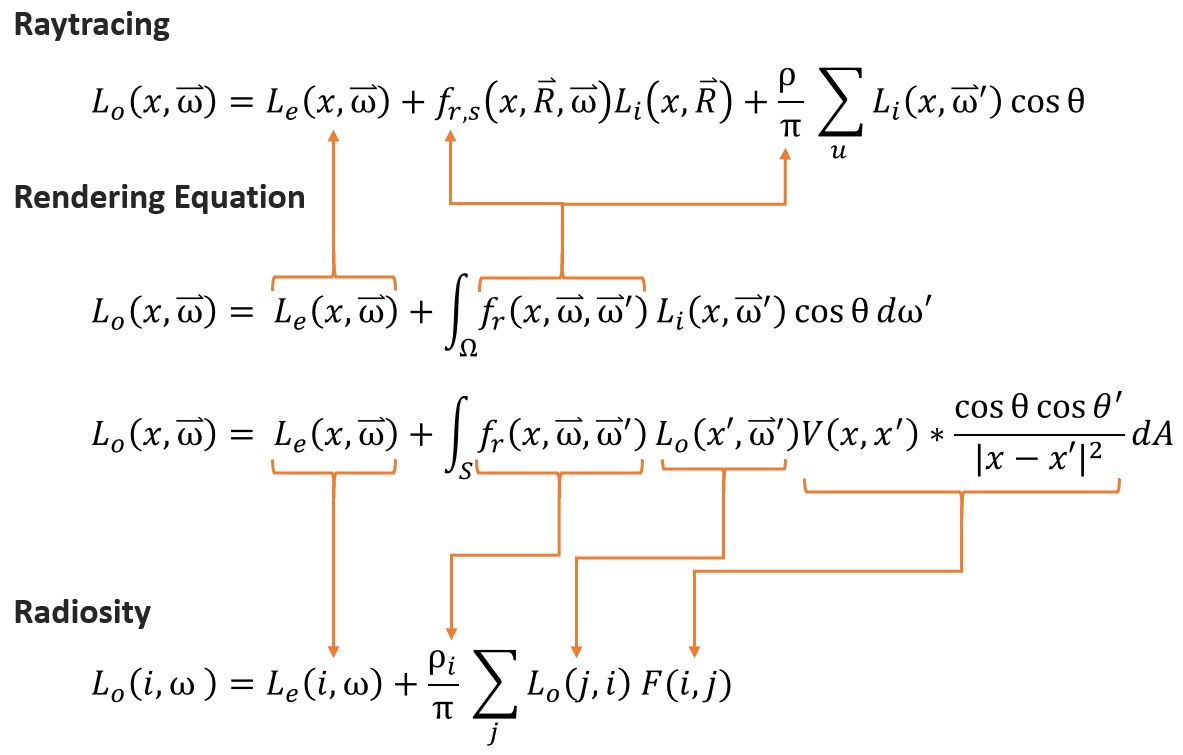}
\decoRule
\caption[]{Approximating the rendering equation through radiosity and raytracing.}
\end{figure}

%% file: Chapters/Chapter4.tex

\chapter{Voxel Cone Tracing} 

\label{Chapter4} 

This chapter will outline the cone tracing algorithm initially introduced by Crassin et al.\cite{Crassin} in 2011.

Unlike the beam tracing method discussed in \ref{BeamTracing}, this approach approximates the solid angle integral of the rendering equation by collapsing it into a sum of cones. These cones sample the incoming radiance from that respective solid angle by reading direct-light values from a three-dimensional bitmap of the scene.

An entry in a regular grid of this form, is termed {\it{voxel}}, hence the name {\it{voxel cone tracing}}.


\section{Cone Tracing}

As described by Bittner\cite{Bittner}, the underlying cone tracing technique constitutes yet another derivative of the raytracing algorithm.

Similarly to beam tracing (see \ref{BeamTracing}), the fundamental process of calculating light-paths originating from the camera through a virtual screen is maintained, but performance is dramatically improved by bundling large sets of rays into three-dimensional cones.


The incident radiance at a point $x$ can thus estimated by partitioning the hemispherical integral in the rendering equation into a sum of $u$ cone-traces:

\begin{equation}
\begin{aligned}
L_o(x, \harpoon\omega) & = L_e(x, \harpoon\omega) + \int_{\Omega} f_r(\harpoon\omega_i, \harpoon\omega, x) L_i(\harpoon \omega_i, x) (\harpoon \omega_i \cdot \harpoon {n_x}) d\omega_i \\
& \approx L_e(x, \harpoon\omega) + \sum^{u} f_r(\harpoon\omega_i, \harpoon\omega, x) L_i(\harpoon \omega_i, x) (\harpoon \omega_i \cdot \harpoon {n_x})
\end{aligned}
\end{equation}

\begin{figure}[th]
\centering
\includegraphics[scale=0.6]{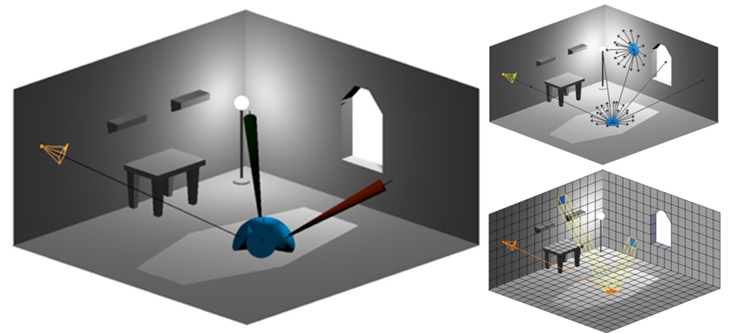}
\decoRule
\caption[]{Cone Tracing (left), Raytracing (top right) and Radiosity (bottom right)}
\end{figure}

\section{Component Merging}

\subsection{Overview}\label{crassin_overveiw}

Rather than determining the entirety of the incident radiance by the use of cone tracing, {\it{voxel cone tracing}} uses phong reflection model techniques (see \ref{PhongModel}) for direct light, and adds on top the indirect light sampled by cone-traces.

For each shader fragment, the Phong model provides two components of direct light: diffuse and glossy, as well as a constant color value that feigns the effect of diffuse interreflection (ambient).

Crassin's voxel cone tracing approach adopts the direct diffuse and direct glossy light components in the same way as they are computed in the Phong model.

However, by introducing the possibility of sampling indirect light via cone-traces, the bare-bones approach at indirect light (Phong's ambient value) can be significantly improved upon.

The nature of these cone-traces in relation to the present fragment closely resembles the categorization used in raytracing, as described in \ref{raytracing}.

Indirect diffuse light is determined via a multitude of wide-aperture cones cast out uniformly in a hemispherical configuration.

On the other hand, the assessment of indirect specular light is conducted through a singular and very narrow cone in the direction of perfect reflectivity.

Furthermore, one occlusion cone per light-source is traced to discern the dimness at the given location. The aperture of these correlates to the hardness of the projected shadows. The resulting value is then used to appropriately diminish the brightness of the fragment's direct light figure.

\begin{figure}[th]
\centering
\includegraphics[scale=0.6]{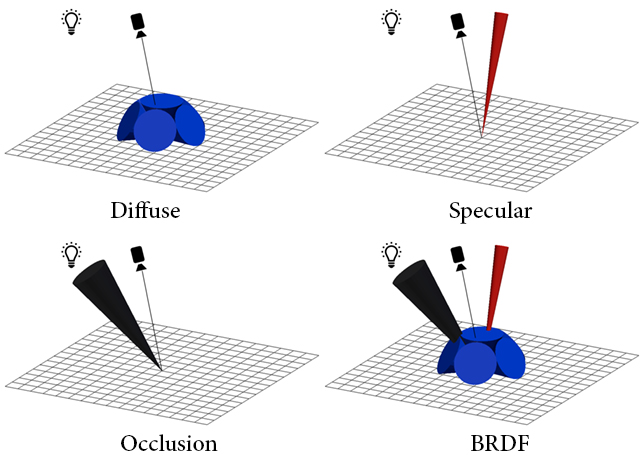}
\decoRule
\caption[]{Required cone traces for indirect illumination}
\end{figure}

\subsection{Approximating the Rendering Equation}

An abstract of the voxel cone tracing approach was provided in the section above.

Defining how exactly it applies to rendering equation, can further benefit the process of delineating the necessary software components required for a successful implementation of this algorithm.

Recall the Phong model rendering equation as defined in (\ref{RenderingEquation_phong_final}):

\begin{equation}\label{four_two}
L_{phong}(x, \harpoon\omega) = c_a i_a + \sum^{n}_{s=0} c_d i_{s,d} (\harpoon L_s \cdot \harpoon N) + c_s i_{s,g} (\harpoon R_s \cdot \harpoon \omega )^{\alpha}
\end{equation}

As described in \ref{crassin_overveiw}, the direct diffuse and specular light components ($c_d i_{s,d} (\harpoon L_s \cdot \harpoon N)$ and $c_s i_{s,g} (\harpoon R_s \cdot \harpoon \omega )^{\alpha}$) are left mostly intact in the voxel cone tracing approach.

Instead, the ambient light value $c_a i_a$ is far expanded.

Let $\Psi_i(\harpoon \omega_i, x)$ be the incident {\it{indirect}} radiance at $x$ from direction $\harpoon \omega_i$. That is, the corresponding total radiance minus the radiance caused directly through the light sources visible from that point:

\begin{equation}
\Psi_i(\harpoon \omega_i, x) = L_i(\harpoon \omega_i, x) - \sum^{n}_{s=0} V(x,x_s)\frac{\Phi}{4 \pi \left | x-x_s \right |^2}
\end{equation}

As $c_a i_a$ in (\ref{four_two}) represents the Phong model's indirect light value, it can be replaced by the hemispherical integral of all indirect light at that point. 

Furthermore, the realism portrayed by the direct light summand can be enhanced by the application of a light-source occlusion factor given by the mutual visibility function $V(x,x_s)$:

\begin{multline}
L_{vxct}(x, \harpoon\omega) = \int_{\Omega} f_r(\harpoon\omega_i, \harpoon\omega, x) \Psi_i(\harpoon \omega_i, x) (\harpoon \omega_i \cdot \harpoon {n_x}) d\omega_i \\ + \sum^{n}_{s=0} V(x, x_s) \Big(c_d i_{s,d} (\harpoon L_s \cdot \harpoon N) + c_s i_{s,g} (\harpoon R_s \cdot \harpoon \omega )^{\alpha} \Big)
\label{multipart_rendereq}
\end{multline}

In this context, the integral term refers to the indirect light and the sum to the direct light.

Let $C_i(x, \harpoon \omega, \gamma)$ be the indirect radiance hitting point $x$ through a circular cone with its apex at $x$ and a centre vector corresponding to $\harpoon \omega$. The cone's opening angle (henceforth called {\it{aperture angle}}) is $\gamma$ and the spherical cap subtended by it corresponds to the overall solid angle $\Omega_i$:

\begin{equation}
C_i(x, \harpoon \omega, \gamma) = \int_{\Omega_i} \Psi_i(\harpoon \omega_i, x) (\harpoon \omega_i \cdot \harpoon {n_x}) d\omega_i
\end{equation}

How this function is implemented on a software-basis, will be investigated in \ref{conetracingproc}.

As already mentioned, the introduction of cone traces enables the approximation of the hemispherical integral by partitioning it into a finite sum of cones.

On this occasion, the integral term in (\ref{multipart_rendereq}) is replaced with a sum of $u$ cone traces.

\begin{multline}
L_{xvct}(x, \harpoon\omega) = \sum_{q=0}^{u} f_r(\harpoon \omega_q, \harpoon \omega, x)C_q(x, \harpoon \omega_q, \gamma_q) \\+ \sum^{n}_{s=0} V(x, x_s) \Big(c_d i_{s,d} (\harpoon L_s \cdot \harpoon N) + c_s i_{s,g} (\harpoon R_s \cdot \harpoon \omega )^{\alpha} \Big)
\end{multline}

It is noteworthy that the BRDF is now only being applied only once per cone, rather than for every individual direction.

Similarly to how the BRDFs were employed in (\ref{phong_eq_brdfs}), $f_r$ is best split into a specular BRDF $f_{spec}$ and a diffuse BRDF $f_{diff}$.

The specular BRDF only needs to be applied for the perfect reflection vector $\harpoon R_\omega$ at a very small angle $\gamma_{spec}$ while the diffuse one still requires a summation over all $u$ incident cones.

\begin{multline}
L_{vxct}(x, \harpoon\omega) = f_{spec}(\harpoon R_{\omega}, \harpoon \omega, x)C(x, \harpoon R_{\omega}, \gamma_{spec}) + \sum_{q=0}^{u} f_{diff}(\harpoon \omega_q, \harpoon \omega, x)C_q(x, \harpoon \omega_q, \gamma_{diff}) \\+ \sum^{n}_{s=0} V(x, x_s) \Big(c_d i_{s,d} (\harpoon L_s \cdot \harpoon N) + c_s i_{s,g} (\harpoon R_s \cdot \harpoon \omega )^{\alpha} \Big)
\label{vxct_rendeq_complete}
\end{multline}

Assuming that the implementation of $V(x,x_s)$ requires a cone-trace to be performed implies that the determination of a single fragment's color necessitates a total of $1+u+n$ cone-traces.

In order to keep performance costs low, an efficient cone tracing algorithm is necessary.

\begin{figure}[th]
\centering
\includegraphics[scale=0.5]{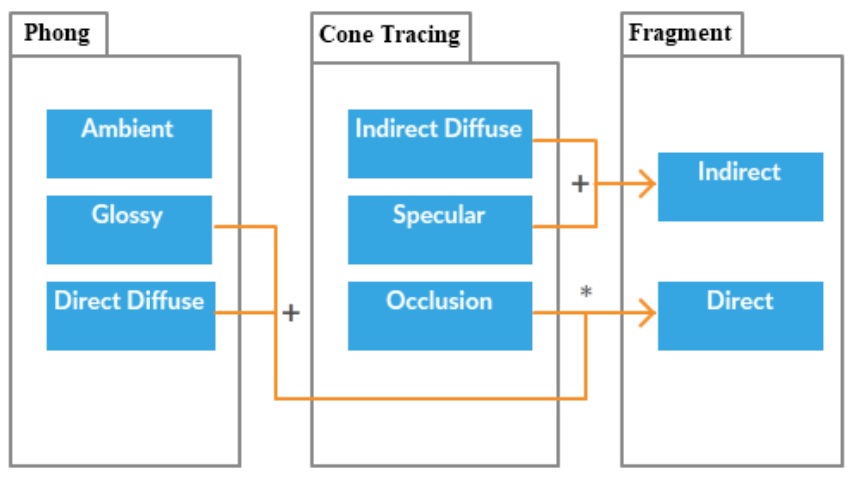}
\decoRule
\caption[]{Flow diagram of different components during voxel cone tracing. The Phong model's glossy component may also be relinquished.}
\end{figure}

\section{Cone Tracing Process}\label{conetracingproc}

\subsection{Prerequisites}

The implementation of equation (\ref{vxct_rendeq_complete}) into a fragment shader is an easy task, so long as a $C_i(x, \harpoon\omega, \gamma)$ function exists, that fulfills the following prerequisites:

For any given direction, position and aperture angle, $C_i$ yields an approximate value of the indirect light incident at $x$ through the cone specified by the angle and direction. Additionally, this value needs to be computed in a timely manner to enable a vast number of executions for every rendered frame.

\subsection{Volume Ray Marching}\label{raymarching}

In a process often referred to as {\it{volume ray marching}}, the cone's propagation direction is iterated over in a series of sampling steps.

With each step along the cone axis, the surrounding volume is sampled for its surfaces and radiance. The scale of the sampled volume remains in proportion to the corresponding cone diameter $d$ at that given point and can be calculated using rudimentary trigonometry on the basis of the aperture angle and distance travelled $t$:

\begin{equation}
d = 2t * \tan (\frac{\gamma}{2})
\end{equation}

\begin{figure}[th]
\centering
\includegraphics[scale=0.35]{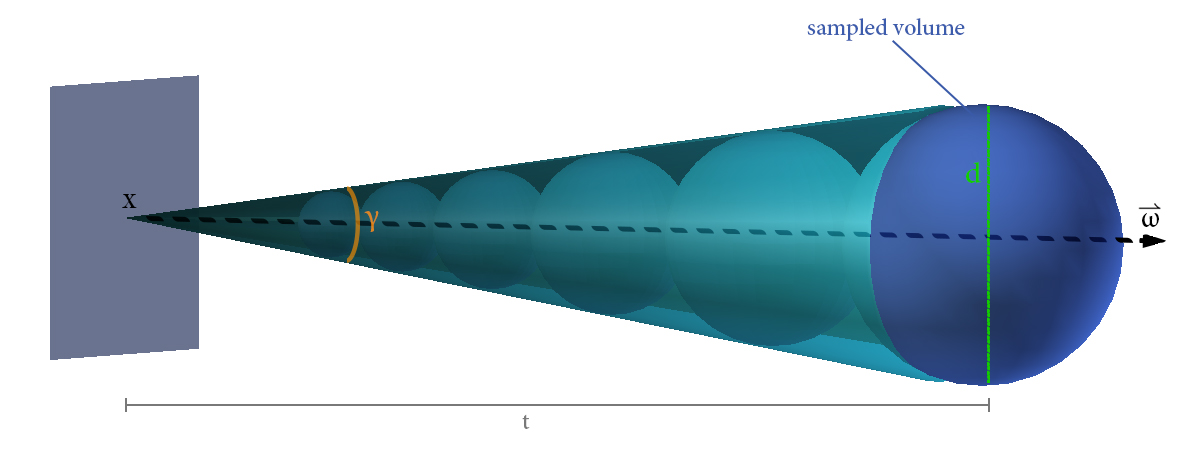}
\decoRule
\caption[]{Volumetric color sampling through volume ray marching. Each sphere corresponds to a taken sample.}
\end{figure}

A key distinction between volume ray marching and ray tracing, lies in that a ray will stop once a surface is hit, while the volumetric variant will "push through" the object, sampling it along its way and continue computation until a specified occlusion threshold is reached.

Given that individual surfaces can occlude only certain parts of the cone while not blocking its path as a whole, the occlusion value will have to be tracked during the tracing process as it signifies how relevant the light in the sampled volume truly is to the cone's origin.

In this view, the occlusion value sampled at any point along the cone axis, represents the fraction of the cone's volume that has been obstructed along its path.

As described by Villegas\cite{villegas}, the final value yielded by a traced cone results from the accumulated color value $c$ and occlusion value $\alpha$ sampled at each step. These variables are tracked for the duration of the process and adequately updated using front-to-back accumulation with each step.

Assuming that an efficient volume sampling function \verb|sample(position, diameter)| exists, the algorithmic overview of a single cone-trace would look as follows:

\begin{lstlisting}[caption={Cone Trace Function (Pseudo-code)}]
t = 0.001                       // Some small initial offset
(!$\alpha$!) = 0
c = (0, 0, 0)
while (!$\alpha$!)<1:
    d = 2*t*tan((!$\gamma$!)/2)              // Sampling diameter
    p = x + t*w                   // Sampling position
    (!$c_{sample}$!), (!$\alpha_{sample}$!) = sample(p, d)
    c = (!$\alpha$!)*c + (1-(!$\alpha$!))*(!$\alpha_{sample}$!)*(!$c_{sample}$!)    // color accumulation
    (!$\alpha$!) = (!$\alpha$!) + (1-(!$\alpha$!))(!$\alpha_{sample}$!)             // occl. accumulation
    t = t + (!$\beta$!)*d
return c, (!$\alpha$!)
\end{lstlisting}

where \verb|x| and \verb|w| correspond to the cone's origin and normalized axis vector accordingly.

Furthermore, the constant $\beta$ characterizes the distance factor used to advance along the cone axis. For $\beta=1$, each step will further the sampling distance by the current diameter of the cone. Lesser values will cause a higher rate of occlusion accumulation, as certain volumes are sampled multiple times, generally leading to shorter cone travel distances.

\subsection{Volume Sampling}

To conclude the voxel cone tracing algorithm, a volume sampling method, as used in \ref{raymarching} is necessary.

Crassin's proposal is to create and employ low-resolution, three-dimensional {\it{mipmaps}} of the scene's direct lighting values and then sample across these by adjusting their respective {\it{level-of-detail}} accordingly.

The following sections will provide a brief overview on the general usage of mipmaps in rendering applications and how this approach provides an avenue for a simple, yet effective volume sampling technique.

\subsubsection{Mipmaps}\label{mipmaps}

The idea behind {\it{mipmaps}} was initially conceived by Lance Williams in his 1983 paper {\it{Pyramidal parametrics}}\cite{w_lance} in which he proposes a method of minimizing the {\it{aliasing-effects}} observed when flat source images are projected onto curved surfaces.

These kinds of aliasing effects emerge when texture samples are available at a higher frequency than required for the point samples put forth by the fragment shader.

\begin{figure}[th]
\centering
\includegraphics[scale=0.7]{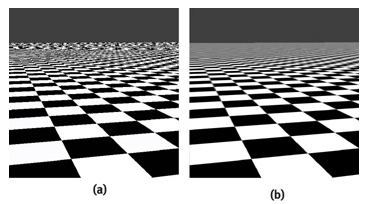}
\decoRule
\caption[]{Unwanted aliasing effects produced by a procedural checkers pattern (a) and the desired image produced through texture filtering (b), as portrayed by Randima\cite{alia}}
\label{aliasing}
\end{figure}

In other words, rendering a high resolution image onto a small screen-area that only occupies a limited amount of pixels leads to a multitude of image-pixels being available for each screen-pixel. Simply relaying one of the many available values onto the rendering screen will produce the unfortunate side-effect observable in fig. \ref{aliasing}.

A manifold of {\it{texture filtering}} methods can minimize, or even completely remove, these incongruities by blending corresponding adjacent texture samples into a single color.

However, applying performance-heavy filters on every texture of every rendered frame can lead to inadequate performance for use in real-time applications. Instead, Williams argues \cite{w_lance}, assembling a set of pre-filtered images, one of which is subsequently rendered, reduces the implied computation time while still eliminating any aliasing effects.

Put simply, for any mipmapped image, the rendering application will pre-compute a set of lower resolution versions of that same image and then render the one that best suits the pixels it occupies on the screen.

In addition to being dogmatic within an OpenGL context, it is also regular, common practice to ascribe an image's mipmaps a {\it{level-of-detail}} (LOD) number where 0 corresponds to the actual image. For each LOD increment of 1, the height and width of the corresponding mipmap is a power of two smaller from the previous level.

\begin{figure}[th]
\centering
\includegraphics[scale=0.4]{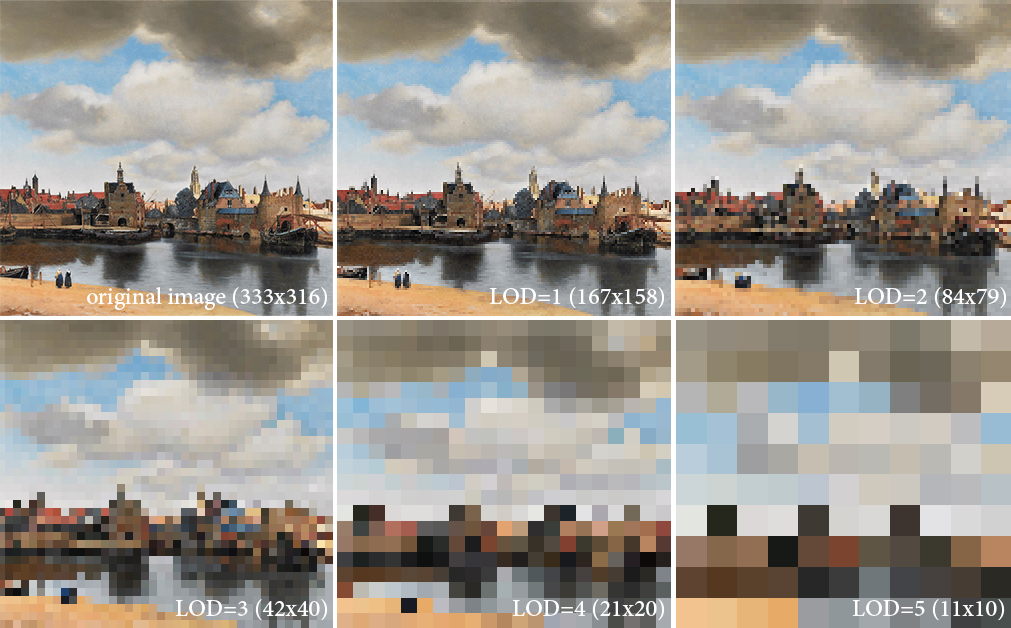}
\decoRule
\caption[]{Mipmapping applied to 'View on Delft', by Johannes Vermeer (1660)}
\end{figure}

OpenGL provides the GLSL function \verb|textureLod(texture, point, lod)| which performs a texture look-up with an explicitly passed LOD value. An important property with regards to this thesis, is that texture coordinates passed to this function do not need to be integer values. For any floating-point coordinates that lie in-between the mipmap's pixels, the function will provide a color value computed in accordance with the currently active {\it{magnification filter}}.

\subsubsection{Texture Filtering}\label{tex_filtering}

Texture filtering methods typically pertain to one of two categories: {\it{magnification filters}}, which fill an images gaps, usually through the interpolation of sparse data, or {\it{minification filters}}, which combine superfluous color values into a single data point.

In the context of this thesis, minification filters will be used for the generation of mipmaps and magnification filters will be used to retrieve floating-point coordinate values from said mipmaps.

The following two commonly employed filtering techniques function identically for either filter type and will be employed for the remainder of this thesis:

\begin{itemize}
    \item Being the crudest and most trivial filtering method, {\textbf{nearest-neighbor interpolation}} simply returns the value of the texture element that is closest to the currently treated pixel.
    \item The significantly more sophisticated {\textbf{bilinear filtering}} method computes an average value of the four closest texture elements, with their respective contributions weighted by distance accordingly.
    
    Using this technique for magnification purposes results in a smooth color-gradient in-between all texture elements.
    
    When used for minification and more specifically mipmapping, the resulting, down-scaled image becomes a somewhat blurry, weighted average of the original. However, if the minifaction ratio exceeds 2.0, meaning that the image is being down-scaled to less than half its original dimensions, some texture elements will have no effect on the resulting outcome.
    
    Given that each mipmap level decreases the corresponding image size by an inverse factor of 2, all elements of a given texture will ultimately contribute to every mipmap generated for it.
    
    More importantly, the averaging nature of this method enables its usage as an inexpensive, GPU-accelerated color-sampler.
\end{itemize}

\begin{figure}[th]
\centering
\includegraphics[scale=0.4]{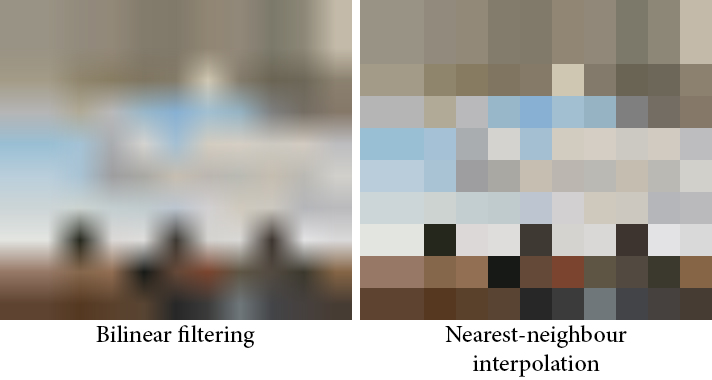}
\decoRule
\caption[]{Nearest-neighbour interpolation and bilinear filtering applied as a magnification filter}
\end{figure}

As implied above, applying a bilinear magnification filter onto a bilinearly computed mipmap yields an average color value for every point on the texture plane, with the averaging-radius correlating to the mipmap's LOD value.

Put differently, as long as both magnification and minification filters are bilinear, calling the OpenGL function \verb|textureLod(texture=t, point=x, lod=l)| will yield the average color value at point \verb|x| on texture \verb|t| in a radius of $2^{\verb|l|}$ pixels.

As one may notice, this sampling method roughly fulfills the requirements set by the specified \verb|sample(...)| function used in \ref{raymarching} and is indeed Crassin's proposed approach towards volumetric color sampling utilized in voxel cone tracing.

However, in order to enable volumetric sampling of indirect light within a scene using mipmaps and texture filters, the scene will have to be stored in a bitmap-like data structure that mipmaps can be generated for in the first place.

\begin{figure}[th]
\centering
\includegraphics[scale=0.4]{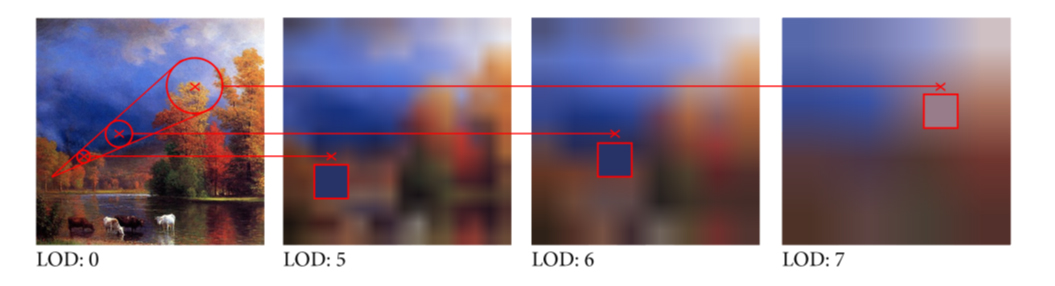}
\decoRule
\caption[]{Example of a 2D cone march using mipmaps of "On the Saco" by Albert Bierstadt (19th century)}
\end{figure}

\subsubsection{3D Texture}

OpenGL allows the texture filtering methods described above not only to be applied to regular, two-dimensional textures but also three-dimensional ones.

Otherwise functioning in exactly same way as regular bitmaps do, {\it{3D textures}} attach an additional spacial dimension of color values to an image, forming a so-called {\it{uniform grid}} of texture elements.

Similarly to how 2D bitmaps are usually represented as an array of pixels or squares, a 3D texture is often visualized as a grid of cubes, adequately termed {\it{voxels}} (combination of the words {\it{volume}} and {\it{pixel}}).\cite{foley}

The fact that all OpenGL functions that handle 2D texture data have their respective counterpart for the utilization with 3D textures, allows the color sampling method described in \ref{tex_filtering} to be used for volumetric sampling.

The only remaining task to complete the voxel cone tracing algorithm thus becomes storing the scene's direct lighting data into a corresponding 3D texture, a process known as {\it{voxelization}}.

\begin{figure}[th]
\centering
\includegraphics[scale=0.5]{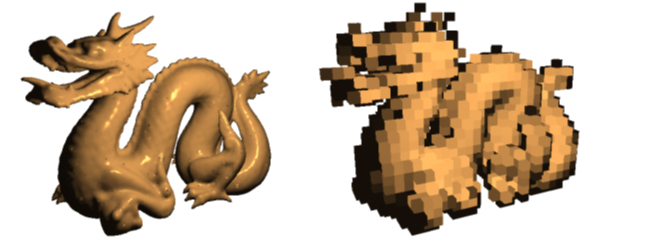}
\decoRule
\caption[]{Example of direct light (left) and corresponding 3D texture (right).}
\end{figure}

\section{Voxelization Process}\label{voxelization_process}

Combining the above described techniques enable an efficient way of sampling a specified volume of a 3D texture for an average color value. 

As per specification, a cone-trace is meant to provide an approximate value of the indirect light coming from the cone's direction. For the sake of simplicity and more importantly, performance, only one bounce of light will be considered, meaning that all indirect light will be caused by the immediate reflection of direct light.

As a result, all indirect light heading through the cone's body towards its origin can be estimated by simply averaging the direct light reflected by any surfaces the cone encounters.

Consequently, the scene's direct, diffuse lighting data (calculated using Phong) will first have to be injected into a corresponding 3D texture of the scene. The task at hand is thus, to convert a scene composed of triangle meshes into a 3d grid representation of their Phong-model colors.

Utilizing the benefits of GPU acceleration by voxelizing the scene directly on the GPU itself through the use of shaders, can greatly improve the time required for this process. 

To start, a 3D texture of RGBA voxels is created on the GPU with all RGBA values being equal to 0.

\subsection{Rasterization}

As described in \ref{frag_shader_general}, a fragment shader is executed at least once for every fragment that maps onto a to-be-rendered triangle. In this sense, by applying an orthographic projection matrix in the vertex shader, each call of the fragment shader will, in essence, correspond to a voxel projected onto said triangle. The size of the corresponding 3D texture is equal to the fragment resolution that is being rendered, or rather, the resolution of the viewport.

\begin{figure}[th]
\centering
\includegraphics[scale=0.4]{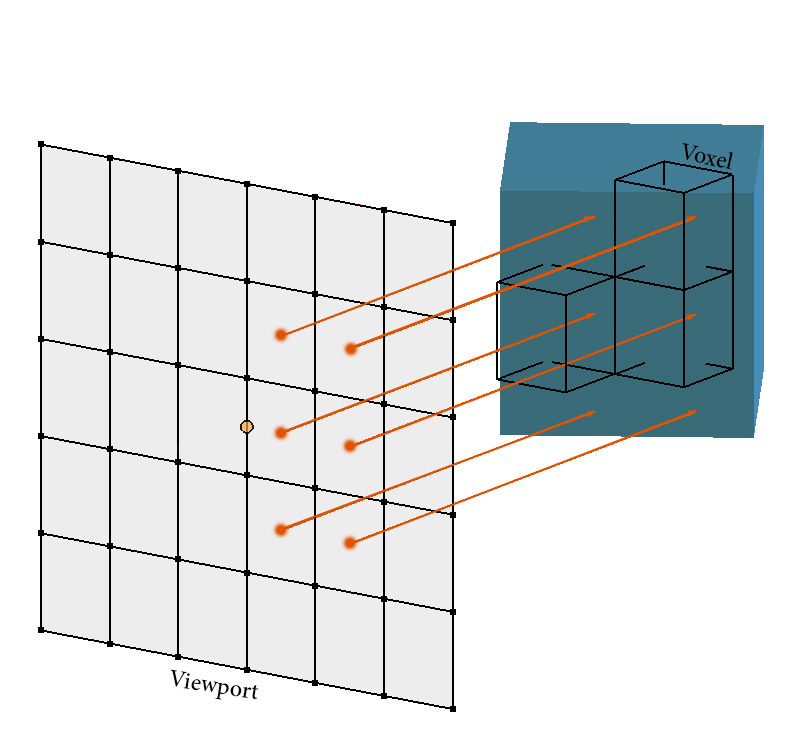}
\decoRule
\caption[]{Voxelizing by projecting onto the viewport.}
\end{figure}

The execution of this process as a graphics pipeline on the GPU enables an almost instantaneous voxelization of any modestly sized scene onto a reasonably small 3D texture (such as 64x64x64).

\begin{figure}[th]
\centering
\includegraphics[scale=0.35]{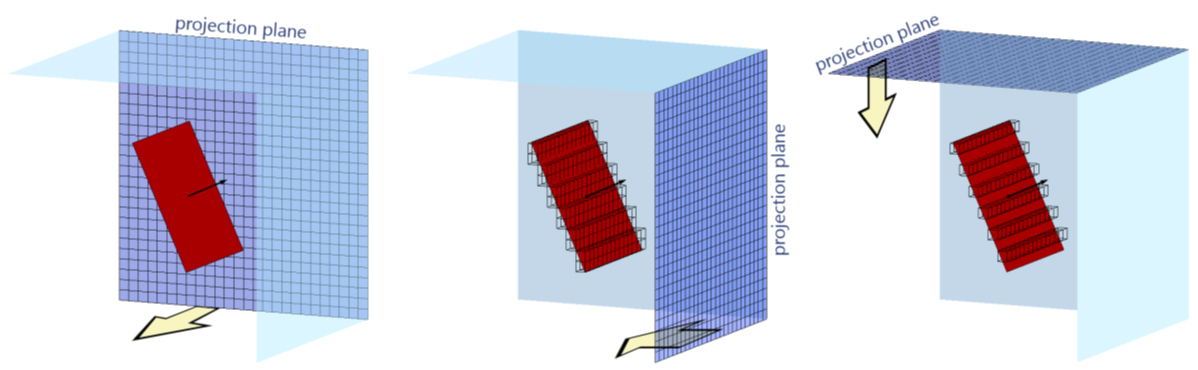}
\decoRule
\caption[]{Resulting voxelization of a surface from three possible directions of rasterization.}
\label{3wayvox}
\end{figure}

However, as described so far, the rasterization process will voxelize the scene only from one direction, which can leave large gaps in the 3D texture if surfaces are not facing the camera, as seen in fig. \ref{3wayvox}.

\subsection{Dominant Axis Selection}\label{dom_axis_selsection}

One possible, albeit inefficient solution to the above stated problem would be to repeat the voxel rasterization process three times, once for each spatial axis, rotating the camera accordingly in-between each cycle.

Alternatively, a far simpler approach is to perform a series of orthogonal rotations on the triangles themselves so that most of their surface is exposed to the camera viewport and subsequently rasterize each triangle only once.

Since the 3D texture constitutes a uniform grid the values of which are meant to represent the corresponding coordinate in space, all rotations performed on a triangle have to be multiples of $\frac{\pi}{2}$rad around each spatial axis.

The simplest way of determining which axis-plane a given triangle projects the most surface area on is to look at its surface normal vector and determine which component is the largest.

Given three vertices $\harpoon v_1, \harpoon v_2$ and $\harpoon v_3$ that constitute a triangle, the surface normal $\harpoon n$ is easily computed through the cross-product of two vectors lying inside the triangle plane:

\begin{equation}
    \harpoon n = (\harpoon v_2 - \harpoon v_1) \times (\harpoon v_3 - \harpoon v_1)
\end{equation}

The {\it{dominant axis}} of this triangle now corresponds to the largest component of $\harpoon n$, meaning that the axis the normal shares its largest component with, will be the direction of voxel rasterization. The triangle will subsequently have to be rotated so that its dominant axis faces the retina. These steps are best performed in the geometry shader.

Assuming that the scene is being rasterized to an orthographic camera looking along the Z axis, the following listing provides an outline of the geometry-shader steps performed where \verb|v1, v2, v3| are the triangle vertices:

\begin{lstlisting}[caption={Dominant Axis Selection (Pseudo-code)}]
n = abs( cross( v2 - v1, v3 - v1 ) )
dom_axis = max( n.x, n.y, n.z )

for v in [v1, v2, v3]:
    // Last coordinate (Z value) corresponds to the triangle's dominant axis.
	if dom_axis == n.x: v.xyz = v.zyx
	if dom_axis == n.y: v.xyz = v.xzy
	if dom_axis == n.z: v.xyz = v.xyz
\end{lstlisting}

\begin{figure}[th]
\centering
\includegraphics[scale=0.15]{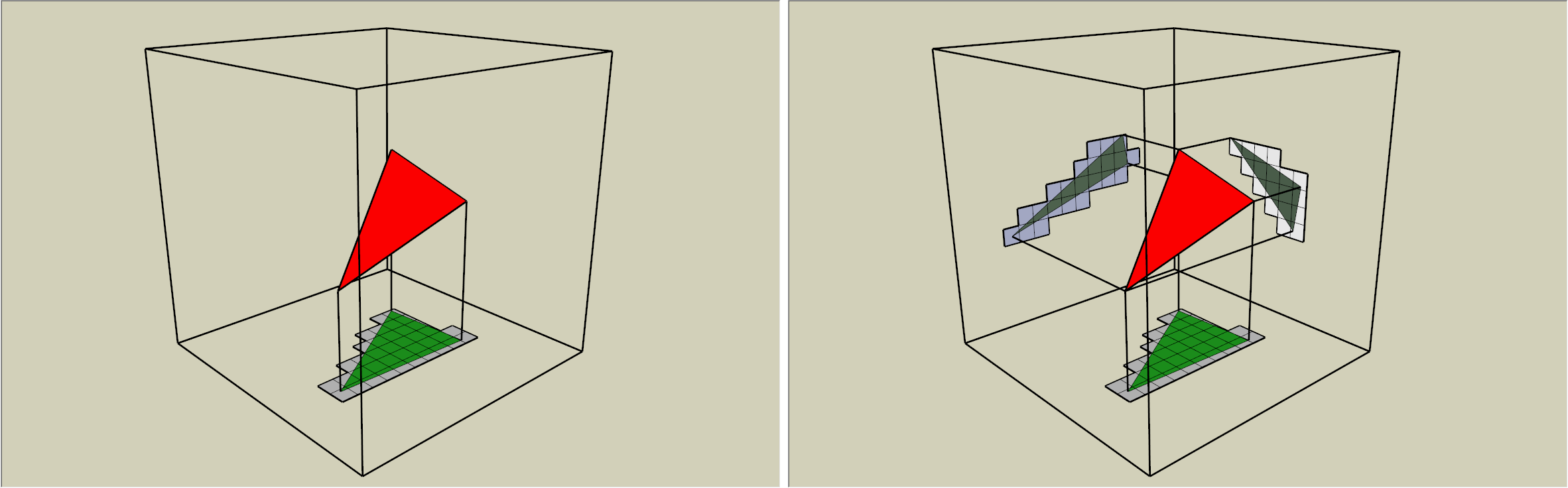}
\decoRule
\caption[]{Dominant axis selection: Three potential directions to project a primitive. (Takeshige\cite{takeshige})}
\end{figure}

To summarize, in the voxelization shader pipeline, the vertex shader applies an orthographic projection matrix to an object, the triangles of which are subsequently rotated towards the camera's forward axis in the geometry shader.

Handing these newly rotated triangles over to the rasterization stage will execute the fragment shader once for every voxel on that surface.

\subsection{3D texture storage}

To conclude the voxelization process, the direct-light color value of a surface point needs to be stored in the corresponding voxel of the 3D texture.

These color values can be computed in exactly the same way as is done in a regular Phong shader (see listing \ref{phong_pseudo}), the only difference being, that the glossy light component can simply be discarded, as it depends on the cameras position.

Furthermore, note that one must not use the triangles {\it{actual}} surface normal for Phong calculations, as it may have been changed in the geometry shader and would thus not provide the correct lighting values. Instead, the surface normal the triangle had before the dominant axis selection stage must be used here.

Once the correct Phong value has been computed, the corresponding voxel on the 3D texture is identified by interpolating from the fragments clip-space coordinates into texture space coordinates.

Once the right voxel has been identified, the GLSL function \\\verb|imageStore(texture, location, color)| can be used to store the calculated color into the 3D texture. As the texture consists of RGBA values, the alpha of a voxel can simply be set to 1 for every fragment rasterized.

This way, the alpha value can be used to represent the 'solidity', or occlusion, of a voxel, meaning that 0 would correspond to empty space and 1 would correspond to an objects surface.

\section{Summary}

Before the voxel cone tracing method can be employed, the scene must first be voxelized.

The voxelization process can be fully executed on the GPU by employing a regular shader pipeline.

However, instead of rendering images onto the screen, the voxelization-pipeline is used to store a 3D texture representation of the scenes direct lighting values.

In order to achieve this, the viewport size must first be set to the equivalent size of the 3D texture. For example, a 64x64x64 sized 3D texture will require a 64x64 large viewport.

The vertex shader will, as per usual, apply a model and view matrix to the objects in question. Additionally, an orthographic projection matrix is applied instead of a regular isometric one.

Subsequently, the geometry shader will select each triangles dominant axis and rotate them to maximize the rasterized surface area.

Each rasterized fragment in the fragment shader now corresponds to a voxel in the 3D texture, which is assigned the corresponding Phong color value.

Afterwards, series of mipmaps are generated for the resulting 3D texture using bilinear filtering methods. The values in the mipmaps now correspond to the average direct light emitted by the surfaces in a volume specified by the given LOD level.

By using a bilinear magnification filter for areas in-between the mipmap elements, a volumetric sampling function that yields both average color and occlusion can be created as follows:

\begin{lstlisting}[caption={Color supersampler (Pseudo-code)}]
sample(position, diameter):
    v_level = (!$log_2$!) (diameter / tex_size)  // Mipmap LOD
    pos_texspace = position / scene_size * tex_size
    color.rgba = textureLod(voxelMap, pos_texspace, v_level)
    return color
\end{lstlisting}

where \verb|tex_size| is the size-vector of the 3D texture, \verb|scene_size| is the size-vector of the scene and \verb|voxelMap| is the 3D texture itself.

This sampling function can be used for a volume ray marching cone-trace algorithm as outlined in \ref{raymarching}. Each cone-trace will deliver an approximation of the indirect light heading towards the cones origin at the cones aperture angle.

The voxel cone tracing fragment shader will isotropically trace a number of cones outwards to determine indirect diffuse light. Furthermore, one cone is traced in the direction of perfect reflection to assess indirect specular light and one cone per light source is traced to estimate occlusion.

Direct diffuse and glossy values are calculated using the Phong model, multiplied with the garnered occlusion value and thereafter combined with the attained indirect diffuse and specular values into the final fragment output.

\begin{figure}[th]
\centering
\includegraphics[scale=0.15]{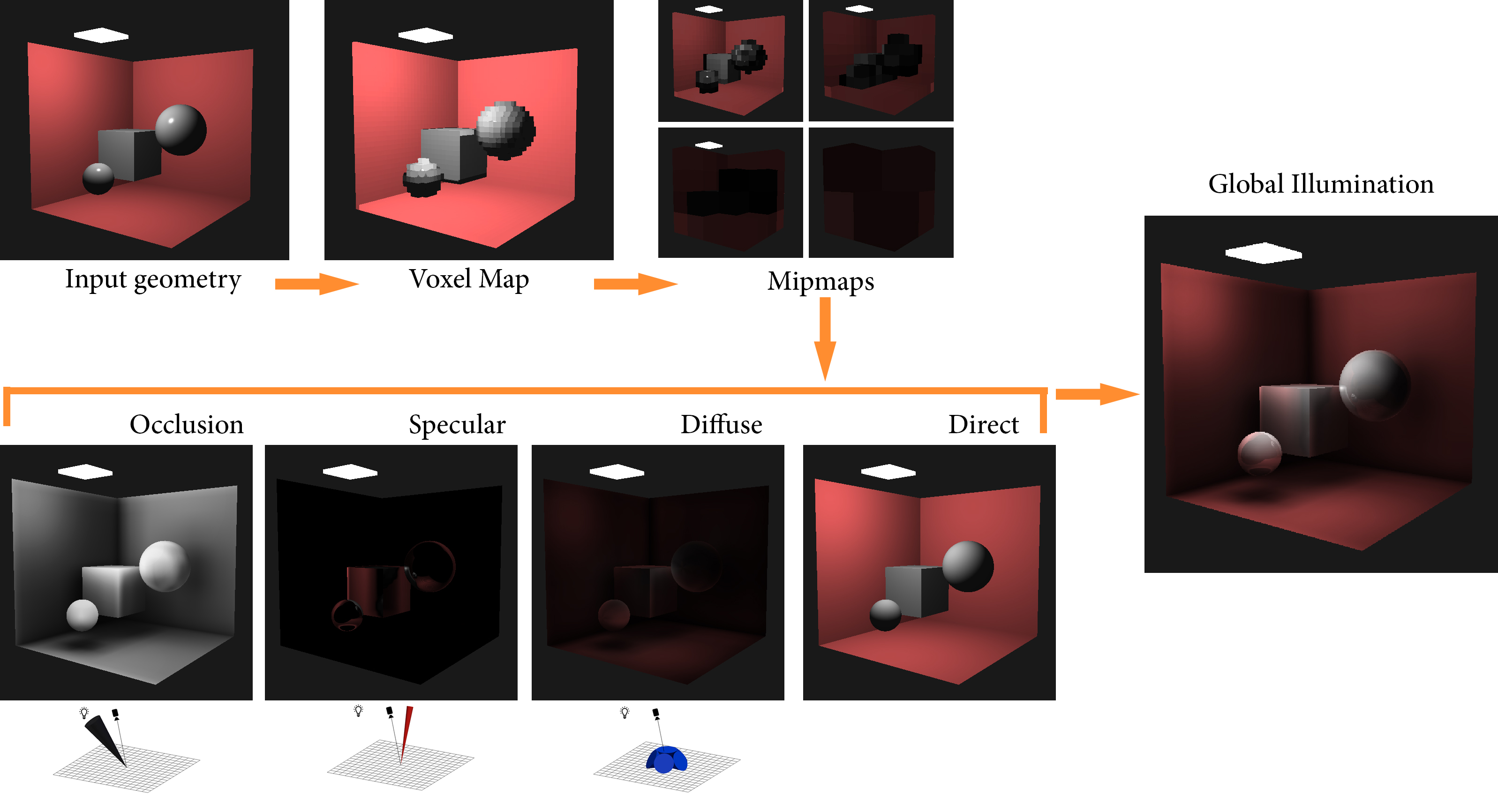}
\decoRule
\caption[]{Overview of the voxel cone tracing algorithm}
\end{figure}

%% file: Chapters/Chapter5.tex

\chapter{Implementation Details} 

\label{Chapter5} 

The rendering algorithm described above was implemented as an OpenGL application. This chapter discusses the details of said program.

The complete source code is available at: \url{https://github.com/Helliaca/VXCT}

\section{Previous Work}

\subsection{Crassin et al.}
The variant presented by Crassin et al.\cite{Crassin} in 2011 has served as the general vanguard of voxel cone tracing algorithms and is undoubtedly a major influencer of the implementation presented here.

Otherwise mostly mirroring the processes described in chapter \ref{Chapter4}, Crassin's adaptation comes with one key distinction to the described algorithm: Instead of simply storing direct lighting data into a regular 3D texture, a {\it{Sparse-Voxel-Octree}} (SVO) is employed instead.

An octree organizes data into a tree-like hierarchy where each internal node possesses exactly eight children. This type of data structure is a regularly used technique to partition three dimensional space using recursive subdivision. In this case, one node represents a voxel and its children, the octants, represent the subdivision of that voxel into eight others.

Since octree nodes can be instantiated dynamically during runtime, this approach comes with the great benefit, that the memory of the entire 3D texture does not need to be allocated at all times, greatly relieving GPU memory strain.

While certainly advantageous for efficiency, storing lighting data in a sparse octree instead of a a regular grid poses a far greater challenge to implement and navigate.

In addition to a different data structure, the voxelization procedure presented by Crassin et al. undergoes three rasterization passes, once for each spatial axis. The improved variant requiring only one rasterization pass, thanks to dominant axis selection (see \ref{dom_axis_selsection}), was introduced later in 2012\cite{crassin_voxelization1}\cite{crassin_voxelization2}.

Further improvements to this process, such as {\it{conservative rasterization}} and others, can provide more adequate voxel representations to a scene, but will not be covered in this thesis.

\subsection{Präntare}
During the development process, the complete source code of a well-documented implementation made available by Fredrik Präntare\cite{friduric} provided additional insight on a code-basis to the inner workings of voxel cone tracing shaders.

The simpler approach of storing direct lighting data into a 3D texture as well as the general fragment shader structure were adopted from here.

\subsection{Villegas}
Jose Villegas \cite{villegas} presents an approach that significantly expands upon rudimentary voxel cone tracing by  merging it with a technique known as {\it{deferred shading}}.

In opposition to {\it{forward rendering}} (the technique contemplated thus far), deferred rendering, as the name implies, delays all lighting computations until every triangle has passed down the rendering pipeline. Subsequently, instead of a fragment shader, a so-called {\it{pixel shader}} computes lighting values on a pixel basis, thus avoiding redundant lighting calculation for overlapping polygons.

This method can provide a particularly enhanced performance with scenes that include a vast amount of light sources.

In the initial stage of voxelization, Villegas generates three different 3D textures from the input geometry. These albedo, surface-normal and light-emission textures are then combined with direct illumination values to generate a global illumination 3D texture within the voxelization stage itself.

This approach allows for the computation of lighting effects such as emissive materials, indirect shadows and multiple light bounces on a voxel cone tracing basis.

Furthermore, the generation of mipmaps is accomplished by utilizing anisotropic filtering methods which, according to Villegas\cite{villegas}, provide more precise results for cone traces.

\section{Program Specification}

The here proposed objective is a program that renders a scene from the perspective of a movable camera using the voxel cone tracing method.

The core components of an OpenGL based implementation can be roughly arranged into three components: The CPU rendering loop, the voxelization pipeline and the cone tracing pipeline.

At program start the CPU loads some given 3D model data and copies the vertex data into GPU memory. Since features such as vertex morphing are not part of the proposed objective, the copied vertex data remains the same for all frames rendered and does not need to undergo regular updates.

However, frame-based camera movement does still need to be accounted for. This is easily accomplished by defining a set of uniform shader variables that are synced between GPU and CPU  before every rendered frame. These include the model matrices and materials for all rendered objects as well as the view and projection matrix of the camera.

Once synced, a voxelization shader pipeline generates a corresponding 3D texture which is soon after mipmapped. This process is executed just once or in regular timesteps depending on whether if any of the scene's objects move.

A separate voxel cone tracing pipeline can then use the data contained within the generated 3D texture to render a global illumination picture.

\begin{figure}[th]
\centering
\includegraphics[scale=0.7]{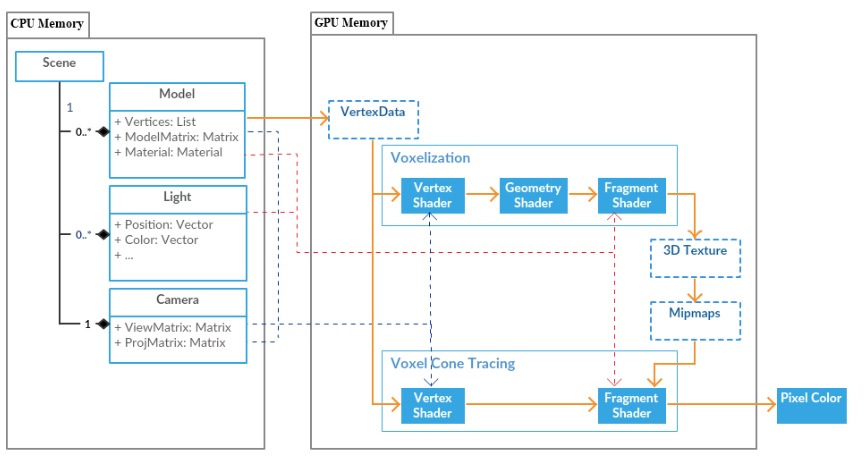}
\decoRule
\caption[]{Voxel cone tracing data flow diagram (dashed lines indicate uniform variables)}
\label{data_flow_diagram}
\end{figure}

Fig. \ref{data_flow_diagram} portrays a generalized data flow diagram of the proposed implementation.

\section{Working Environment}

The here presented implementation was performed with a C++ 17 compiler using Visual Studio 2017 and OpenGL 4.5.

The underlying hardware consisted of an AMD Ryzen 1700 CPU and an NVIDIA GeForce GTX 1070 GPU with 8088MB of display memory.

\subsection{Dependencies}

\subsubsection{3D Model format}

For the initial prototyping phase of this project, a set of primitive shapes such as a plane, triangle or cube were manually declared as arrays of coordinates.

Once this phase was satisfied with a rudimentary phong rendering setup, the ability to load and render increasingly complex objects became necessary to accurately assess the adequacy of the used algorithms.

The {\it{wavefront OBJ}} format is a widely popular data-format for the storage of polygon based 3D models. 
In gratitude to the multitude of software packages that allow for effortless importing and manipulation of OBJ data in C++ code, this was the general format opted for with regards to 3D models.

Highly detailed 3D models for testing purposes were obtained from the Stanford 3D Scanning Repository\cite{stanford}. Given that some models consisted of up to 116 million triangles, these were adequately decimated using open source software such as Blender and MeshLab to lower polygon counts.

\subsubsection{3D Model loading}\label{model_loading}

For the purpose of importing or loading wavefront OBJ models into the rendering environment, two different libraries were employed.

As the name implies, {\it{tinyobjloader}}\cite{tinyobjloader} is a light-weight C++ header file which is easily incorporated into a C++ project. While being remarkably convenient in the initial outlining of the voxel cone tracing algorithm,  the imported coordinate format provided by tinyobjloader considerably differed from the format used in the remainder of the program.

The vast majority of the program stored an object's vertices in a float-array with a stride of 6. The first three values of each stride would correspond to the position of the vertex and the latter three to its normal vector.

A further {\it{indeces}} array would reference these vertices by their index and arrange them into groups of three, forming a triangle.

Unfortunately, tinyobjloader imports 3D models into two separate position and normal-vector arrays. Each entry in the indeces array contains two separate indeces for each of the vertex and normal arrays.

This resulted in the need for a lengthy conversion process where normal-vector values had to be matched with their respective vertex values.

As this process could, on occasion, take up to several seconds, tinyobjloader was ultimately replaced by {\it{Assimp}}\cite{assimp}. 

Although a far larger software package, the Open Asset Import Library (Assimp) can accommodate a far greater number of formats than just wavefront OBJ. Additionally, the data is provided in exactly the same vertex, indeces -array layout as described above, leading to a great decrease in model loading time.

\subsubsection{OpenGL conext}

As discussed in \ref{opengl}, OpenGL merely sets specifications for a set of functions for GPU manufactures to provide. In order to utilize these functions in C++ program, an {\it{OpenGL context}} needs to be created.

A manifold of programming libraries like SFML, SDL or GLUT easily satisfy this purpose, and also provide additional helper classes and wrapper objects to facilitate the challenge that working directly with OpenGL would otherwise pose.

Implementing an adept rendering algorithm like voxel cone tracing requires working very closely with OpenGL functions, making most helper functions provided by libraries like SDL redundant.

For this reason, GLFW\cite{glfw} (Graphics Library Framework) was the chosen OpenGL development library for the purposes of this thesis, as it provides a high degree of control over OpenGL context creation.

As described by de Vries\cite{learnopengl}, GLFW is written in C and is equipped with only the most bare-bones necessities required for running OpenGL based applications. These include input handling and OpenGL context creation.

In addition, GLAD\cite{glad} was used as a {\it{loader-generator}} to retrieve the OpenGL-driver specific location of functions and reference these through respective function pointers for their use in the program. This allows for OpenGL functions to be called without the otherwise necessary manual retrieval of their specific locations.

\subsubsection{OpenGL Mathematics}

Any common shader based rendering process requires a significant amount of vector and matrix based mathematics.

Performing these calculations with manually declared data-types is a cumbersome and completely unnecessary procedure.

{\it{OpenGL Mathematics}} (GLM)\cite{glm} is a header-only C++ library which provides a plethora of convenient mathematical data-types and functions.

The rotation of a model, for instance, can be carried out by simply utilizing the \verb|glm::rotate| function on a model matrix, rather than performing a manual matrix multiplication.

Furthermore, GLM was designed specifically for usage with OpenGL in mind. As a result, the provided functions and data-types closely resemble those available in the GLSL shading language.

\section{Program Structure/Milestones/Overview}

The following section will cover the most significant milestones reached during the implementation process.

Note that the order in which the milestones are presented does not necessarily correspond to the exact chronological order of implementation, but instead provides a hierarchical overview in the programs complexity.

\subsection{Base Framework}

The core of the application consists of a general rendering setup constructed through regular object-oriented programming practices.

\subsubsection{Error Accountability}

Firstly, to aid the development process, an underlying abstract class called {\it{IoObject}} serves as the primary fundament that all other classes inherit from. It defines a {\it{name}} attribute for the identification of all objects as well as a custom \verb|print(IoObject, string)| function, which allows all produced output to be traced back to the instance that caused it.

Furthermore, it includes a general \verb|checkErrors(...)| method that relays OpenGL related errors through the aforementioned \verb|print| function. Whenever an error occurs, OpenGL simply sets an error flag, which can then be retrieved using the \verb|glGetError| function. As multiple errors of the same flag cannot be raised, it is recommended OpenGL practice to call \verb|glGetError| multiple times for each rendering pass until all flags are reset.\cite{opengl_docu}

Doing this on an individual instance basis additionally facilitates any required debugging, as an OpenGL related error can quickly be traced to the object and method that produced it.

Any class method that utilizes OpenGL functions must simply include a finishing \verb|checkErrors| call with an appropriate string that describes the current context passed as a parameter.

For instance, calling \verb|checkErrors("Initialization")| at the bottom of a VoxelMap class constructor will produce the following output if a non-existent filtering method is set for a texture's magnification filter:

\begin{lstlisting}[caption={Erroneous VoxelMap Initialization Output}]
voxel_map >> OpenGL Error:
voxel_map >>     CONTEXT: Initialization
voxel_map >>     TYPE: Invalid Enum
\end{lstlisting}

\subsubsection{Base Classes}

\begin{figure}[th]
\makebox[\textwidth][c]{\includegraphics[width=1.3\textwidth]{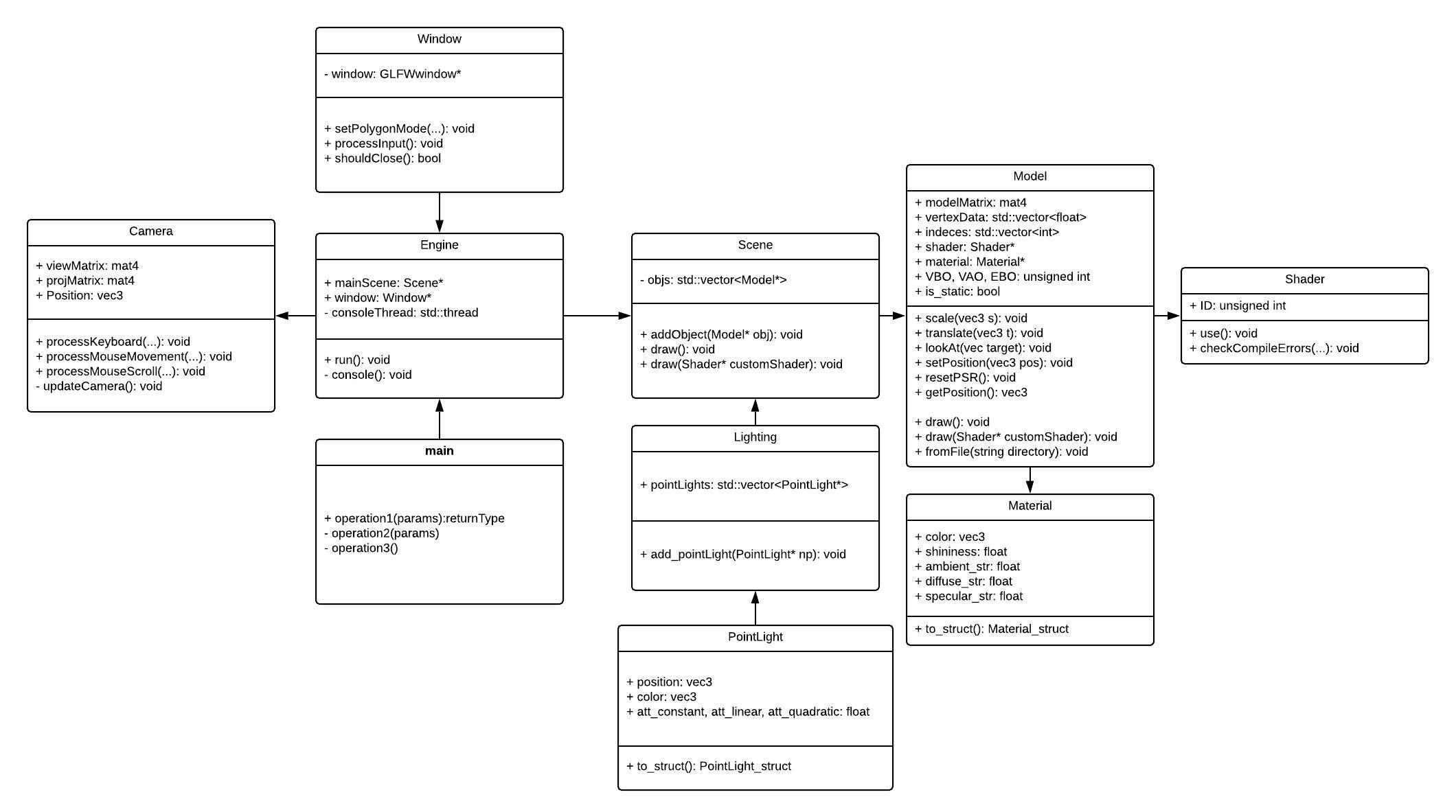}}
\caption[]{UML class diagram of proposed implementation}
\label{uml}
\end{figure}

Fig. \ref{uml} shows an UML class diagram of the base rendering framework used in this project. All classes portrayed directly inherit from the IoObject class and are thus equipped with the above listed functionalities. Keep in mind that, for the sake of visibility, some attributes and methods like contructors are not listed here.

Note that in the actual implementation, objects of the {\it{Lighting}} and {\it{Camera}} classes are not solely included by the {\it{Scene}} and {\it{Engine}} classes respectively, but are instead referenced in a \verb|globals.h| header file and namespace that is accessible across most of the program.

This design choice was made because camera and light dependent constructs such as view matrices and light positions are necessary in many areas of the application and are thus easier to access via globally defined pointers.

The intended features and purposes of each class are as follows:

\begin{itemize}
    \item {\textbf{Engine:}} An object of the Engine class is instantiated in the program's \verb|main| function. Subsequently, the respective \verb|run| method is executed which sets the bulk of the application in motion.
    
    An OpenGL context is created with a corresponding {\it{rendering loop}}, which continuously clears and draws into the context-window. Initially a default scene is loaded through the {\it{Scene}} class and rendered using rudimentary Phong shading techniques (see \ref{impl_phong}).
    
    In parallel, a separate thread, called \verb|consoleThread|, reads input from a command shell through the \verb|console()| function, allowing the user to execute actions and adjust runtime variables through a set instructions.
    \item {\textbf{Camera:}} The general implementation of a camera class written by de Vries\cite{learnopengl} was used as an underlying basis that was adjusted to fit the surrounding code-environment of the project.
    
    As the name implies, the class represents a camera or retina that images are rendered for. Functions like \verb|ProcessKeyboard()| and \verb|ProcessMouseMovement| will adjust the camera's position and rotation accordingly whenever mouse-movement or key-presses are detected.
    
    Every time this occurs, a publicly accessible view- and projection matrix are updated using the camera's directional axes.
    
    Once rendering takes place, these values are then sent to the GPU for the vertex shader to apply.
    \item {\textbf{Window:}} The Window class serves as a wrapper object for the GLFWwindow class and represents the rendering window of the application.
    
    While active, input from the keyboard and mouse are processed here, calling the respective functions of the globally defined scene camera object accordingly.
    
    Furthermore, some rendering settings such as {\it{wireframe}} mode or double-sided triangles are set through the \verb|setPolygonMode| function.
    
    Once the conditions of the \verb|shouldClose| function are met, the parent engine will terminate the application.
    \item {\textbf{Scene:}} A Scene object contains a list, more specifically a \verb|std::vector|, of pointers referencing {\it{Model}} objects. Once the \verb|draw| method is invoked, all carried objects will be rendered using their own specified shaders. Alternatively, a different shader can be passed as a function argument, leading to all objects labeled as {\it{static}} to be drawn with this shader instead.
    
    This feature can, for instance, be used to switch between phong and voxel cone tracng shaders during runtime.
    \item {\textbf{Lighting:}} Similarly to the Camera, a Lighting object which contains a list of point-lights in the scene is accessible in most parts of the program.
    
    As done in the Scene class, PointLight objects are referenced through a \verb|std::vector| of pointers.
    \item {\textbf{PointLight:}} A Pointlight entity contains all necessary data for the definition of a point light, as described in \ref{lightsource_types}. This includes the position and emitted light color as well as three attenuation constants that describe the light's radial falloff in accordance to \ref{Attenuation}.
    
    Given that class-based objects cannot be utilized on the GPU, a conversion method \verb|to_struct| exists that produces an equivalent C struct which can be passed on to a fragment shader for lighting calculations.
    \item {\textbf{Model:}} The Model class is used to represent all 3D models in a scene. The \verb|vertexData| and \verb|indeces| attributes contain the complete polygonal data as described in \ref{model_loading}.
    
    Data can be loaded into these vectors by the utilization of the \verb|fromFile| method, which uses the aforementioned Assimp library to do so.
    
    Working under the assumption that the total amount of rendered models remains relatively low, three integers identifying an object's individual VAO, VBO and EBO are defined.
    
    The \verb|modelMatrix| attribute is a transformation matrix that determines the relative rotation, scale and position of the 3D model. These values can be manipulated through by calling functions such as \verb|scale(...)|, \verb|translate(...)| or \verb|lookAt(...)|. The \verb|resetPSR()| method will undo all of these modifications by instating an identity matrix.
    
    Similarly to the Scene class, the \verb|draw()| method will render the object to the screen using the instance's self-referenced shader, if no alternative shader is supplied.
    
    A further key component of this class is that each model instance contains several dictionaries (\verb|std::map|) of shader references. Before rendering can occur, all uniform variables of the shader pipeline, including light positions, MVP matrices etc., need to be set correctly. This is best done by allowing each model to keep a reference for every necessary shader uniform.
    
    Matrices would, for instance be referenced in a \verb|std::map<std::string, glm::mat4>| variable, where the string denotes the shader uniform's name. Before every rendering cycle, all the containing values would first be iterated over, setting their respective shader uniforms accordingly.
    
    To avoid unnecessary clutter, these reference dictionaries were omitted in fig.\ref{uml}, given that a total of six different data-types where used.
    \item {\textbf{Shader:}} Representing a complete shader pipeline on the GPU, each instance of this class carries a generated ID number that identifies the corresponding shader program on the GPU memory.
    
     The class template presented by de Vries\cite{learnopengl} was used as a point of departure and significantly expanded upon.
    
    The source codes of corresponding vertex, fragment and optionally geometry shaders are read from specified files, transferred to GPU memory and subsequently compiled into a completed shader program, distinguished by it's ID.
    
    This process is done inside the class constructor, with the \verb|checkCompileErrors| method allowing for shader debugging after compilation.
    
    Although omitted in fig.\ref{uml}, the shader class also provides functions for setting uniform variables of a wide range of data-types. These functions are used by the Model objects to ascribe all uniform variables their desired values before rendering takes place.
    \item {\textbf{Material:}} Functioning similarly to the PointLight class, the Material class only functions as container the necessary data to describe the reflective behaviour of an object's surfaces. This data can then be converted into a struct format with the \verb|to_struct| function.
    
    The four different floating-point numbers stored here are equivalent to their counterparts defined by the Phong reflection model in (\ref{material_components_phong}). Additionally an RGB vector denotes the material's surface color.
    
    Working under the assumption that every model consists of only one material, the respective classes undergo a direct, one-to-one associative relationship.
\end{itemize}

\subsection{Rendering Process}

The output produced by shaders can be influenced through the adjustment of uniform variables, which need to be assigned before every rendering cycle.

As described in the class descriptions above, the Model class is given reference dictionaries of string uniform names and pointer variables to their corresponding values. Once the object's draw method is called, these dictionaries are iterated over and all the values are assigned accordingly.

Since pointers reference a memory address rather than a variable's own value, they will automatically and continuously be updated even when accessed by different parts of the program.

Should further uniforms be required by a shader, as for instance with a 3D texture for voxelization, they will need to be manually assigned.

\subsection{Phong Shading}\label{impl_phong}

The current framework allows for a wide range of different shading techniques to be run, including the likes of Phong. For a simple Phong shading model only a vertex and fragment shader are necessary.

The vertex positions and normal vectors handed to the vertex shader were named \verb|pos_vs| and \verb|nrm_vs| respectively. The ultimate task of vertex shader is to compute the corresponding screen-space positions by applying the view-, model- and projection-matrices to the received parameters.

The first step consists in extrapolating the vertex position an normal vector in {\it{world space}}. This is achieved by multiplying the $4\times4$ model matrix $M$ with the position vector $\harpoon p$ and the transposed inverse of the model's rotation matrix with the normal vector $\harpoon n$. The rotation matrix is equivalent to the upper-left, 3x3 sub-matrix of the model matrix.

\begin{equation}
\harpoon p_{world} = M * \begin{pmatrix}p_{x} \\ p_{y} \\ p_{z} \\ 1\end{pmatrix}
\end{equation}

\begin{equation}
\harpoon n_{world} = 
\left(
\begin{pmatrix}M_{1,1} & M_{2,1} & M_{3,1} \\ M_{1,2} & M_{2,2} & M_{3,2} \\ M_{1,3} & M_{2,3} & M_{3,3} \end{pmatrix}^{-1}
\right)
^T * \harpoon n
\end{equation}

These values can now be transferred over to the fragment shader for lighting computation.

The final, screen-space vertex position results from the multiplication of the projection and view matrices (in that order) with the world position vector.

\begin{equation}
\harpoon p_{screen} = P * V * \begin{pmatrix}p_{world, x} \\ p_{world, y} \\ p_{world, z} \\ 1\end{pmatrix}
\end{equation}

The information of where to draw which vertex is relayed to the GPU by assigning it to \verb|gl_Position|. 

The main body of the corresponding vertex shader looks as follows:

\begin{lstlisting}[caption={Phong Vertex Shader (GLSL)}]
void main()
{
    // pos_fs and nrm_fs are passed to the fragment shader
    pos_fs = vec3(model_u * vec4(pos_vs, 1.0));
    nrm_fs = mat3(transpose(inverse(model_u))) * nrm_vs;  
    
    gl_Position = proj_u * view_u * vec4(pos_fs, 1.0);
}
\end{lstlisting}

The subsequent fragment shader is mostly identical with the sample code displayed in \ref{phong_pseudo}, the only distinction being, that the same process is repeated for all lights in the scene with the values being added into the final result.

\begin{figure}[th]
\centering
\includegraphics[scale=0.7]{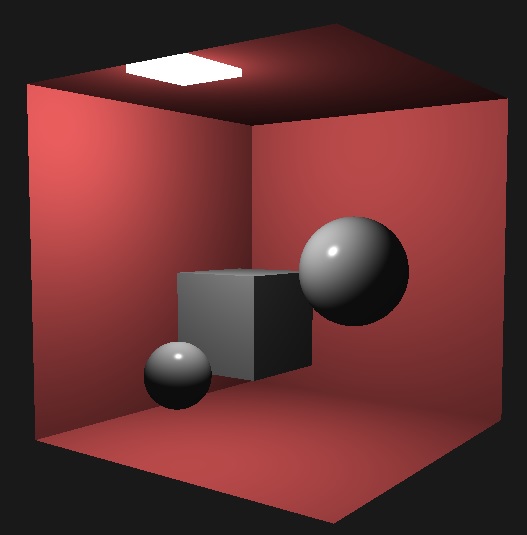}
\decoRule
\caption[]{Phong-based direct illumination rendering of primitive shapes}
\end{figure}

\subsection{Indirect Illumination}

\subsubsection{Additional components}

As laid out in \ref{raymarching}, the nature of a cone trace mainly depends on three parameters: An initial offset value $t_0$, an aperture angle $\gamma$ and a distance factor $\beta$.

To permit real-time adjustments to these values to be made, a \verb|VoxSettings| class was established. 

An object of this class is accessible via the \verb|globals| namespace and includes three separate instances of the above listed values for diffuse-, specular- and shadow-cones respectively. A series of booleans allow the user to toggle the various lighting components, further aiding the process of adequately fine-tuning these parameters.

Well into the development process, two additional settings, \verb|shadow_str| and \\ \verb|shininess_falloff|, were included to regulate occlusion strength as well as the employed specular BRDF.

In addition to the \verb|VoxSettings| class, a \verb|VoxelMap| class was introduced, representing the utilized 3D texture.

Following common OpenGL texture practices, the class includes a \verb|textureID| integer and is specified on GPU memory using the \verb|glTexStorage3D| function. An additional \verb|use| method activates the texture as a uniform image or sampler variable, allowing a shader to read or write values respectively.

Before (re-) voxelization can occur, a \verb|clear()| method assigns every voxel of the texture a $(0, 0, 0, 0)$ color. The actual voxelization process is executed by the \verb|Engine| class.

In addition, the \verb|VoxelMap| class also includes storage pointers as well as \verb|retrieveData| and \verb|updateMemory| methods which copy the 3D texture from GPU memory back to CPU memory, allowing the user to visualize the texture through the \verb|visualize| method.

\subsubsection{Voxelization}

As is outlined above, the model class is outfitted with a \verb|draw(customShader)| method that allows for rendering with a specifically passed shader.

This feature can be utilized to render the entire scene with the desired voxelization pipeline.
For this purpose, the \verb|Engine| class is given an additional method called \verb|voxelize()|.

The nature of this function is quite simple and performs the following steps:

\begin{itemize}
    \item All previous data contained in the scenes voxel map is cleared and all elements of the 3D texture are set to 0.
    \item The appropriate OpenGL settings for the voxelization procedure are set. This includes enabling double sided triangles, disabling depth tests and most importantly, setting the viewport to the size of the voxel map, \verb|vox_size|.
    \item The 3D texture corresponding to the \verb|voxelMap| object is set as the shader uniform variable that the voxelization shader writes to.
    \item The entire scene is rendered with the voxelization shader.
    \item Up to 7 levels of mipmaps are generated of the resulting 3D texture.
    \item If the option is set, all data written into the 3D texture is copied back on to the CPU, allowing the program to visualize resulting 3D textures. (See \ref{vox_vis}).
    \item All settings from step 2 are reverted back to their previous values.
\end{itemize}

The underlying voxelization shader follows the steps listed in \ref{voxelization_process}.
Here, the vertex shader is virtually identical with the one employed for phong shading, except that the camera's orthographic projection matrix is passed as a uniform instead.

The geometry shader performs the process of dominant axis selection and rotates the triangles to maximize the rasterized area.

The fragment shader then calculates the fragment's direct light value and inserts it into the adequate 3D texture element by calling \\\verb|imageStore(tex3D, location, vec4(result, 1.0f));|, where \verb|location| is the fragment's clip space position linearly interpolated into a range of $[0, vox_{}size]$.

Once this procedure concludes, the voxel cone tracing shader can be employed.

\begin{figure}[th]
\centering
\includegraphics[scale=0.25]{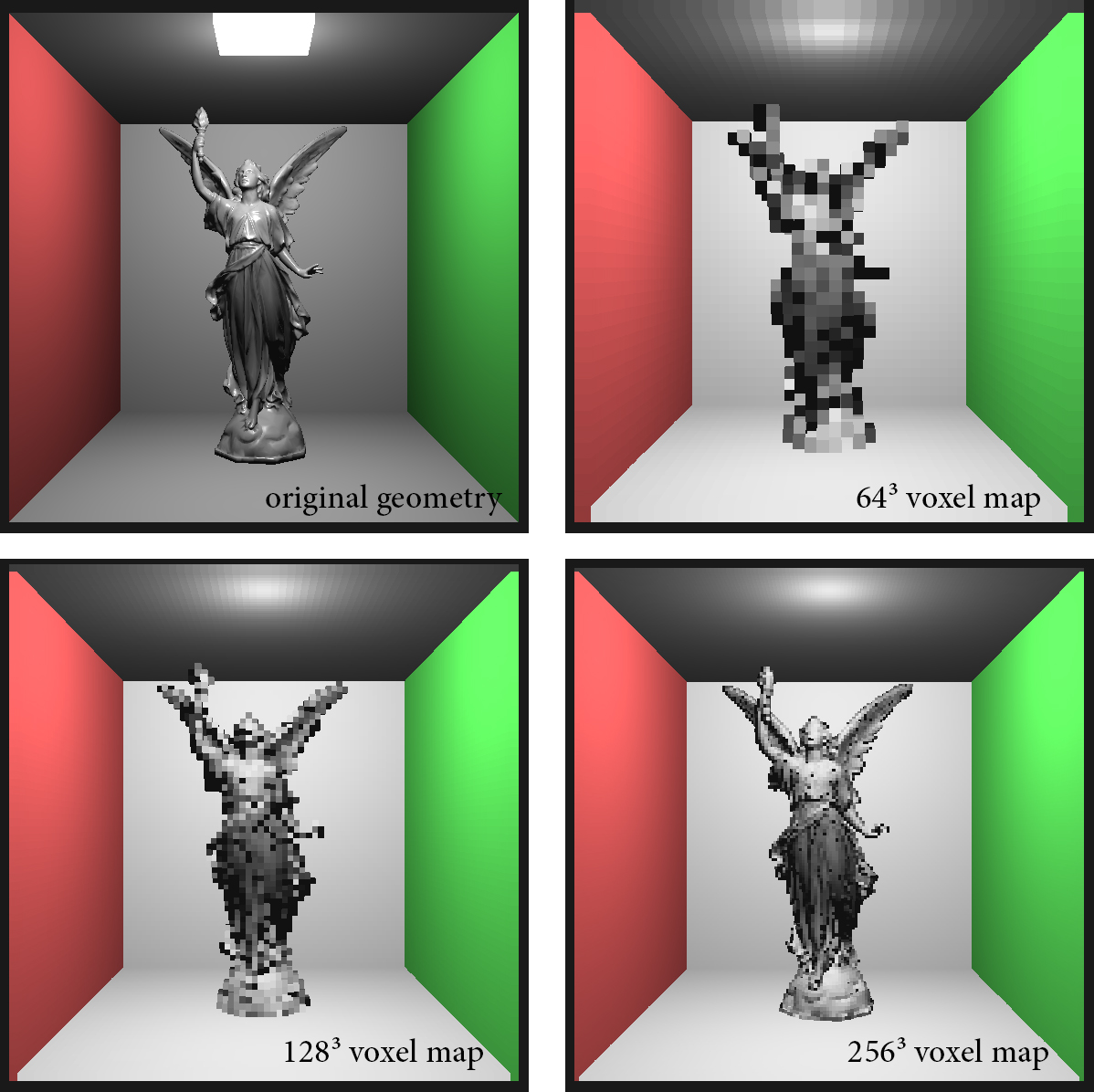}
\decoRule
\caption[]{Stanfords 'Lucy' in a Cornell box voxelized into 3 different resolutions.}
\end{figure}

\subsubsection{Cone Tracing}

With a 3D texture of the scene's direct light values in place, the cone tracing algorithm can commence. 

As per usual, the vertex shader simply performs the coordinate conversion process and then dispatches the vertice's world space coordinates which are adequately interpolated during the rasterization process.

The fragment shader is where most of the calculations take place. In addition to the fragment's world space coordinates and normal vectors, a plethora of additional parameters is required for lighting calculations.

Each fragment is dispatched with the position of the camera, the voxelized scene as a 3D sampler object, the surface material in question, an array of point-lights and a \verb|VoxSettings| struct.

The cone traces each follow an algorithm mostly equivalent to the one presented in \ref{raymarching}. However, by a long process of experimentation the individual functions were mildly modified to whichever attempt would yield the most aesthetically pleasing results.

Where the cone tracing algorithms differ from the model established in \ref{raymarching} is listed below:

\begin{itemize}
    \item {\textbf{Occlusion cones:}} To receive reasonably realistic shadows, the occlusion cone's aperture angle was chosen relatively large. However, due to the wide spread, shadow cones originating from corners tend to immediately collide with the neighbouring wall if the light-source is positioned at a narrow angle.
    
    This behaviour is generally sought-after, as it provides a degree of {\it{ambient occlusion}} to the scene. Unfortunately, the effect was somewhat too intense and led to very darkened corners.
    
    As a result, to mitigate the impact of short-distance occlusion, every sampled value was decreased in proportion to the square root of the distance travelled. Further smoothing was accomplished by performing a Hermite interpolation between said value and the maximum possible distance.
    
    The final front-to-back accumulation looked as follows:
    \begin{lstlisting}
    float occ_r = voxel.a * smoothstep(0.0f, max_dist, sqrt(current_dist)*shadow_str);
    occ = occ + (1 - occ) * occn_r;
    \end{lstlisting}
    
    Where \verb|occ| is the current occlusion, \verb|voxel.a| is the sampled occlusion and \verb|smoothstep| is a Hermite interpolation function.
    
    \begin{figure}[th]
    \centering
    \includegraphics[scale=0.4]{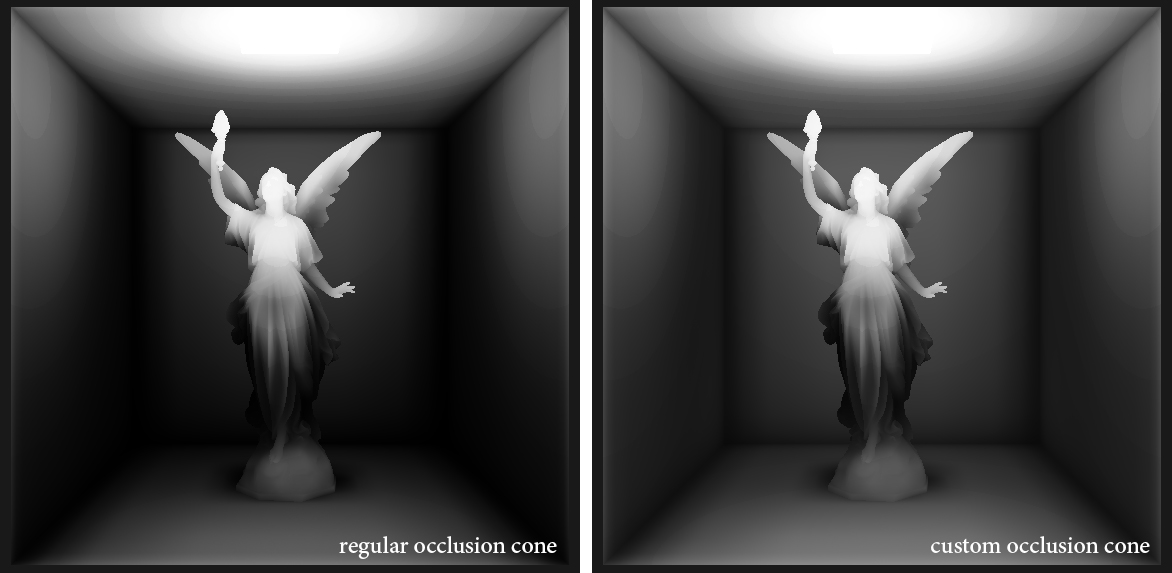}
    \decoRule
    \caption[]{Occlusion values as traced by a regular occlusion cone (left) and the customized occlusion cone (right).}
    \end{figure}
    
    \item {\textbf{Specular cones:}} As can be seen on the lower part of the statue in fig. \ref{spec_brdf_cones}, tracing a specular cone from a highly curved surface can cause it to collide with the opposite side of said surface. Despite this being a highly occluded area, the direct lighting of the surface hit is returned by the cone-trace, leading to unrealistic specular highlights.
    
    One conceivable option to solve this issue, is to multiply the specular component with the given occlusion. But having no specular highlights at all in dark areas did not seem like the desired effect either.
    
    Instead, a highly specialized BRDF depending on the angle towards the light-source was employed:
    
    Assuming $\harpoon \omega$ to be the travelling direction of the cone and $\harpoon l$ to be direction of the light-source, the angle $\gamma$ between them is calculated like so:
    \begin{equation}
        \gamma = \arccos (\harpoon \omega \cdot \harpoon l)
    \end{equation}
    
    An {\it{angular grace}} value $p$, then subtracts from the given angle to allow for some leeway for specular reflection to occur.
    \begin{equation}
        \gamma_n = \max (0, \gamma - p)
    \end{equation}
    The value $\gamma_n$ is large for undesired reflections and low for highly intensive ones.
    Dividing it by the maximum possible angle $\pi$, reciprocating the resulting fraction and then taking it to some high exponent $q$ results in the final BRDF:
    \begin{equation}
    \begin{aligned}
        f_{spec}(\harpoon\omega, \harpoon \omega_i, x) & = (1 - \frac{\gamma_n}{\pi})^q \\
        & = \Big(1 - \frac{\max (0, \gamma - p)}{\pi}\Big)^q \\
        & = \Big(1 - \frac{\max (0, \arccos (\harpoon \omega \cdot \harpoon l) - p)}{\pi}\Big)^q
    \end{aligned}
    \end{equation}
    
    \begin{figure}[th]
    \centering
    \includegraphics[scale=0.4]{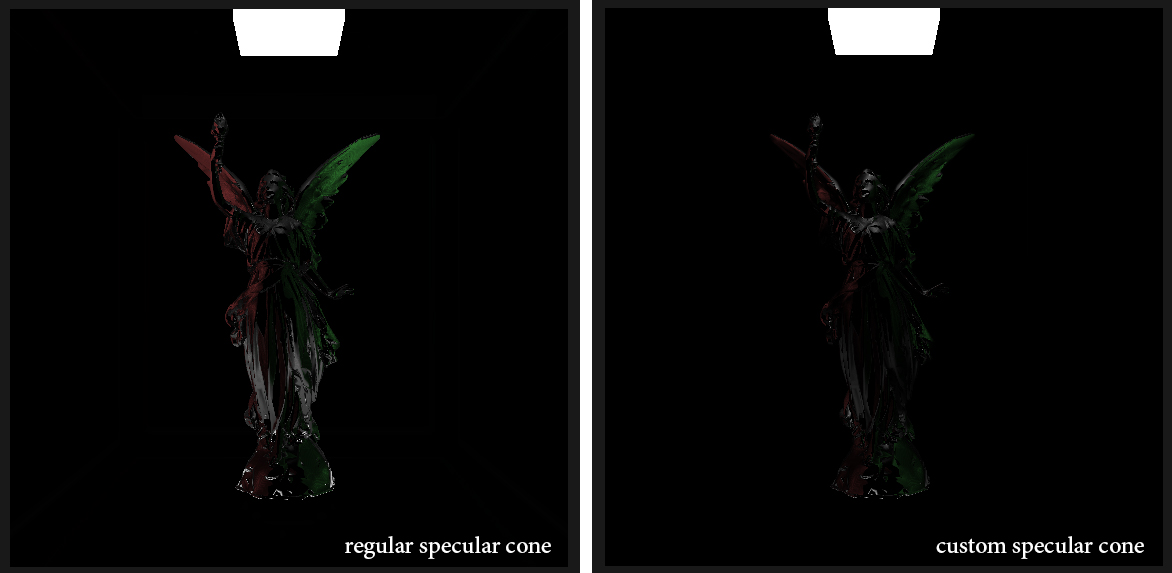}
    \decoRule
    \caption[]{Specular values as traced by a regular specular cone (left) and the customized specular cone (right). Note in particular the unrealistic grey highlight on the left variant.}
    \label{spec_brdf_cones}
    \end{figure}
    
    The fact that the proposed BRDF does not depend on the viewing direction $\harpoon\omega_i$, might appear as unusual at first glance, but since it is only applied for a cone travelling in the direction of specular reflection, this vector becomes redundant.
    
    Changing the parameter $p$ seemed to have a similar effect to that of changing the shininess value $\alpha$ in the phong shading model. Thus $p$ was defined to be some factor of $\alpha$:
    \begin{equation}
        p = const. * \alpha
    \end{equation}
    The constant was chosen to be $0.008*\pi$, since this provided sensible results for a shininess between 0 and 256, which are regularly chosen figures in the phong model.
    
    The parameter $q$ was termed \verb|shininess_falloff| and included separately in the \verb|VoxSettings| object. Sensible results would arise for $p$ values between 0 and 20, roughly correlating with the overall strength of the specular component.
    
    \item {\textbf{Diffuse cones:}} While mostly satisfactory, the tracing of diffuse cones often resulted in very dim indirect diffuse light.
    
    Surprisingly, a very crude approach utilized in the prototyping phase of this project seemed to provide the most aesthetically pleasing results.
    
    The approach in question immediately stops the cone-trace once a sampled volume has occlusion value larger than $\frac{1}{100}$ and returns the color sampled at that point.
    
    Albeit a very blunt method, the resulting images, rather appropriately for the purposes of this thesis, very clearly display the effects of indirect diffuse light.
    
    \begin{figure}[th]
    \centering
    \includegraphics[scale=0.4]{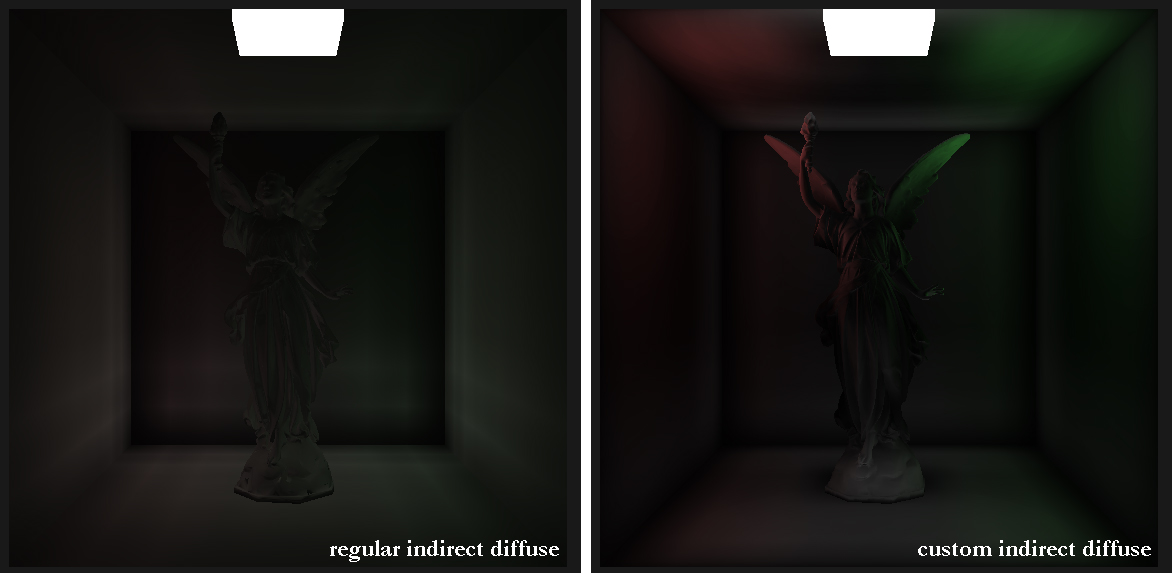}
    \decoRule
    \caption[]{Indirect diffuse component as traced by regular diffuse cones (left) and customized diffuse cones (right).}
    \end{figure}
\end{itemize}

The main body of the fragment shader computes the Phong-based, direct lighting value and multiplies it with a factor of occlusion yielded by tracing a shadow-cone towards the corresponding light source.

The classical voxel cone tracing algorithm proposed in chapter \ref{Chapter4} would add the indirect diffuse and indirect specular values atop.

However, due to the above listed modification to the cone-tracing algorithms, a slightly different approach is necessary:

Since tracing specular cones now depends on the direction of a given light-source $\harpoon l$, specular cones need to be traced for all given light-sources.

Furthermore, the employed diffuse cone-tracing function essentially returns the color of the first surface encountered. For this reason, the result must be appropriately obscured by the given occlusion value. Since occlusion values depend on light sources, but the diffuse component does not, the minimum of the recorded occlusion values is adopted. (Note that, due to multiplication, smaller occlusion values produce a stronger shadow.)

To recapitulate, recall the rendering equation for cone tracing from (\ref{vxct_rendeq_complete}):

\begin{multline}
L_{vxct}(x, \harpoon\omega) = f_{spec}(\harpoon R_{\omega}, \harpoon \omega, x)C(x, \harpoon R_{\omega}, \gamma_{spec}) + \sum_{q=0}^{u} f_{diff}(\harpoon \omega_q, \harpoon \omega, x)C_q(x, \harpoon \omega_q, \gamma_{diff}) \\+ \sum^{n}_{s=0} V(x, x_s) \Big(c_d i_{s,d} (\harpoon L_s \cdot \harpoon N) + c_s i_{s,g} (\harpoon R_s \cdot \harpoon \omega )^{\alpha} \Big)
\end{multline}

The here implemented approach calculates the specular component with respect to each light source. Furthermore, the indirect diffuse component is multiplied with the minimum value produced by the occlusion cone traces.

Defining $C_{spec}$, $C_{diff}$ and $C_{occ}$ to be specular, diffuse and occlusion cone traces respectively (as described above), the equation utilized in the present fragment shader would correspond to the following:

\begin{multline}
L_{impl}(x, \harpoon\omega) = o_{min} * \sum_{q=0}^{u} C_{diff}(x, \harpoon \omega_q, \gamma_{diff}) + \sum^{n}_{s=0} f_{spec}(\harpoon R_{\omega}, \harpoon \omega, x)C_{spec}(x, \harpoon R_{\omega}, \gamma_{spec}) \\ + \sum^{n}_{s=0} C_{occ}(x, x_s) \Big(c_d i_{s,d} (\harpoon L_s \cdot \harpoon N) + c_s i_{s,g} (\harpoon R_s \cdot \harpoon \omega )^{\alpha} \Big)
\end{multline}

where $o_{min}$ is the minimum value returned by $C_{occ}$ for that fragment. Note also, that $f_{spec}$ depends on the direction of the point light in question, since the sum $\sum^{n}_{s=0}$ would otherwise not be required.

Ultimately, a total of $2n+u$ cone traces are performed per fragment as opposed to the $n+u+1$ of a more classical implementation, implying a slightly worse performance for scenes with a large amount of light-sources.

\subsubsection{Cone Configurations}

The initial proposition made by Crassin et al. \cite{Crassin} does not specify an ideal cone configuration to be used for the indirect diffuse component.

While some investigations\cite{matus} have been made in regards to this matter, the subject of cone configurations remains mostly untouched, leaving developers to simply employ whichever distribution works best.

For the purposes of this thesis, all diffuse cones were given an aperture angle of 0.55rad and separated into three sets:

\begin{itemize}
    \item The {\it{front cone}} simply propagates in the direction of the surface normal.
    \item The {\it{intermediate cones}} are a set of four cones with their center axis being at an angle of 45deg to the front cone and an angle of 90 deg to each other.
    \item Similarly, the {\it{side cones}} are orthogonal to each other as well as the fornt cone and at a 45deg angle to the intermediate cones.
\end{itemize}

\begin{figure}[th]
\centering
\makebox[\textwidth][c]{\includegraphics[width=1.3\textwidth]{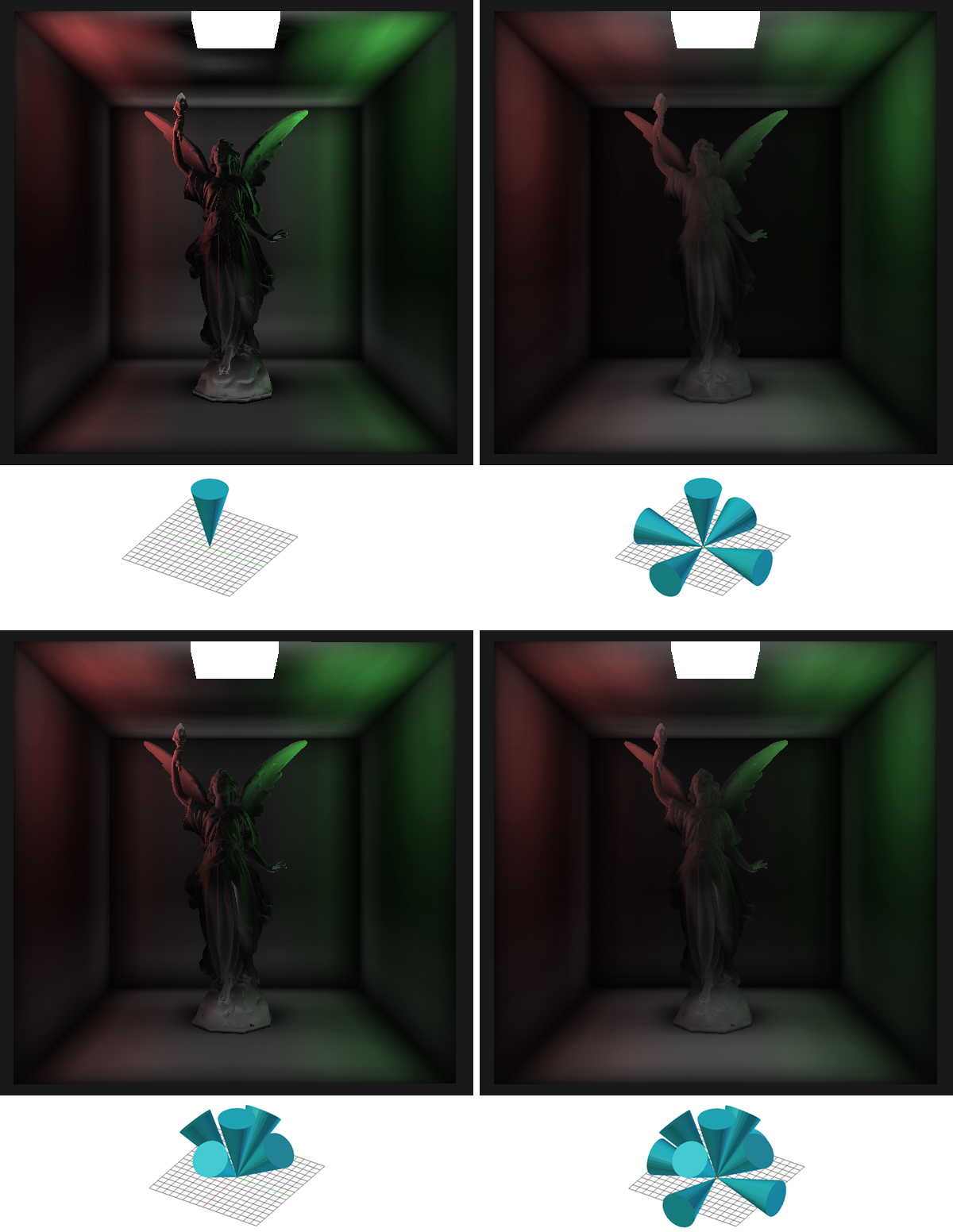}}%
\caption[]{Indirect diffuse components with their respective cone configurations underneath.}
\label{cone_configs}
\end{figure}

As can be seen in fig.\ref{cone_configs}, utilizing only a front cone produces undesired dark smudges and is thus not adequate for the intended purpose.

On the other hand, the result obtained when using a front cone and four side cones causes the side cones to collide with the fragment's own surface, leading to the highly grey areas on the statue seen in the top right image in fig. \ref{cone_configs}.

The most adequate render appears to result either from a combination of intermediate cones and a front cone or a blend of all three sets. Due to the potential increase in computation cost, the presented implementation disables the side cones by default but allows for easy adjustments by including boolean values to toggle any of three sets in \verb|VoxSettings|.

\subsection{Extensions}

\subsubsection{Scene Parser}

In the initial stages of development, the rendered scene (including models, lights, materials etc.) was declared on a code basis as a series of object instances. The unfortunate consequence caused by this approach, is that the entire program needed to be recompiled after any changes to a scene were made.

To rectify this issue a \verb|SceneParser| was introduced, which parses a description of a scene from an independent text file and automatically builds it.

The chosen format for the scene-describing text files resembles C based syntax and is scanned by a lexical analyzer that separates it into a chain of tokens. These are then organized into a syntax tree where every node is either a parent or a primitive datatype.

\begin{figure}[th]
\makebox[\textwidth][c]{\includegraphics[width=1.3\textwidth]{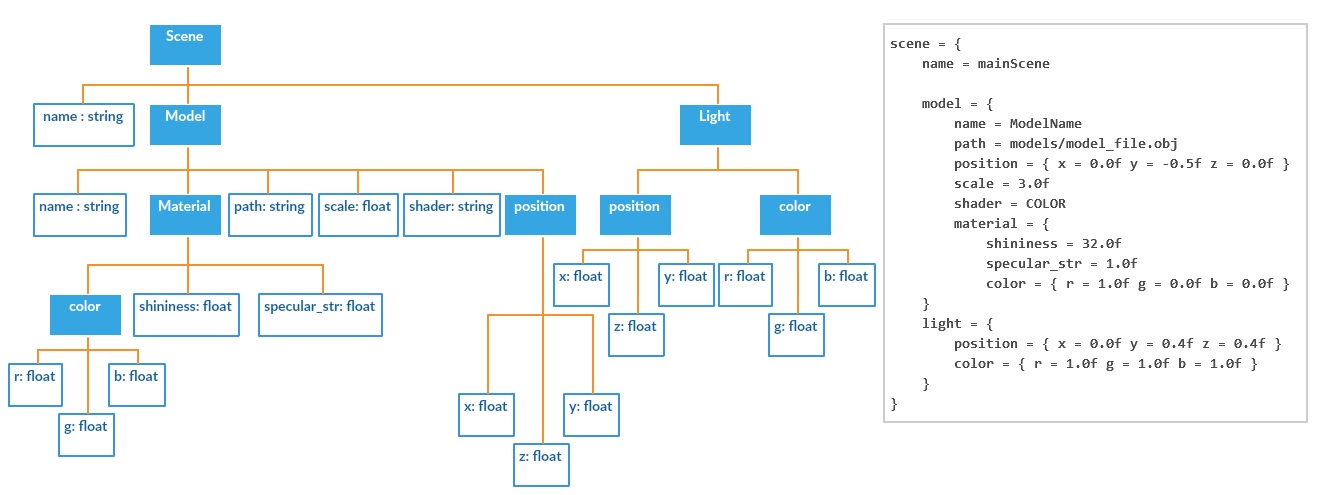}}%
\caption[]{Syntax tree produced by the scene parser (left) and corresponding scene description (right).}
\label{parser}
\end{figure}

An example of a scene description together with the corresponding generated syntax tree are portrayed in fig.\ref{parser}. A regular depth-first traversal of the tree extrapolates all required data and instantiates the listed entities into a scene.

\subsubsection{Voxel Visualization}\label{vox_vis}

The OpenGL \verb|glGetTexImage| function allows image data to be copied back to CPU memory. This allows 3D textures generated by the voxelization pipeline to be analyzed and rendered.

For this purpose, the \verb|VoxelMap| class is equipped with a \verb|retrieveData(int lod)| method which stores the GPU 3D texture into an array in CPU memory.

The \verb|visualize| method, as the name implies, renders every voxel with a non 0 occlusion value onto the screen. This is accomplished by simply re-sizing a cube to the size of a voxel and then placing it at the appropriate location with the corresponding color and then rendering it.

Completing this procedure for every element of a 3D texture is a performance heavy task which causes low frame-rates while voxel visualization is active. (E.g. visualizing a small 64x64x64 3D texture may lead to over 260.000 cubes being placed, colored and rendered individually)

Despite the low frame-rates, visualizing voxelized scenes was highly beneficial towards debugging and comprehension purposes.

\subsubsection{Commandline}

The user can interact with the rendered scene by providing commands through the \verb|std::input| of a command shell.

Running on a separate thread, the \verb|console()| method manipulates mutex-secure parameters which may in turn sway the course of the program respectively.

Approximately 50 different commands are available, each with their respective purposes. A complete list is available on the github repository of the project.

%% file: Chapters/Chapter6.tex

\chapter{Evaluation} 

\label{Chapter6} 

\section{Goal Recapitulation}

Local illumination algorithms describe only how individual surfaces reflect light in conjunction with a material description and light-source positions.

For more accurate images, global illumination algorithms need to be employed. Unfortunately, these take into account the ways in which light is transferred between surfaces and thus convert every potential surface into a secondary light source.

The vast amount of recursive calculations required to compute indirect lighting values is difficult to accomplish with interactive frame-rates. Applying generous approximations to the rendering equation speeds up the calculation process but also impairs the physical fidelity of the image.

The objective at hand is to find an algorithm which maximizes the possible fidelity while maintaining a reasonable complexity.

\section{Voxel Cone Tracing}

\subsection{Summary}

Chapter \ref{Chapter4} presented the rendering algorithm of voxel cone tracing while chapter \ref{Chapter5} provided insight into an actual implementation of it.

The proposed algorithm computes indirect diffuse, indirect specular and occlusion values in addition to the regular direct light components. This is accomplished by tracing a series cones which use texture filtering methods on a voxelized scene to extrapolate a rough estimate of indirect light.

\begin{figure}[th]
\centering
\includegraphics[scale=0.45]{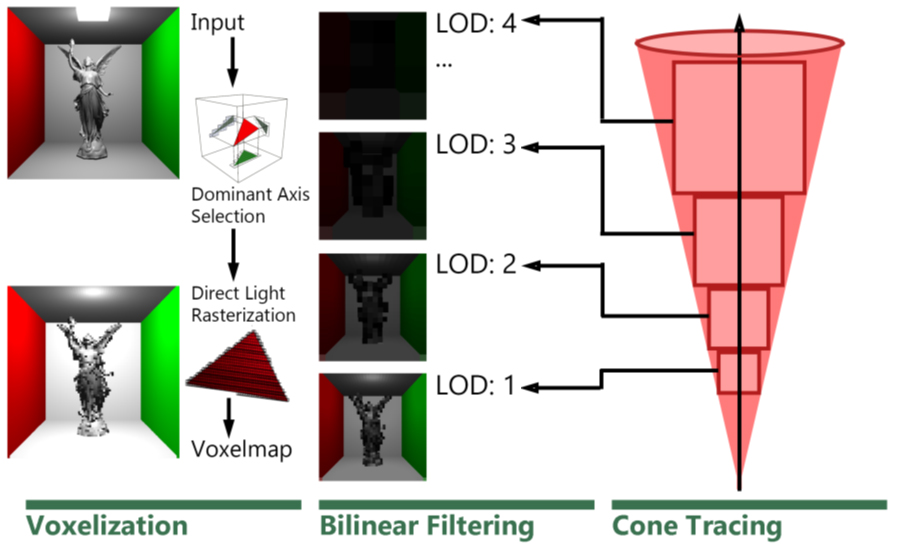}
\decoRule
\caption[]{Voxel cone tracing summary}
\end{figure}

\subsection{Algorithm Limitations}

Voxel cone tracing suffers from several limitations as a result from the procured simplifications. Namely, the reliance on hardware acceleration for fast voxelization and cone traces cause poor performance on systems with low-end or integrated GPUs.

Furthermore, the variant presented in this thesis is only capable of computing one bounce of indirect light. Incorporating additional bounces would require a further 3D texture including surface normal vectors as its values with recursive cones being traced in respect to the average surface normal hit.

This would, however, further increase GPU memory requirements which are already significant if no octreee is employed.

In addition, the presented core concepts of voxelization and mipmapped cone tracing suffer from inaccuracies with respect to the theoretical overview:

\begin{itemize}
    \item {\textbf{Power of two constraint}}: Approximating a cone with a volume ray march already poses some geometric simplification, which is further enhanced by the simple fact that the radii of samples taken are constrained by the power-of-two size increments of mipmaps.
    
    As a result, the dimensions of each sample along a cone are either equal or double the size of their predecessor, causing either superfluous volume to be sampled or parts of the cone left un-sampled.
        \begin{figure}[th]
        \centering
        \includegraphics[scale=0.52]{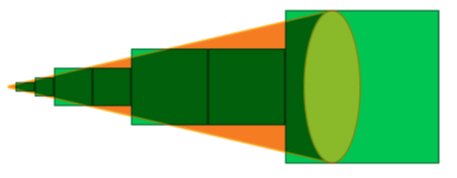}
        \decoRule
        \caption[]{Inaccuracies resulting from the mipmaps power of two constraint.}
        \end{figure}
    \item {\textbf{Color leaks}}: Some negligible lighting inaccuracies may occur in consequence of the nature of bilinear filtering, with certain occluded colors leaking through their respective occluders.
    
    In fig. \ref{greenbunny} a green bunny is hidden behind a red wall. However, higher LOD values will cause the green voxels to merge with the red ones, potentially leading a cone travelling in this direction to pick up on green color even though none is visible.
        \begin{figure}[th]
        \centering
        \includegraphics[scale=0.65]{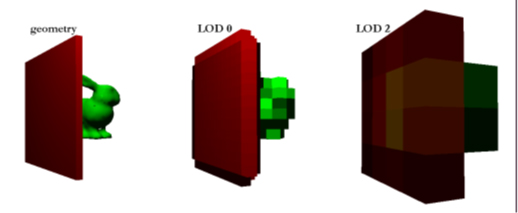}
        \decoRule
        \caption[]{Occluded color leaks due to mipmapping.}
        \label{greenbunny}
        \end{figure}
    \item {\textbf{Offset and self-collision}}: Cone traces are launched from the location of the fragment but are conducted on a 3D texture. A voxel may find itself occupying more space than just the corresponding fragment which may lead the cone to immediately collide with its own starting position.
    
    To counteract this, an offset to the cone's origin of at least a voxel diameter is typically employed. As a result, mutually facing surfaces lying very close to each other may cause cone-traces to originate from behind the other surface. In fig. \ref{specproblems} the left-side spherical cap is displayed as completely black due to cones being traced from behind the red wall.
    \item {\textbf{Specular reflection inaccuracies}}: The specular component measured via voxel cone traces only displays a reflection of the scene's direct light. As a result self-occluded, dark corners may appear as bright in a specular reflection. Additionally, detailed geometry may become lost due to only the voxel map being reflected.
    
    Voxel cone tracing can provide specular reflections that are quite believable on first sight, but upon further inspection fail to portray the surrounding environment. As a result, the specular component might best be reserved for specular highlights instead of genuine mirror images, which are better achieved with techniques such as cubemapping.
        \begin{figure}[th]
        \centering
        \includegraphics[scale=0.75]{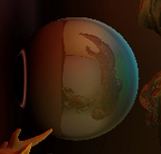}
        \decoRule
        \caption[]{Inaccurate specular reflections and cone-offset problematics.}
        \label{specproblems}
        \end{figure}
\end{itemize}

\subsection{Global Illumination Algorithm Comparison}

While raytracing rules supreme in the realm of photo-realism, it is also conceivably one of the slowest techniques available. A compelling solution to this problem is to simply perform all calculations in advance and store them in lighting textures, as is done in the radiosity method. Although otherwise providing great physical fidelity at high frame-rates, this approach proves to be inadequate for specular reflections and dynamic scenes, as all lighting is computed independently of the viewer's perspective.

The voxel cone tracing approach attempts to merge both methods into one by utilizing pre-computed light maps and tracing large bundles of rays (cones) into the directions of potential incident light.

Since a cone trace requires significantly less computation time and the pre-computed lightmaps can be generated at lightning speeds, voxel cone tracing appears to outclass raytracing in terms of performance while also solving the dynamic scene problem of radiosity.

However, radiosity vastly outperforms voxel cone tracing in terms of computation time and raytracing supersedes it in photo-realism. Providing a fine balance between the two, voxel cone tracing is best employed for rendering highly dynamic scenes.

This begs the question whether if a combination of radiosity for static objects and voxel cone tracing for dynamic ones would provide best of both worlds.

\begin{figure}[th]
\centering
\makebox[\textwidth][c]{\includegraphics[width=1.35\textwidth]{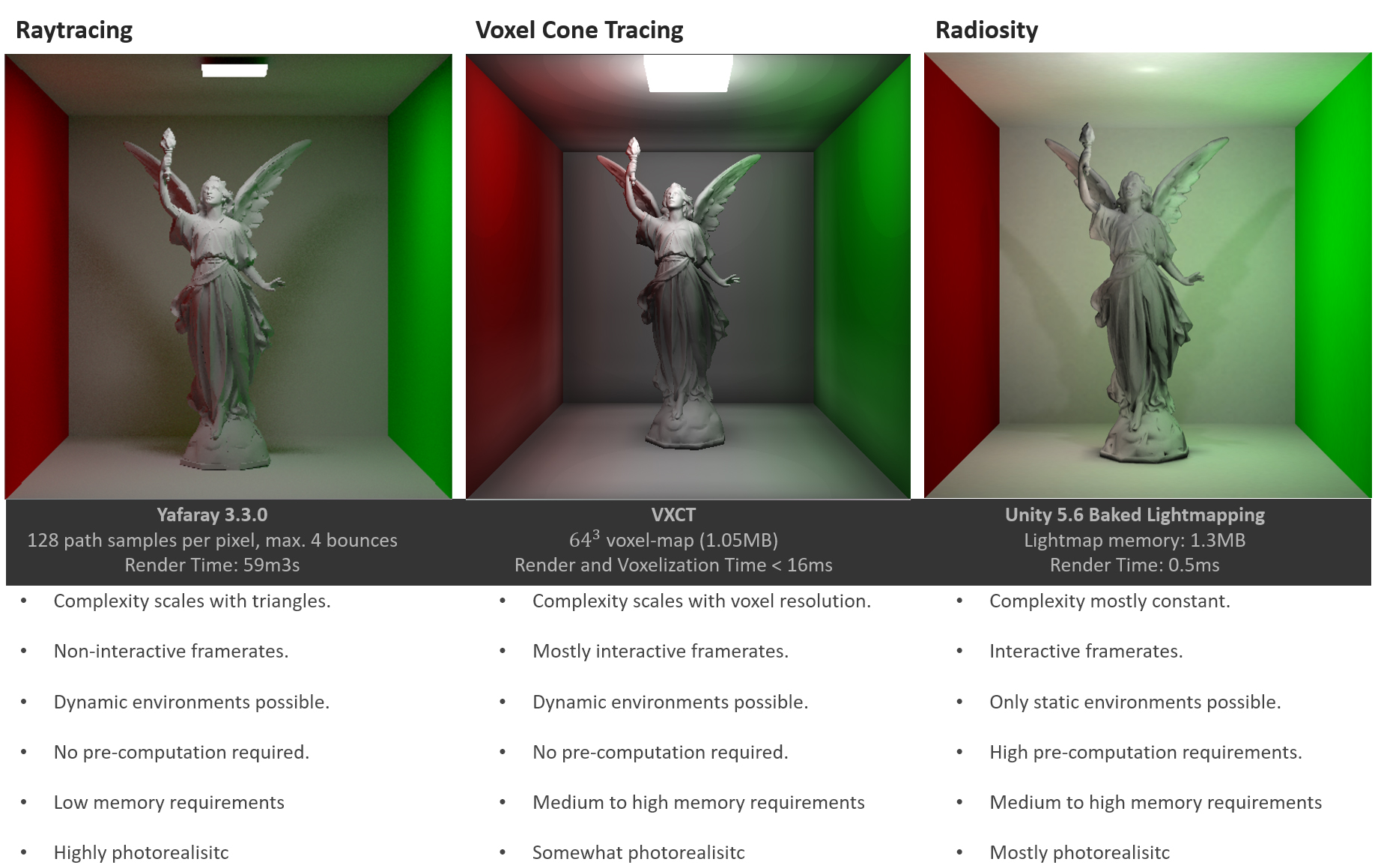}}
\caption[]{Global illumination algorithm comparison}
\end{figure}

\subsection{Possibilities for Expansion}

The most well-defined part of voxel cone tracing is the process of voxelization, which has merely been scraped in the contents of this thesis. A great many deal of advanced techniques such as conservative rasterization can provide improved results from the ones produced here.

On the other hand, the cone-tracing process itself follows a rough specification but still leaves many doors open. These include but are not limited to:

\begin{itemize}
    \item Investigating optimal cone configurations on a mathematical basis (which directions and aperture angles best approximate a hemisphere) as well as through rudimentary testing.
    \item Attempting to generate phenomena usually computed via raytracing like caustics or refraction by using cone traces with specialized BRDFs.
    \item Including more data in cone traces by inserting indirect light, or even raytraced values, directly into an additional 3D texture, similarly to deferred voxel cone tracing\cite{villegas}.
\end{itemize}

\section{Implementation Results}

\subsection{Encountered Problems}

Implementing voxel cone tracing as an OpenGL based shader pipeline posed many uncertainties which had to be overcome. Ultimately, two resulting problems are of particular importance:

\subsubsection{Floating Point Voxelization Inaccuracy}

A voxel's spatial coordinates are constrained by the resolution of the respective 3D texture but the coordinates of individual vertices are not.

\begin{figure}[th]
\centering
\includegraphics[scale=0.4]{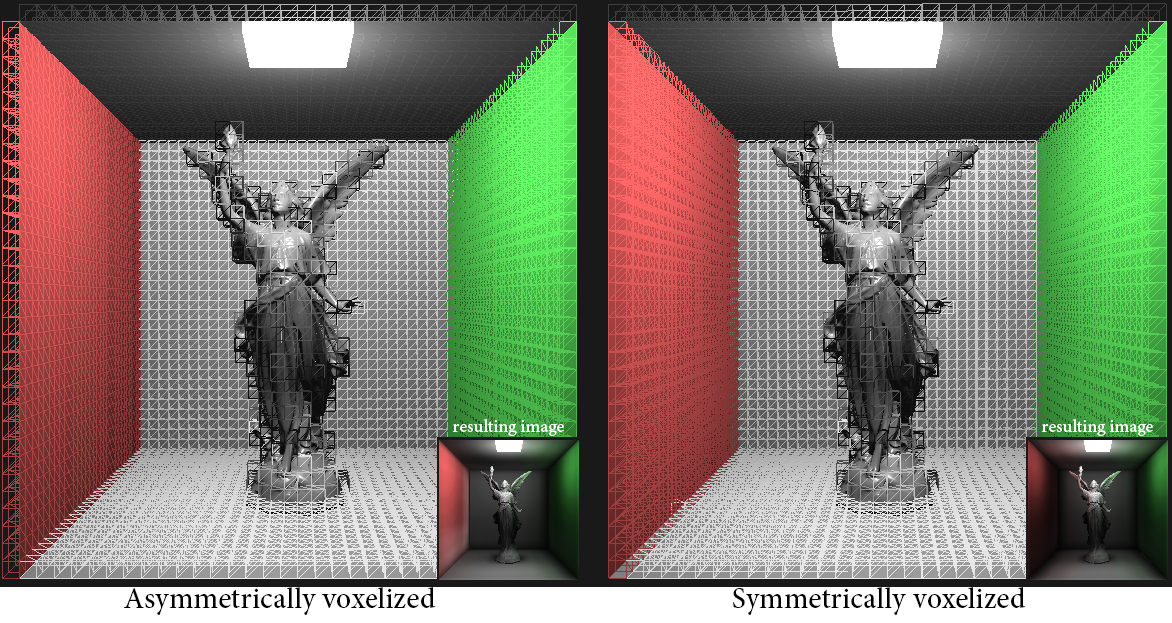}
\decoRule
\caption[]{Inconsistent voxelization due to floating point coordinates. Note how the voxelmap overlay on the red wall shifts due to minimal changes in the wall's position, leading to large scale changes in the resulting lighting.}
\label{flaot_error}
\end{figure}

Say, for instance that the coordinates of a wall are given as \verb|4.9999f| instead of a rounded \verb|5.0f|, which is a common occurrence in 3D model data. As can be observed in fig.\ref{flaot_error}, the voxels corresponding to the surface, shift from location to another.

The cone traces, however, are still performed from the object's surface, not the voxel. This results in the left wall having an inherent offset built in for cone-traces while the right one does not, leading to an unfortunate difference in ambient occlusion.

Whether if the severity of this problem can be mitigated by employing improved voxelization techniques or storage methods such as octrees is unknown.

\subsubsection{Non-smooth Circular Degradation}

An unexplained phenomenon specific to the here presented implementation came in the form of an unusual, wave-like degradation occurring only in certain constellations of objects and light-sources.

Upon further investigation it became clear, that these circles were caused by the occlusion as they were only visible in the occlusion component.

Unfortunately, the exact cause is still unknown and problem remains unsolved.

\begin{figure}[th]
\centering
\includegraphics[scale=1.0]{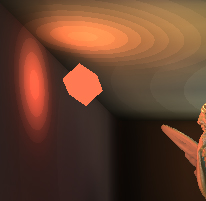}
\decoRule
\caption[]{Lighing attenuating in a wave-like fashion.}
\end{figure}

\subsection{Performance}

The OpenGL \verb|glFinish| function blocks the current CPU process until all GL functions are complete and thus allows for a rough performance measurement to be made.

A high-polygon scene with multiple light-sources (\verb|scene_complex.txt| on the github repository) was rendered using three different sizes of 3D textures.

Below are the recorded averages over 100 consecutive frames for time required per frame:

\begin{table}[h!]
\centering
 \begin{tabular}{||c c ||} 
 \hline
 voxel map size & avg. frametime \\ [0.5ex] 
 \hline\hline
 64x64x64 & 0.01666s  \\ 
 128x128x128 & 0.0313s  \\ 
 256x256x256 & 0.1274s  \\ 
 \hline
\end{tabular}
\caption{Avg. frametime over 100 recorded frames without continuous revoxelization.}
\end{table}

Additionally, the same tests were repeated with \verb|vox_freq| set to 0, meaning that the scene is voxelized anew with every frame.

\begin{table}[h!]
\centering
 \begin{tabular}{||c c ||} 
 \hline
 voxel map size & avg. frametime \\ [0.5ex] 
 \hline\hline
 64x64x64 & 0.02039s  \\ 
 128x128x128 & 0.06301s  \\ 
 256x256x256 & 0.33401s  \\ 
 \hline
\end{tabular}
\caption{Avg. frametime over 100 recorded frames with continuous revoxelization enabled.}
\end{table}

Of the resulting frame-rates the only one bordering on unusable is a 256-sized voxelMap with continuous voxelization enabled.

Note that a repeated voxelization on every frame is somewhat excessive as lower frequencies tend to suffice for reasonable results.

\section{Verdict}

In conclusion, being the newcomer of global illumination algorithms, voxel cone tracing provides highly promising results. The voxelization and cone-tracing procedures both make great use of hardware-acceleration enabling a modestly complex algorithm to be executed with a considerable performance.

However, the lack of large industry-based applications of this method call into question whether if the performance can truly compete with well established techniques such as cubemapping, radiosity or raytracing.

Mediocre frame-rates combined with high memory requirements as well as geometric incongruities have left the algorithm to be discarded by real-time rendering engines such as Unreal Engine or Unity.

Further inquiries into the potential improvements and applications of voxel cone tracing are necessary.

\begin{figure}[th]
\centering
\makebox[\textwidth][c]{\includegraphics[width=1.3\textwidth]{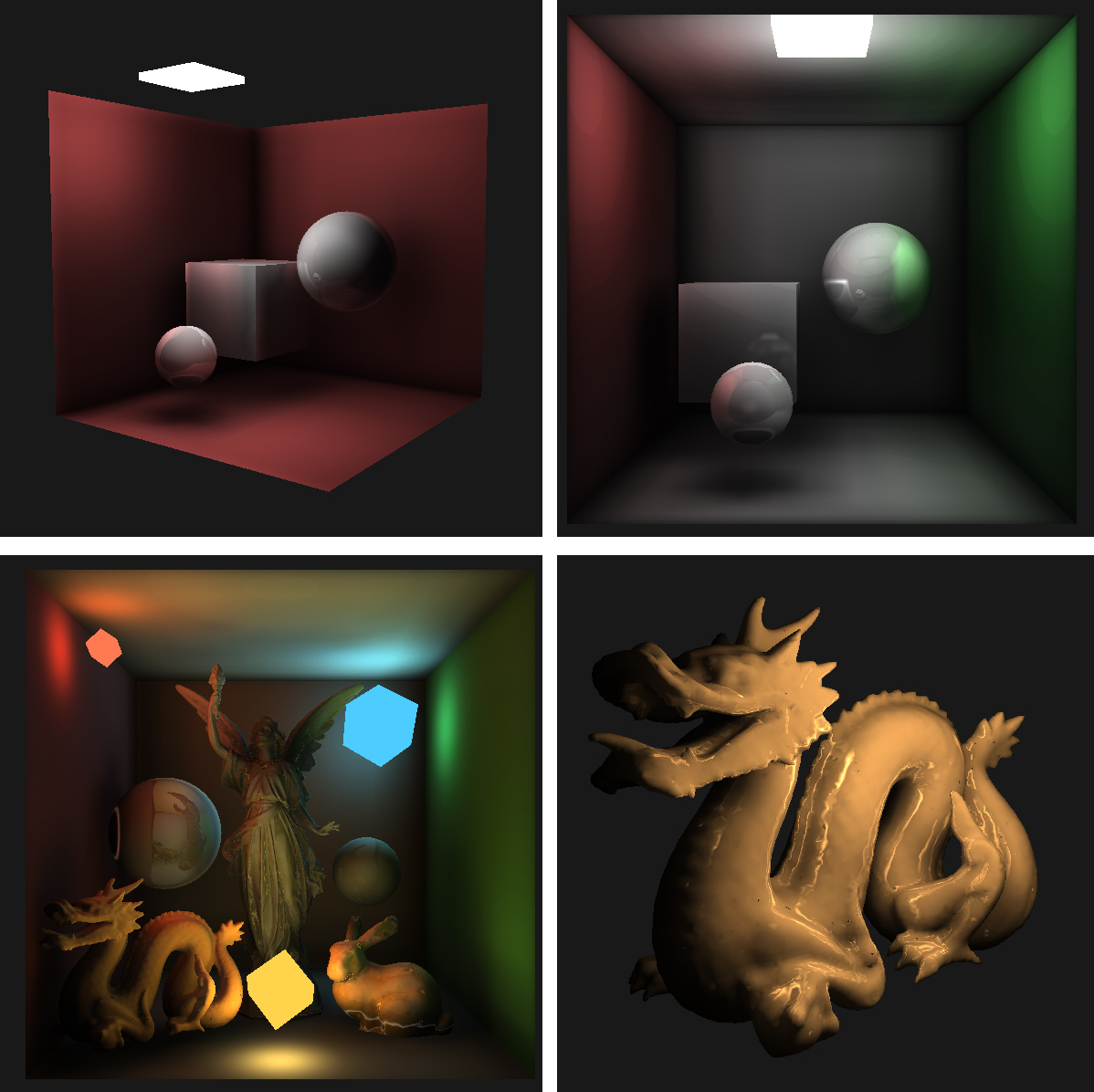}}
\caption[]{Examples of pictures rendered with the presented implementation.}
\end{figure}